\documentclass[11pt,a4paper]{article}
\pdfoutput=1
\usepackage{jcappub}
\usepackage{url}

\usepackage{amsfonts,amsmath,amssymb}

\allowdisplaybreaks

\def\beq{\begin{equation}}
\def\eeq{\end{equation}}
\def\Lcdm{\Lambda {\rm CDM}}
\def\Mpl{M_{\rm pl}}
\def\H0{\mathcal{H}_{0}}

\def\dpi{\pi'}
\def\ddpi{\pi''}
\def\pif{{\pi^{(1)}}}

\def\pis{{\pi^{(2)}}}

\def\dphi{\phi'}
\def\ddphi{\phi''}
\def\dphif{{\phi^{(1)}}'}
\def\ddphif{{\phi^{(1)}}''}

\def\om{\omega}

\def\domf{{\omega^{(1)}}'}

\def\a{\alpha}
\def\b{\beta}
\def\g{\gamma}

\def\lcdm{\Lambda\textrm{CDM}}

\title{Matter bispectrum in cubic Galileon cosmologies}
\author[a,b]{Nicola Bartolo,}
\author[a,b]{Emilio Bellini,}
\author[c]{Daniele Bertacca,}
\author[a,b]{and Sabino Matarrese}

\affiliation[a]{Dipartimento di Fisica e Astronomia ``G. Galilei'', Universit\`a degli Studi di Padova, via Marzolo 8, I-35131, Padova, Italy}
\affiliation[b]{INFN, Sezione di Padova, via Marzolo 8, I-35131, Padova, Italy}
\affiliation[c]{Physics Department, University of the Western Cape, Cape Town 7535, South Africa}

\emailAdd{nicola.bartolo@pd.infn.it}
\emailAdd{emilio.bellini@pd.infn.it}
\emailAdd{daniele.bertacca@gmail.com}
\emailAdd{sabino.matarrese@pd.infn.it}

\begin{document}

\abstract{In this paper we obtain the bispectrum of dark matter density perturbations in the frame of covariant cubic Galileon theories. This result is obtained by means of a semi-analytic approach to second-order 
perturbations in
Galileon cosmologies, assuming Gaussian initial conditions. In particular, we show that, even in the presence of large deviations of the linear growth-rate w.r.t.\ the $\lcdm$ one, at the bispectrum level such deviations are reduced to a few percent.}


\maketitle

\section{Introduction}

Several observations, such as those referring to the magnitude-redshift relation for type-Ia Supernovae (SNIa) \cite{Perlmutter:1998np,Riess:1998cb}, Cosmic Microwave Background (CMB) temperature anisotropies 
\cite{Spergel:2003cb,Komatsu:2010fb,Hinshaw:2012fq} and Baryonic Acoustic Oscillations (BAO) features in galaxy clustering \cite{Eisenstein:2005su,Percival:2009xn}, suggest that the universe is currently 
undergoing an accelerated expansion phase, caused by the presence of a positive cosmological constant or 
a more general Dark Energy (DE) component or a suitable modified gravity model. Assuming that the matter distribution is dominated by Cold Dark Matter (CDM), the simplest model that reproduces this effect and fits present data is the so-called $\Lcdm$ one, based on the existence of a cosmological constant term that fills the gap between the matter energy density and the critical one. 
Even though the presence of a cosmological constant term $\Lambda$ is fully consistent with General Relativity, its value appears too small to be explained by fundamental physics \cite{Weinberg:2000yb}. 
Consequently, alternative models have been explored such as, for instance, quintessence \cite{Caldwell:1997ii,Zlatev:1998tr}, $f(R)$ \cite{DeFelice:2010aj}, massive 
gravity \cite{Hinterbichler:2011tt} (see \cite{Tsujikawa:2010sc} and references therein).

In this paper we focus on a modified gravity model obtained in the context of the Galileon scalar-tensor theory \cite{Nicolis:2008in}. This theory is obtained by taking the decoupling limit of the Dvali-Gabadadze-Porratti model (DGP) \cite{Dvali:2000hr}. 
The Galileon is the most general theory containing second-order derivatives in the scalar field with some properties: specifically, in a flat space-time it preserves the Galilean shift symmetry ($\partial_\mu\pi\rightarrow\partial_\mu\pi+b_\mu$), while in curved space-times, adding suitable coupling terms between gravity and the Galileon field \cite{Deffayet:2009mn,Deffayet:2009wt}, the model avoids the Ostrogradski instability \cite{Ostrogradski:1850in}. In addition, on non-linear scales the self-interactions of the Galileon field screen the fifth force through the Vainshtein mechanism \cite{Vainshtein:1972sx}. The essence of this mechanism lies in the non-standard kinetic terms 
(i.e.\ $\Box\pi \partial_\mu\pi\partial^\mu\pi$), which decouple the scalar field from gravity at small scales ($r\ll r_V$, where $r_V$ is a characteristic scale around a matter source, named ``Vainshtein radius''). Many literature has recently appeared on these models and their generalizations \cite{deRham:2010eu,Deffayet:2010zh,Padilla:2010de,VanAcoleyen:2011mj,Khoury:2011da,Deffayet:2011gz,Burrage:2011bt,Liu:2011ns,deRham:2011by,Germani:2011bc,deRham:2012az}. 
Galileon models have been extensively studied at late-times \cite{Chow:2009fm,Silva:2009km,Kobayashi:2009wr,Kobayashi:2010wa,Gannouji:2010au,DeFelice:2010pv,Ali:2010gr,DeFelice:2010nf,DeFelice:2010as,Kaloper:2011qc,Appleby:2011aa,Bellini:2012qn,Barreira:2012kk,Curtright:2012ph,Leon:2012mt,Babichev:2012re}, during inflation \cite{Creminelli:2010ba,Kobayashi:2010cm,Mizuno:2010ag,Burrage:2010cu,Gao:2011qe,Wang:2012bq,Ohashi:2012wf,Creminelli:2012my,Kamada:2010qe,Kamada:2012se}, and a subclass of these models has been already compared by observations \cite{Nesseris:2010pc,Appleby:2012ba,Okada:2012mn}.

The main goal of this paper is to study the late-time non-Gaussianities (NG) of the matter distribution arising from gravitational instability in the cubic covariant Galileon theory. 
It is well known that NG can be classified in primordial and late-time. The primordial ones come from non-linearities encoded in the inflationary perturbations \cite{Bartolo:2004if}; these are imprinted in the CMB and in the Large-Scale Structure (LSS) of the universe \cite{Bartolo:2005xa,Sefusatti:2006pa,Bartolo:2009rb,Sefusatti:2010ee,Liguori:2010hx,Verde:2010wp,Figueroa:2012ws}, and should be constrained by present and future surveys \cite{Planck,Amendola:2012ys}. The late-time non-Gaussianity in the LSS is generated classically by gravitational instability, when cosmological perturbations enter non-linear scales. While a Gaussian universe can be completely described by the power-spectrum, the deviations from Gaussianity are encoded in higher-order statistics, such as the bispectrum and the trispectrum \cite{Peebles1980,Bernardeau:2001qr}. 

The interest in studying the dark matter bispectrum in the Galileon model comes from the possibility to measure the signature of modifications from standard gravity.\footnote{For other works on the dark matter bispectrum within other modified gravity models see~\cite{Borisov:2008xn,Tatekawa:2008bw,Bernardeau:2011sf,GilMarin:2011xq}.} If this is the case, the bispectrum can be used to lift degeneracies among different models giving rise to the same observed power spectrum and the same background cosmology.
We choose Gaussian initial conditions, in order to extract only the late-time non-Gaussianity.
In particular, we will focus on the dark matter bispectrum calculated at tree-level (second-order perturbations), since it gives the leading contribution in the weakly non-linear regime.
Even though we consider models with important modifications in the background and in the growth rate w.r.t.\ $\lcdm$, we will show that the matter bispectrum deviations that we obtain are less than $5\%$. 
We think that this suppression is connected with a compensation effect when the equation of state is $w\lesssim-0.8$. Our results are obtained by using a semi-analytic technique both at first and second-order in perturbations.

The paper is organized as follows. In Sec.\ \ref{SEC:Action} we introduce the action for the Galileon model we focus on and the resulting equations of motion. In Sec.\ \ref{SEC:Background} we solve the equations of motion in a Friedmann-Lema\^{i}tre-Robertson-Walker (FLRW) universe. In Sec.\ \ref{SEC:Perturbations} we introduce the setup used to calculate the perturbed equations. In Sec.\ \ref{SEC:LinearTheory} we analyze the first-order field equations. We also find two integral solutions for the growing  and the decaying modes that simplify many analytic calculations. In Sec.\ \ref{SEC:second} second-order perturbations are considered. In Sec.\ \ref{SEC:bispectrum} we compute the power-spectrum and the 
bispectrum for the DM perturbations. In Sec.\ \ref{SEC:Conclusions} we draw our conclusions and provide some comments. In Appendix \ref{SEC:AppendixB} we derive an equation for the first-order DM perturbations in different gauges. 
In Appendix \ref{SEC:AppendixA} we provide the source terms of the second-order field equations.
In Appendix \ref{SEC:AppendixC} we show the coefficients of the kernel for the second-order DM fluctuations. 

Throughout the paper we adopt units $c = \hbar= G = 1$, except where explicitly indicated; our signature is $(-, +, +, +)$. Greek indices run over $\{0, 1, 2, 3\}$, denoting space-time coordinates, whereas Latin indices run over $\{1, 2, 3\}$, labelling spatial coordinates.

\section{Action and field equations}\label{SEC:Action}

In this paper we study the simplest form of the covariant Galileon model with a potential, which reads \cite{Nicolis:2008in,Deffayet:2009mn}
\beq\label{EQ:Action}
S=\frac{1}{2}\int {\rm d}^4 x \sqrt{-g}\,\left[ \Mpl^2 R+ 2{\cal L}_{\pi} + 2 {\cal L}_{M}\right]\,,
\eeq
where
\beq
{\cal L}_{\pi}\equiv \frac{c_1 M^3}{2} \pi +\frac{c_2}{2} (\nabla \pi)^2 +\frac{c_3}{2 M^3}(\square \pi) (\nabla \pi)^2\,.
\eeq
Here, $c_i$ are dimensionless constants, $\Mpl$ is the reduced Planck mass and $M$ is a constant with dimensions of mass. ${\cal L}_{M}$ is the Lagrangian of a pressure-less perfect fluid (DM) with density $\rho$ and four-velocity $u^\mu$.

Varying Eq.\ (\ref{EQ:Action}) w.r.t.\ the metric $g_{\mu \nu}$ we obtain the Einstein equations
\beq \label{EQ:Einstein}
G_{\mu\nu}=\Mpl^{-2}\left[T^{(m)}_{\phantom{(m)}\mu\nu}+T^{(\pi)}_{\phantom{(\pi)}\mu\nu}\right] \,,
\eeq
where
\begin{align}
T^{(m)}_{\phantom{(m)}\mu\nu}=&\rho_m u_\mu u_\nu\,,\nonumber\\
T^{(\pi)}_{\phantom{(m)}\mu\nu}=&\frac{c_1 M^3}{2} g_{\mu\nu}\,\pi-c_2\pi_{;\mu}\, \pi_{;\nu}+\frac{c_2}{2} g_{\mu\nu}\, (\nabla \pi)^2\nonumber\\
&-\frac{c_3}{M^3}\left[\pi_{;\mu}\,\pi_{;\nu}\, \square\pi-\pi_{;\{\mu}\,\pi_{;\nu\}\a}\,\pi^{;\a}+g_{\mu\nu}\,\pi^{;\a}\,\pi_{;\a\b}\,\pi^{;\b}\right]\,.
\end{align}
The equation of motion for the scalar field $\pi$ is
\beq \label{EQ:Galileon}
\frac{c_1 M^3}{2}-c_2\square\pi+\frac{c_3}{M^3} \left[-(\square\pi)^2 +R_{\mu\nu}\, \pi^{;\mu}\, \pi^{;\nu} +\pi_{;\mu\nu}\, \pi^{;\mu\nu} \right]=0\,.
\eeq
The stress-energy tensor continuity equation for the DM component, reads
\beq\label{EQ:StressEnergyConservation}
\nabla_\mu T^{(m)\mu}_{\phantom{(m)\mu}\nu}=0\,.
\eeq

\section{Background evolution}\label{SEC:Background}

From Eqs.\ (\ref{EQ:Einstein}) and (\ref{EQ:Galileon}) we can study the background evolution in an expanding FLRW universe with scale-factor $a(\tau)$,
\begin{align}
ds^{2}={a(\tau)}^2\left[-d\tau^{2} +\delta_{ij}dx^i dx^j\right]\,,
\end{align}
where $\tau$ is the conformal time. Let $\pi\equiv\pi(\tau)$ be the Galileon field 
at the background level and $\rho_m(\tau)$ and $\rho_\pi(\tau)$ the background matter and the Galileon energy density respectively. The $(0,0)$ and the $(i,i)$ components of the Einstein equations read
\begin{align}
\frac{3\Mpl^2 \mathcal{H}^2}{a^2}&=\rho_m+\rho_{\rm \pi}\,,\label{EQ:Friedmann1}\\
\frac{\Mpl^2}{a^2}\left(\mathcal{H}^2+2\mathcal{H}'\right)&=-p_\pi\,,\label{EQ:Friedmann2}
\end{align}
where $\mathcal{H}\equiv a'(\tau)/a(\tau)$ is the Hubble parameter, primes represent derivatives w.r.t.\ the conformal time $\tau$ and
\begin{align}
\rho_{\rm \pi} \equiv& -\frac{c_1 M^3}{2}\pi-\frac{c_2}{2 a^2}\dpi^2+\frac{3c_3}{M^3 a^4}\mathcal{H} \dpi^3\,,\\
p_\pi \equiv& \frac{c_1 M^3}{2}\pi-\frac{c_2}{2 a^2}\dpi^2-\frac{c_3}{M^3 a^4}\dpi^2\left(\ddpi -\mathcal{H}\dpi\right)\,,
\end{align}
are the scalar field density and pressure, respectively. The equation of motion for the Galileon, Eq.\ (\ref{EQ:Galileon}), becomes
\begin{align}
&\frac{c_1 M^3}{2}+\frac{c_2}{a^2}\left[\ddpi+2 \mathcal{H} \dpi\right]-\frac{3c_3}{M^3 a^4}\dpi \left[2 \mathcal{H} \ddpi+ \mathcal{H}'\dpi\right]=0\,,\label{EQ:BackgroundGalileon}
\end{align}
where, without loss of generality, we have defined $M^3\equiv\Mpl \H0^2$. Here $\H0$ is the value of the Hubble parameter $\mathcal{H}(\tau)$ in a FRLW universe today.

\begin{figure}[!ht] %  figure placement: here, top, bottom, or page
   %\centering
\includegraphics[width=0.5\textwidth]{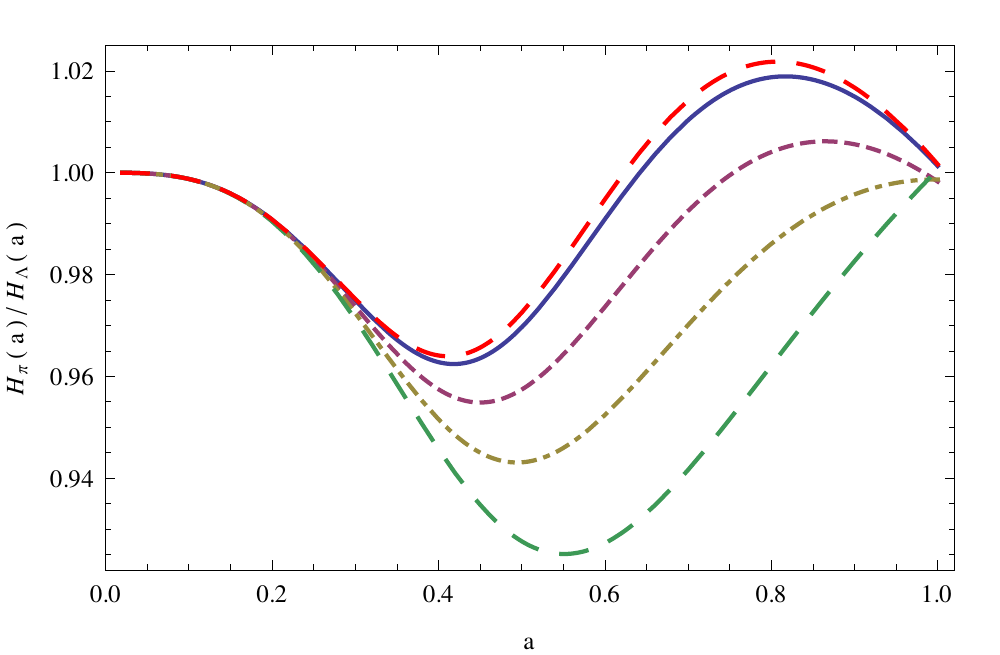} \includegraphics[width=0.5\textwidth]{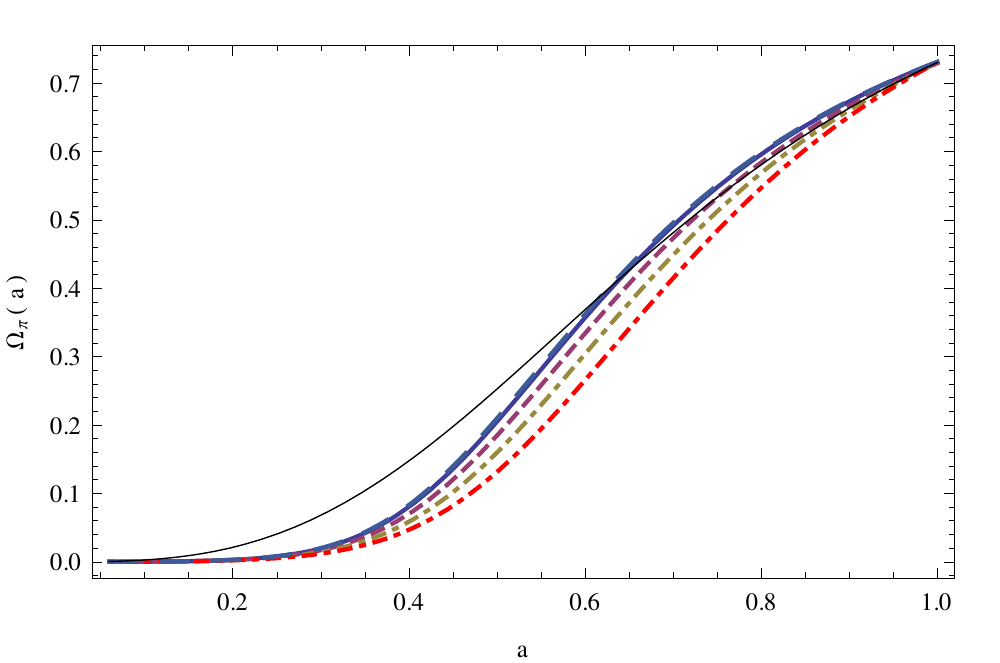}\\ \includegraphics[width=0.5\textwidth]{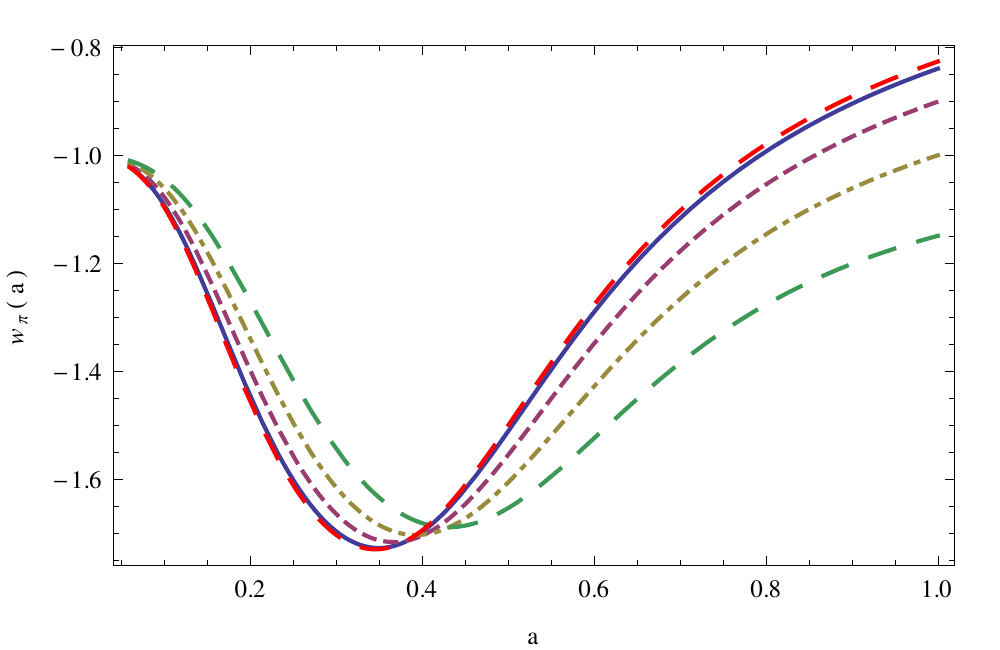}
\caption{Background evolution of the Galileon model. In the top panels we plot the deviations of the Hubble parameter w.r.t.\ the Hubble parameter in $\Lcdm$ and the evolution of the DE density. In the bottom panel we plot $w_\pi(a)\equiv\rho_\pi(a)/p_\pi(a)$. The parameter values are: $c_1=1.6$, $c_2=0.04$, $c_3=10^{-3}$ (red line); $c_1=1.5$, $c_2=0.04$, $c_3=10^{-3}$ (blue line); $c_1=11$, $c_2=3.8$, $c_3=1$ (purple line); $c_1=6$, $c_2=3.6$, $c_3=1$ (yellow line); $c_1=10^{-4}$, $c_2=3.3$, $c_3=1$ (green line); $\Lcdm$ (black line).}      \label{FIG:background}
\end{figure}

We have studied the background evolution solving Eqs.\ (\ref{EQ:Friedmann2}) and (\ref{EQ:BackgroundGalileon}). Following \cite{Appleby:2011aa}, we have chosen a parameter region in which ghosts and Laplace instabilities are avoided. The initial conditions are determined fixing a negligible initial vacuum energy density $\rho_\pi(a_i)$ w.r.t.\ $\rho_m(a_i)$ and using the background equations. In particular in this regime Eq.\ (\ref{EQ:Friedmann1}) reduces to $\mathcal{H}^2\propto a^{-1}$. In our analysis we have chosen $a_i=10^{-5}$ and $\rho_\pi(a_i)= 10^{-5} \rho_m(a_i)$ as in \cite{Appleby:2011aa}. To fix $\pi(a_i)$ and $\pi'(a_i)$ we have used this condition and Eqs.\ (\ref{EQ:Friedmann2}) and (\ref{EQ:BackgroundGalileon}) properly combined in order to eliminate $\pi''$. The DM and the DE energy densities today are $\Omega_m(\tau_0)\equiv\rho_m(\tau_0)/(3 \Mpl^2 \H0^2)=0.27$ 
and $\Omega_\pi(\tau_0)\equiv\rho_\pi(\tau_0)/(3 \Mpl^2 \H0^2)=0.73$ respectively \cite{Komatsu:2010fb}. To reach these values today we tuned the parameter $c_2$ by a trial and error approach. We take into account the parameter $c_1\neq 0$, which is the most general potential term preserving the Galilean shift symmetry. It acts as a cosmological constant in the case $\dpi\rightarrow 0$.

In Fig.\ \ref{FIG:background} we show the evolution of $\mathcal{H}(a)$, $\Omega_\pi(a)$  and the equation of state $w_\pi(a)\equiv\rho_\pi(a)/p_\pi(a)$ for the models we are considering. In the limit $c_1\rightarrow 0$ (green line) we have noted that the evolution of the background is $c_2$ and $c_3$ independent~\cite{DeFelice:2010pv}. This behavior is expected because if $c_2$ (or $c_3$) is absorbed through a redefinition of the Galileon field, $c_3$ (or $c_2$) is constrained by the condition $\Omega_\pi(\tau_0)=0.73$. In order to have a free parameter that allows to decrease the difference between $\Lcdm$ and our Galileon models it is crucial to impose $c_3\sim c_1\neq 0$.

\section{Cosmological perturbations}\label{SEC:Perturbations}

In this section we give some definitions needed to analyze the evolution of the DM perturbations on sub-horizon scales \cite{Matarrese:1997ay}. Without choosing any gauge the metric can be written as
\beq
ds^{2}={a(\tau)}^2\left[-(1+2\psi)d\tau^{2}+2\hat{\omega}_i dx^i d\tau+\left[(1-2\phi)\delta_{ij}+\hat{\chi}_{ij}\right]dx^i dx^j\right]\,.
\eeq
Here the dependence of all the perturbations on both the conformal time $\tau$ and the spatial coordinates $\vec{x}$ is implicit. The symmetric trace-free perturbation $\hat{\chi}_{ij}$ and $\hat{\omega}_i$ can be decomposed as
\begin{align}
\hat{\omega}_i \equiv& \omega_i+\partial_i \omega\,,\\
\hat{\chi}_{ij} \equiv& \chi_{ij}+\partial_i\chi_j+\partial_j\chi_i+D_{ij}\chi\,,
\end{align}
where $\omega_i$ and $\chi_i$ are transverse vectors (i.e.\ $\delta^{ij}\partial_i\omega_j=0$), $\chi_{ij}$ is a trace-free transverse symmetric tensor ($\delta^{ij}\chi_{ij}=\delta^{ij}\partial_i\chi_{jk}=0$) and $D_{ij}$ is a trace-free operator defined by $D_{ij}\equiv \partial_i\partial_j-(1/3) \delta_{ij} \nabla^2$. Perturbations of the energy-density and the four-velocity of the DM fluid can be written as
\begin{align}
\rho(\vec{x},\tau) \equiv&\, \rho^{(0)} (\tau)\left[1+\delta(\vec{x},\tau)\right]\,,\\
u^\mu(\vec{x},\tau) \equiv&\,\frac{1}{a}\left[\delta^{\mu}_{\phantom{\mu}0}+v^\mu(\vec{x},\tau)\right]\,.
\end{align}
We can expand any perturbation up to the desired order in this way
\beq
\pi\simeq\pi^{(0)}+ \pif+\frac{1}{2} \pis+\ldots+\frac{1}{n!} \pi^{(n)}\,.
\eeq
In the following we will drop the suffix ``$0$''. At first order we can safely neglect vector and tensor perturbations. In fact the first-order vector perturbations have decreasing amplitudes and are not generated by the presence of a scalar field. 
Moreover, the first-order tensor perturbations give a negligible contribution to second-order perturbations. This result cannot be generalized to second-order perturbations, since second-order vector and tensor perturbations are generated 
by products of first-order scalars.

Perturbing the well-known relation $u^\mu u_\mu=-1$, these useful equations can be obtained
\begin{align}
v^{(1)0} &= - \psi^{(1)}\,,\\
v^{(2)0} &=-  \psi^{(2)}+3 {\psi^{(1)}}^{2} + 2 \partial_{i}\omega^{(1)} \partial^{i}v^{(1)} + \partial_{i}v^{(1)} \partial^{i}v^{(1)}\,.
\end{align}

\section{Linear perturbation theory}\label{SEC:LinearTheory}

At the linear level from Eq.\ (\ref{EQ:Einstein}) we obtain four independent equations, the $(0,0)$, the $(0,i)$, the trace and the traceless of $(i,j)$ parts. These are, respectively,
\begin{align}
&2 {\Mpl}^{2} \nabla^2 \phi^{(1)} + \frac{1}{3} {\Mpl}^{2} \nabla^2 \nabla^2 \chi^{(1)} - \left(6 {\Mpl}^{2} \mathcal{H} -  \frac{3 c_3 {\pi^{\prime}}^{3}}{{M}^{3} {a}^{2}}\right) {\phi^{(1)}}^{\prime}- \left(2 {\Mpl}^{2} \mathcal{H} -  \frac{c_3 {\pi^{\prime}}^{3}}{{M}^{3} {a}^{2}}\right) \nabla^2 \omega^{(1)}\nonumber \\ 
&={a}^{2}\rho_m \delta^{(1)} + \left(c_2 {\pi^{\prime}}^{2} + 6 {\Mpl}^{2} {\mathcal{H}}^{2} -  \frac{12 c_3 {\pi^{\prime}}^{3} \mathcal{H}}{{M}^{3} {a}^{2}}\right) \psi^{(1)} -  \frac{c_3 {\pi^{\prime}}^{2} \nabla^2 \pi^{(1)}}{{M}^{3} {a}^{2}} -  \frac{1}{2} c_1 {M}^{3} {a}^{2}\pi^{(1)}\nonumber \\ 
&+ \left(- c_2 \pi^{\prime} + \frac{9 c_3 {\pi^{\prime}}^{2} \mathcal{H}}{{M}^{3} {a}^{2}}\right) {\pi^{(1)}}^{\prime}\,,\label{EQ:linear1}
\end{align}

\begin{align}
&- 2 {\Mpl}^{2} {\phi^{(1)}}^{\prime} -\left( 2 {\Mpl}^{2} \mathcal{H} - \frac{c_3 {{\pi}^{\prime}}^{3}}{{M}^{3} {a}^{2}}\right) \psi^{(1)} -  \frac{1}{3} {\Mpl}^{2} \nabla^2 {\chi^{(1)}}^{\prime} = {a}^{2}\rho_m v^{(1)}+{a}^{2}\rho_m \omega^{(1)} \nonumber \\ 
& +\left(  c_2 {\pi}^{\prime} - \frac{3 c_3 {{\pi}^{\prime}}^{2} \mathcal{H}}{{M}^{3} {a}^{2}}\right) \pi^{(1)} +\frac{c_3 {{\pi}^{\prime}}^{2} {\pi^{(1)}}^{\prime}}{{M}^{3} {a}^{2}}\,,\label{EQ:linear2}
\end{align}

\begin{align}
&\frac{2}{3} {\Mpl}^{2} \nabla^2\left(\psi^{(1)} - \phi^{(1)} + {\omega^{(1)}}^{\prime} + 2 \mathcal{H} \omega^{(1)} - \frac{1}{6} \nabla^2 \chi^{(1)}\right) + 2 {\Mpl}^{2} {\phi^{(1)}}^{\prime\prime} + 4 {\Mpl}^{2} \mathcal{H}{\phi^{(1)}}^{\prime}\nonumber \\ 
& + \left(2 {\Mpl}^{2} {\mathcal{H}}^{2} + 4 {\Mpl}^{2} {\mathcal{H}}^{\prime} -c_2 {{\pi}^{\prime}}^{2} -  \frac{4 c_3 {\pi}^{\prime\prime} {{\pi}^{\prime}}^{2}}{{M}^{3} {a}^{2}} + \frac{4 c_3 {{\pi}^{\prime}}^{3} \mathcal{H}}{{M}^{3} {a}^{2}}\right)\psi^{(1)} \nonumber \\ 
& + \left(2 {\Mpl}^{2} \mathcal{H} -  \frac{c_3 {{\pi}^{\prime}}^{3}}{{M}^{3} {a}^{2}}\right){\psi^{(1)}}^{\prime} =- \left(c_2 {\pi}^{\prime} + \frac{2 c_3 {\pi}^{\prime\prime} {\pi}^{\prime}}{{M}^{3} {a}^{2}} -  \frac{3 c_3 {{\pi}^{\prime}}^{2} \mathcal{H}}{{M}^{3} {a}^{2}}\right){\pi^{(1)}}^{\prime} \nonumber \\ 
&+\frac{1}{2} c_1 {M}^{3} {a}^{2} \pi^{(1)} - \frac{c_3 {{\pi}^{\prime}}^{2} {\pi^{(1)}}^{\prime\prime}}{{M}^{3} {a}^{2}} \,,\label{EQ:linear3}
\end{align}

\begin{align}
 {\chi^{(1)}}^{\prime\prime} + 2 \mathcal{H} {\chi^{(1)}}^{\prime} + \frac{1}{3} \nabla^2 \chi^{(1)} - 2 {\omega^{(1)}}^{\prime} - 4 \mathcal{H} \omega^{(1)} + 2 \phi^{(1)} - 2 \psi^{(1)}=0\,.\label{EQ:linear4}
\end{align}
The equation of motion for the linear perturbation of the Galileon field, Eq.\ (\ref{EQ:Galileon}), reads
\begin{align}
&\left(c_2 -  \frac{6 c_3 {\pi}^{\prime} \mathcal{H}}{{M}^{3} {a}^{2}}\right) {\pi^{(1)}}^{\prime\prime} + \left(2 c_2 \mathcal{H} -  \frac{6 c_3 {\pi}^{\prime} {\mathcal{H}}^{\prime}}{{M}^{3} {a}^{2}} -  \frac{6 c_3 {\pi}^{\prime\prime} \mathcal{H}}{{M}^{3} {a}^{2}}\right) {\pi^{(1)}}^{\prime} - \left(c_2 - \frac{2 c_3 \left({{\pi}^{\prime\prime} + \mathcal{H}\pi}^{\prime}\right)}{{M}^{3} {a}^{2}}\right) \nabla^2 \pi^{(1)}\nonumber \\ 
& + \left(-2 c_2 {\pi}^{\prime\prime} - 4 c_2 {\pi}^{\prime} \mathcal{H} + \frac{12 c_3 {{\pi}^{\prime}}^{2} {\mathcal{H}}^{\prime}}{{M}^{3} {a}^{2}} + \frac{24 c_3 {\pi}^{\prime\prime} {\pi}^{\prime} \mathcal{H}}{{M}^{3} {a}^{2}}\right) \psi^{(1)} + \left(- c_2 {\pi}^{\prime} + \frac{9 c_3 {{\pi}^{\prime}}^{2} \mathcal{H}}{{M}^{3} {a}^{2}}\right) {\psi^{(1)}}^{\prime} \nonumber \\ 
& + \frac{c_3 {{\pi}^{\prime}}^{2} \nabla^2 \psi^{(1)}}{{M}^{3} {a}^{2}} + \left(-3 c_2 {\pi}^{\prime} + \frac{6 c_3 {\pi}^{\prime\prime} {\pi}^{\prime}}{{M}^{3} {a}^{2}} + \frac{9 c_3 {{\pi}^{\prime}}^{2} \mathcal{H}}{{M}^{3} {a}^{2}}\right) {\phi^{(1)}}^{\prime} + \frac{3 c_3 {{\pi}^{\prime}}^{2} {\phi^{(1)}}^{\prime\prime}}{{M}^{3} {a}^{2}}\nonumber \\ 
& + \left(- c_2 {\pi}^{\prime} + \frac{2 c_3 {\pi}^{\prime\prime} {\pi}^{\prime}}{{M}^{3} {a}^{2}} + \frac{3 c_3 {{\pi}^{\prime}}^{2} \mathcal{H}}{{M}^{3} {a}^{2}}\right) \nabla^2 \omega^{(1)} + \frac{c_3 {{\pi}^{\prime}}^{2} \nabla^2 {\omega^{(1)}}^{\prime}}{{M}^{3} {a}^{2}}=0\,.\label{EQ:linear5}
\end{align}
From the time and the space components of the stress-energy tensor continuity equation, Eq.\ (\ref{EQ:StressEnergyConservation}), we obtain
\begin{align}
{\delta^{(1)}}^{\prime} = 3 {\phi^{(1)}}^{\prime} - \nabla^2 v^{(1)}\,,\label{EQ:linear6}
\end{align}
\begin{align}
{\omega^{(1)}}^{\prime} + {v^{(1)}}^{\prime} + \mathcal{H} \omega^{(1)} + \psi^{(1)} + \mathcal{H} v^{(1)}=0\,.\label{EQ:linear7}
\end{align}

There are many ways to decouple these equations . First of all, it is convenient to work in Fourier space. From Eqs.\ (\ref{EQ:linear4}) we can immediately obtain $\psi^{(1)}$. 
In the sub-horizon ($k^2\gg \mathcal{H}^2$) and quasi-static ($\left|\ddphi\right|\lesssim\mathcal{H}\left|\dphi\right|\ll k^2 \left|\phi\right|$) approximation, the relevant equations we need are (\ref{EQ:linear1}), (\ref{EQ:linear5}) and the derivative of (\ref{EQ:linear6})
\begin{align}\label{EQ:linear8}
2 {\Mpl}^{2} {k}^{2} \left(\phi^{(1)} - \frac{1}{6} {k}^{2} \chi^{(1)}\right) - {k}^{2} \omega^{(1)} \left(2 {\Mpl}^{2} \mathcal{H} - \frac{c_3 {{\pi}^{\prime}}^{3}}{{M}^{3} {a}^{2}}\right) =- \rho_m \delta^{(1)} {a}^{2}-\frac{c_3 {{\pi}^{\prime}}^{2} {k}^{2} \pi^{(1)}}{{M}^{3} {a}^{2}}\,,
\end{align}
\begin{align}\label{EQ:linear9}
& \left(c_2 -  \frac{2 c_3 {\pi}^{\prime\prime}}{{M}^{3} {a}^{2}} -  \frac{2 c_3 {\pi}^{\prime} \mathcal{H}}{{M}^{3} {a}^{2}}\right)\pi^{(1)} + \left(c_2 {\pi}^{\prime} -  \frac{2 c_3 {\pi}^{\prime\prime} {\pi}^{\prime}}{{M}^{3} {a}^{2}} -  \frac{c_3 {{\pi}^{\prime}}^{2} \mathcal{H}}{{M}^{3} {a}^{2}}\right)\omega^{(1)} \nonumber \\ 
&= \frac{c_3 {{\pi}^{\prime}}^{2}}{{M}^{3} {a}^{2}} \left(\phi^{(1)} -\frac{1}{6}{k}^{2} \chi^{(1)}\right)\,,
\end{align}
\begin{align}\label{EQ:linear10}
 {\delta^{(1)}}^{\prime\prime} + {\delta^{(1)}}^{\prime} \mathcal{H} + {k}^{2} \left(\phi^{(1)} - \frac{1}{6} {k}^{2} \chi^{(1)}\right) -  \mathcal{H} {k}^{2} \omega^{(1)}=0\,.
\end{align}
Combining Eqs.\ (\ref{EQ:linear8}) and (\ref{EQ:linear9}) to eliminate $\phi^{(1)}$ and $\chi^{(1)}$ it is straightforward to obtain
\begin{align}\label{EQ:linear11}
& \left(c_2 {\Mpl}^{2} + \frac{{c_3}^{2} {{\pi}^{\prime}}^{4}}{2 {M}^{6} {a}^{4}} -  \frac{2 c_3 {\Mpl}^{2} \left({\pi}^{\prime\prime}+\mathcal{H}{\pi}^{\prime}\right)}{{M}^{3} {a}^{2}}\right) {k}^{2} \left[\pi^{(1)}+{\pi}^{\prime}\omega^{(1)}\right]=-\frac{c_3 \rho_m {{\pi}^{\prime}}^{2} \delta^{(1)}}{2 {M}^{3}}\,.
\end{align}
Finally, using Eqs.\ (\ref{EQ:linear8}), (\ref{EQ:linear10}) and (\ref{EQ:linear11}) we are able to single out an equation for the DM perturbation $\delta^{(1)}$
\begin{align}\label{EQ:lineardelta}
\delta^{(1)\prime\prime} + \mathcal{H} \delta^{(1)\prime} = 4 \pi G\left(1-\frac{{c_3}^2 {\pi^\prime}^4}{2 c_2 M^6 \Mpl^2 a^4\a}\right) a^2 \rho_m\delta^{(1)}\,,
\end{align}
where $G\equiv {(8\pi \Mpl^2)}^{-1}$ is Newton's constant and we have defined
\beq
\a(\tau)\equiv 1-\frac{2 c_3}{c_2 M^3 a^2}\left(\pi^{\prime\prime}+\mathcal{H}\pi^{\prime}\right)+\frac{{c_3}^2 {\pi^{\prime}}^4}{2 c_2 M^6 \Mpl^2 a^4}\,.
\eeq
The crucial difference between Eq.\ (\ref{EQ:lineardelta}) and the one obtained in the $\Lcdm$  model is that the Galileon acts modifying the Newton's constant at late-times. To recover the standard Newton's constant it is sufficient to set $c_3=0$. On the left hand side of Eq.\ (\ref{EQ:lineardelta}) the other modification lies inside the friction term ($\mathcal{H} \delta^{(1)\prime}$) due to the evolution of the Hubble parameter. As shown in Fig.\ \ref{FIG:background} these differences cannot be neglected and should play an important role in the growth of structures.

Eq.\ (\ref{EQ:lineardelta}), which describes the dynamics of DM perturbations on sub-horizon scales, together with Eqs.\ (\ref{EQ:linear6}), (\ref{EQ:linear7}), (\ref{EQ:linear9}) and (\ref{EQ:linear11}) forms our complete set of equations that allow to solve the dynamics of the fluctuations at first order. In Appendix \ref{SEC:AppendixB} we show how to obtain the same result in the Poisson, spatially flat and synchronous gauges. In particular it is important to pay attention doing the sub-horizon approximation in the synchronous gauge, due to the residual gauge freedom.

Eq.\ (\ref{EQ:lineardelta}) can be divided in the linear combination of two independent solutions
\begin{align}
\delta^{(1)}(\vec{k},\tau)=c_+D_+(\tau) \delta^{(1)}(\vec{k}) + c_-D_-(\tau) \delta^{(1)}(\vec{k})\,,
\end{align}
where $\delta^{(1)}(\vec{k})$ is the primordial amplitude of the density contrast perturbation. We have also added explicitly two integration constants, $c_+$ and $c_-$. $D_{+}(\tau)$ and $D_{-}(\tau)$ are the growing and the decaying modes and they depend on the coefficients $c_i$. In the next subsection we will find an integral solution for these modes.

\subsection{Integral solutions for the growing and the decaying modes of DM perturbations}

To solve Eq.\ (\ref{EQ:lineardelta}) it is convenient to redefine

\beq
A(\tau)\equiv \frac{4 \pi G}{\mathcal{H}^2}\left(1-\frac{{c_3}^2 {\pi^{\prime}}^4}{2 c_2 M^6 \Mpl^2 a^4\a}\right) \rho_m\,.
\eeq
We can also use the scale factor as the new time variable
\begin{align}
\frac{d^2\delta(a)}{d a^2} + \left(\frac{2}{a}+\frac{1}{\mathcal{H}(a)}\frac{d\mathcal{H}(a)}{d a}\right) \frac{d\delta(a)}{d a} = A(a) \delta(a)\,.
\end{align}
After that we can perform the change of the variable
\beq
\delta(a)=u(a)\sqrt{\frac{\mathcal{H}_0}{a^2\mathcal{H}(a)}}\,.
\eeq
After a straightforward calculation we shall obtain Eq.\ (\ref{EQ:lineardelta}) in its normal form 
\beq\label{EQ:deltanorm}
\frac{d^2 u(a)}{d a^2}-I(a) u(a)=0\,,
\eeq
where $(-I(a))$ is often called the \textit{invariant} of the equation
\beq
I(a)=A(a)+\frac{1}{a \mathcal{H}(a)}\frac{d\mathcal{H}(a)}{da}-\frac{1}{4 {\mathcal{H}(a)}^2}{\left(\frac{d\mathcal{H}(a)}{da}\right)}^2+\frac{1}{2 \mathcal{H}(a)}\frac{d^2 \mathcal{H}(a)}{da^2}\,.
\eeq
Now, suppose we have to solve
\beq\label{EQ:y}
\frac{d^2 y(a)}{da^2}+g(a) \frac{d y(a)}{d a} +\frac{d g(a)}{d a} y(a)=0\,.
\eeq
After the substitution
\beq
Y(a)=y(a) e^{+\frac{1}{2}\int^a_{a_m} da' g(a')}\,,
\eeq
where $a_m$ is some initial time deep inside the matter dominated era, we obtain
\beq\label{EQ:ynorm}
\frac{d^2 Y(a)}{d a^2} +\frac{1}{2}\left[\frac{d g(a)}{da} -\frac{1}{2}{g^2(a)}\right]Y(a)=0\,.
\eeq
We can choose $g(a)$ to be a solution of  
\beq\label{EQ:f}
\frac{d g(a)}{d a} -\frac{1}{2}g^2(a) +2 I(a)=0\,,
\eeq
which is a particular Riccati equation. In this case Eqs.\ (\ref{EQ:deltanorm}) and (\ref{EQ:ynorm}) become equals. Thus, we can relate the solutions of Eq.\ (\ref{EQ:y}) with the ones of Eq.\ (\ref{EQ:lineardelta}) through
\beq
\delta(a)=\frac{y(a)}{a}\sqrt{\frac{\mathcal{H}_0}{\mathcal{H}(a)}} e^{+\frac{1}{2}\int^a_{a_m} da' g(a')}\,.
\eeq
It is straightforward to integrate Eq.\ (\ref{EQ:y}) the first time
\beq\label{EQ:g}
\frac{d y(a)}{d a} +g(a)y(a)=a_2,
\eeq
where $a_2$ is the first integration constant. A second integration is also possible, giving us the solutions for $y(\tau)$ in their integral form
\beq\label{EQ:ysol}
y(\tau)=\kappa_1 \gamma^2(a)+\kappa_2 \gamma^2(a)\int^a_{a_m}\frac{da'}{\gamma^2(a')}\,,
\eeq
where
\beq\label{EQ:gamma}
\gamma^2(a)=e^{-\int^a_{a_m} da' g(a')}\,.
\eeq
From Eq.\ (\ref{EQ:ysol}) we have two independent solutions of Eq.\ (\ref{EQ:lineardelta}) in their integral form 
\begin{align}\label{EQ:growdec}
D_{1}(a) =&\frac{\gamma(a)}{a}\sqrt{\frac{\mathcal{H}_0}{\mathcal{H}(a)}} \nonumber\\
D_{2}(a) =&\frac{\gamma(a)}{a}\sqrt{\frac{\mathcal{H}_0}{\mathcal{H}(a)}}\int^a_{a_m}\frac{da'}{\gamma^2(a')}\,.
\end{align}
To determine the growing and the decaying modes it is important to note that there is an additional degree of freedom due to the boundary condition in Eq.\ (\ref{EQ:f}). 
If we want to separate them we have to choose carefully the behavior of $g(a)$ at early times. As shown in Fig.\ \ref{FIG:background}, during the matter dominated 
epoch (MD) the contribution of the Galileon field can be neglected. Indeed, during this epoch we expect that 
${D_{MD}}^+(a)\propto a$, and ${D_{MD}}^-(a)\propto \mathcal{H}(a)/a\propto a^{-3/2}$. We can impose $D_1(a)= {D_{MD}}^+(a)=a$, obtaining $g(a)=-7/(2 a)$. 
Taking into account the right coefficients, we can extend this result to the general solution, i.e. valid also after the matter-dominated epoch
\begin{align}\label{EQ:growdec2}
D_+(a)&= {a_m}^{7/4} D_1(a)=\frac{{a_m}^{7/4} \gamma(a)}{a}\sqrt{\frac{\mathcal{H}_0}{\mathcal{H}(a)}}\nonumber\\
D_-(a)&= {a_m}^{-3/4} D_1(a)-\frac{5}{2 {a_m}^{7/4}} D_2(a)\nonumber\\
&=\frac{\gamma(a)}{{a_m}^{3/4} a}\sqrt{\frac{\mathcal{H}_0}{\mathcal{H}(a)}}\left(1-\frac{5}{2 a_m}\int^a_{a_m}\frac{da'}{\gamma^2(a')}\right)\,.
\end{align}

\begin{figure}[tb]
 \begin{center}
  \includegraphics[width=0.6\textwidth]{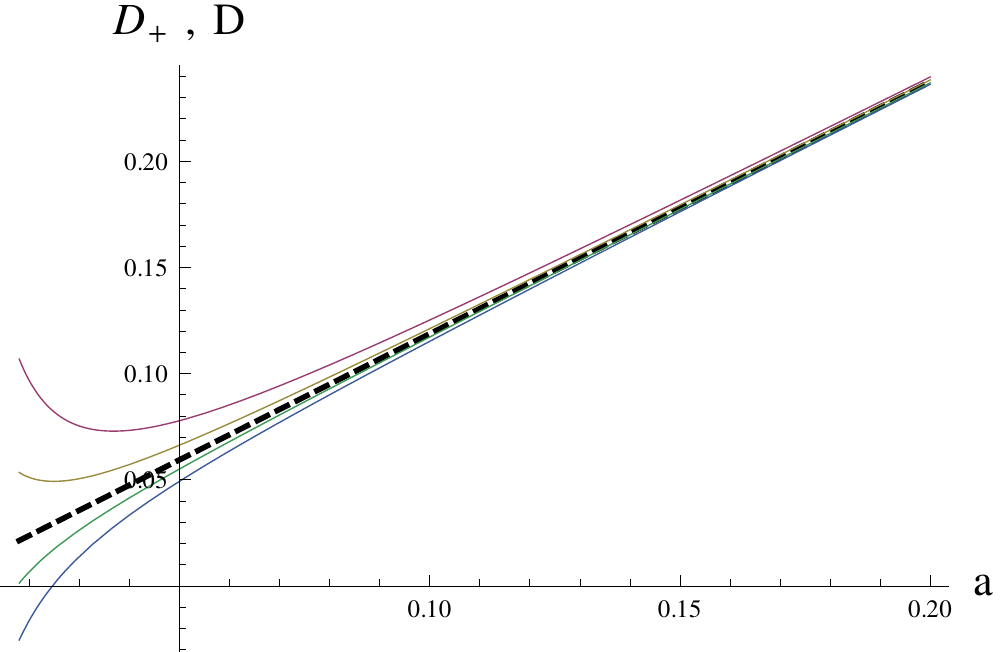}
 \end{center}
 \caption{\textit{Evolution of $D_+(a)$ (black dashed line), Eq.\ (\ref{EQ:growdec2}), and $D(a)$ (the other lines), solutions of Eq.\ (\ref{EQ:lineardelta}) with different initial conditions (for a fixed background corresponding to $c_1=1.5$, $c_2=0.04$ and $c_3=10^{-3}$).}}\label{FIG:semianalyticD}
 \end{figure}

\begin{figure}[tb]
 \begin{center}
  \includegraphics[width=0.6\textwidth]{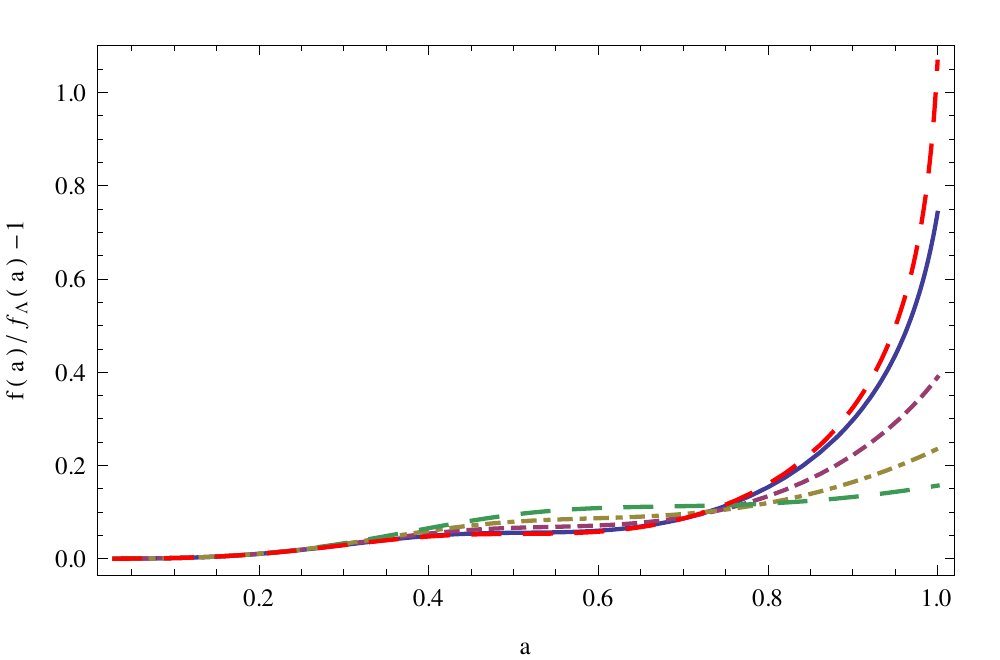}
 \end{center}
 \caption{\textit{Growth rate $f(a)$ of the Galileon compared with the growth rate of $\Lcdm$. The values for the parameters $c_1$, $c_2$ and $c_3$ are the same as in Fig.\ \ref{FIG:background}.}}\label{FIG:growthrate}
 \end{figure}

These solutions are important because they are valid in every modified gravity theory in which the evolution of first order DM perturbations, Eq.\ (\ref{EQ:lineardelta}), is scale-independent. In Fig.\ \ref{FIG:semianalyticD} we show 
the evolution of our integral solution, Eq.\ (\ref{EQ:growdec2}), vs. the numerical solution of Eq.\ (\ref{EQ:lineardelta}) for various and arbitrary initial conditions. It is important to note that every numerical solution approaches 
$D_+(a)$, this proves that the first line of Eq.\ (\ref{EQ:growdec2}) is the pure growing mode of Eq.\ (\ref{EQ:lineardelta}). In Fig.\ \ref{FIG:growthrate} we plot the deviations of the Galileon growth rate, $f(a)\equiv d \ln D/d\ln a$, w.r.t.\ the growth rate of the $\Lcdm$ model. For models in which the value of $c_3$ is negligible w.r.t.\ the value $c_1$ the deviations are large (up to about $100\%$), while, increasing $c_3$ the deviations decrease reaching $\simeq10\%$. We noted that when $c_3\gg c_1$ the modifications w.r.t.\ the $\lcdm$ model are dominated by the friction term. This is in agreement with the fact that the deviations we find in Fig.\ \ref{FIG:growthrate} have the same magnitude as the deviations in the first panel of Fig.\ \ref{FIG:background}. On the other hand, if $c_3\simeq c_1$ or $c_3\ll c_1$ we expect that the modifications of Eq.\ (\ref{EQ:lineardelta}) are both important.

\section{Second-order perturbations}\label{SEC:second}

By perturbing the Einstein and the Galileon field equations, Eqs. (\ref{EQ:Einstein}) and (\ref{EQ:Galileon}), at second order we can study the dynamics of the DM fluctuations in the weakly non-linear regime. 
The structure of these equations is the same as in the linear case, up to additional source terms formed by product of first-order scalar quantities that we will indicate with $S^{(n)}$ (their explicit expression in a general gauge can be found in Appendix \ref{SEC:AppendixA}). From the Einstein equations we obtain, respectively, the $(0,0)$, $(0,i)$ the trace and the traceless part of $(i,j)$
\begin{align}
&2 {\Mpl}^{2} \nabla^2 \phi^{(2)} + \frac{1}{3} {\Mpl}^{2} \nabla^2 \nabla^2 \chi^{(2)} - \left(6 {\Mpl}^{2} \mathcal{H} -  \frac{3 c_3 {\pi^{\prime}}^{3}}{{M}^{3} {a}^{2}}\right) {\phi^{(2)}}^{\prime}- \left(2 {\Mpl}^{2} \mathcal{H} -  \frac{c_3 {\pi^{\prime}}^{3}}{{M}^{3} {a}^{2}}\right) \nabla^2 \omega^{(2)}\nonumber \\ 
&={a}^{2}\rho_m \delta^{(2)} + \left(c_2 {\pi^{\prime}}^{2} + 6 {\Mpl}^{2} {\mathcal{H}}^{2} -  \frac{12 c_3 {\pi^{\prime}}^{3} \mathcal{H}}{{M}^{3} {a}^{2}}\right) \psi^{(2)} -  \frac{c_3 {\pi^{\prime}}^{2} \nabla^2 \pi^{(2)}}{{M}^{3} {a}^{2}} -  \frac{1}{2} c_1 {M}^{3} {a}^{2}\pi^{(2)}\nonumber \\ 
&+ \left(- c_2 \pi^{\prime} + \frac{9 c_3 {\pi^{\prime}}^{2} \mathcal{H}}{{M}^{3} {a}^{2}}\right) {\pi^{(2)}}^{\prime}-S^{(1)} \,,\label{EQ:second1}
\end{align}
\begin{align}
&- 2 {\Mpl}^{2} \nabla^2{\phi^{(2)}}^{\prime} -\left( 2 {\Mpl}^{2} \mathcal{H} - \frac{c_3 {{\pi}^{\prime}}^{3}}{{M}^{3} {a}^{2}}\right) \nabla^2\psi^{(2)} -  \frac{1}{3} {\Mpl}^{2} \nabla^2\nabla^2 {\chi^{(2)}}^{\prime} = {a}^{2}\rho_m \nabla^2v^{(2)} \nonumber \\ 
&+{a}^{2}\rho_m \nabla^2\omega^{(2)} +\left(  c_2 {\pi}^{\prime} - \frac{3 c_3 {{\pi}^{\prime}}^{2} \mathcal{H}}{{M}^{3} {a}^{2}}\right) \nabla^2\pi^{(2)} +\frac{c_3 {{\pi}^{\prime}}^{2}}{{M}^{3} {a}^{2}}\nabla^2{\pi^{(2)}}^{\prime}+S^{(2)}\,,\label{EQ:second2}
\end{align}
\begin{align}
&\frac{2}{3} {\Mpl}^{2} \nabla^2\left(\psi^{(2)} - \phi^{(2)} + {\omega^{(2)}}^{\prime} + 2 \mathcal{H} \omega^{(2)} - \frac{1}{6} \nabla^2 \chi^{(2)}\right) + 2 {\Mpl}^{2} {\phi^{(2)}}^{\prime\prime} + 4 {\Mpl}^{2} \mathcal{H}{\phi^{(2)}}^{\prime}\nonumber \\ 
& + \left(2 {\Mpl}^{2} {\mathcal{H}}^{2} + 4 {\Mpl}^{2} {\mathcal{H}}^{\prime} -c_2 {{\pi}^{\prime}}^{2} -  \frac{4 c_3 {\pi}^{\prime\prime} {{\pi}^{\prime}}^{2}}{{M}^{3} {a}^{2}} + \frac{4 c_3 {{\pi}^{\prime}}^{3} \mathcal{H}}{{M}^{3} {a}^{2}}\right)\psi^{(2)} \nonumber \\ 
& + \left(2 {\Mpl}^{2} \mathcal{H} -  \frac{c_3 {{\pi}^{\prime}}^{3}}{{M}^{3} {a}^{2}}\right){\psi^{(2)}}^{\prime} =- \left(c_2 {\pi}^{\prime} + \frac{2 c_3 {\pi}^{\prime\prime} {\pi}^{\prime}}{{M}^{3} {a}^{2}} -  \frac{3 c_3 {{\pi}^{\prime}}^{2} \mathcal{H}}{{M}^{3} {a}^{2}}\right){\pi^{(2)}}^{\prime} \nonumber \\ 
&+\frac{1}{2} c_1 {M}^{3} {a}^{2} \pi^{(2)} - \frac{c_3 {{\pi}^{\prime}}^{2} {\pi^{(2)}}^{\prime\prime}}{{M}^{3} {a}^{2}}+S^{(3)} \,,\label{EQ:second3}
\end{align}
\begin{align}
 \nabla^2\nabla^2\left({\chi^{(2)}}^{\prime\prime} + 2 \mathcal{H} {\chi^{(2)}}^{\prime} + \frac{1}{3} \nabla^2 \chi^{(2)} - 2 {\omega^{(2)}}^{\prime} - 4 \mathcal{H} \omega^{(2)} + 2 \phi^{(2)} - 2 \psi^{(2)}\right)=2 S^{(4)}\,.\label{EQ:second4}
\end{align}
Eq.\ (\ref{EQ:Galileon}), for the Galileon field fluctuations, becomes
\begin{align}
&\left(c_2 -  \frac{6 c_3 {\pi}^{\prime} \mathcal{H}}{{M}^{3} {a}^{2}}\right) {\pi^{(2)}}^{\prime\prime} + \left(2 c_2 \mathcal{H} -  \frac{6 c_3 {\pi}^{\prime} {\mathcal{H}}^{\prime}}{{M}^{3} {a}^{2}} -  \frac{6 c_3 {\pi}^{\prime\prime} \mathcal{H}}{{M}^{3} {a}^{2}}\right) {\pi^{(2)}}^{\prime} - \left(c_2 - \frac{2 c_3 \left({{\pi}^{\prime\prime} + \mathcal{H}\pi}^{\prime}\right)}{{M}^{3} {a}^{2}}\right) \nabla^2 \pi^{(2)}\nonumber \\ 
& + \left(-2 c_2 {\pi}^{\prime\prime} - 4 c_2 {\pi}^{\prime} \mathcal{H} + \frac{12 c_3 {{\pi}^{\prime}}^{2} {\mathcal{H}}^{\prime}}{{M}^{3} {a}^{2}} + \frac{24 c_3 {\pi}^{\prime\prime} {\pi}^{\prime} \mathcal{H}}{{M}^{3} {a}^{2}}\right) \psi^{(2)} + \left(- c_2 {\pi}^{\prime} + \frac{9 c_3 {{\pi}^{\prime}}^{2} \mathcal{H}}{{M}^{3} {a}^{2}}\right) {\psi^{(2)}}^{\prime} \nonumber \\ 
& + \frac{c_3 {{\pi}^{\prime}}^{2} \nabla^2 \psi^{(2)}}{{M}^{3} {a}^{2}} + \left(-3 c_2 {\pi}^{\prime} + \frac{6 c_3 {\pi}^{\prime\prime} {\pi}^{\prime}}{{M}^{3} {a}^{2}} + \frac{9 c_3 {{\pi}^{\prime}}^{2} \mathcal{H}}{{M}^{3} {a}^{2}}\right) {\phi^{(2)}}^{\prime} + \frac{3 c_3 {{\pi}^{\prime}}^{2} {\phi^{(2)}}^{\prime\prime}}{{M}^{3} {a}^{2}}\nonumber \\ 
& + \left(- c_2 {\pi}^{\prime} + \frac{2 c_3 {\pi}^{\prime\prime} {\pi}^{\prime}}{{M}^{3} {a}^{2}} + \frac{3 c_3 {{\pi}^{\prime}}^{2} \mathcal{H}}{{M}^{3} {a}^{2}}\right) \nabla^2 \omega^{(2)} + \frac{c_3 {{\pi}^{\prime}}^{2} \nabla^2 {\omega^{(2)}}^{\prime}}{{M}^{3} {a}^{2}}=S^{(5)}\,.\label{EQ:second5}
\end{align}
The stress-energy tensor continuity equation reads
\begin{align}
{\delta^{(2)}}^{\prime} = 3 {\phi^{(2)}}^{\prime} - \nabla^2 v^{(2)}+S^{(6)}\,,\label{EQ:second6}
\end{align}
\begin{align}
\nabla^2\left({\omega^{(2)}}^{\prime} + {v^{(2)}}^{\prime} + \mathcal{H} \omega^{(2)} + \psi^{(2)} + \mathcal{H} v^{(2)}\right)=S^{(7)}\,.\label{EQ:second7}
\end{align}
In Eqs.\ (\ref{EQ:second2}), (\ref{EQ:second4}) and (\ref{EQ:second7}) second-order vector and tensor perturbations were present. In order to decouple scalar from vector and tensor perturbations we have used the operator $\partial_i$ in Eqs.\ (\ref{EQ:second2}) and (\ref{EQ:second7}), while we have used $\partial_i\partial_j$ in Eq.\ (\ref{EQ:second4}). Once the equations of motion for the scalar perturbations are obtained the steps to obtain the evolution for $\delta$ are the same as in the linear case. The result is
\begin{align}\label{EQ:seconddelta}
&\delta^{(2)\prime\prime} + \mathcal{H} \delta^{(2)\prime} - 4 \pi G\left(1-\frac{{c_3}^2 {\pi^{\prime}}^4}{2 c_2 M^6 \Mpl^2 a^4\a}\right) a^2 \rho_m \delta^{(2)} =S^{(\delta)}\,,
\end{align}
where
\begin{align}\label{EQ:sourceg}
S^{(\delta)}=& - \left(1 - \frac{{c_3}^{2} {\pi^\prime}^{4}}{2 c_2 {M}^{6} {\Mpl}^{2} {a}^{4}\alpha}\right) \left[\frac{S^{(1)}}{2 {\Mpl}^{2}}-\frac{S^{(4)}}{{k}^{2}}\right] + \frac{c_3 {\pi^\prime}^{2} S^{(5)}}{2 c_2 {M}^{3} {\Mpl}^{2} {a}^{2}\alpha} \nonumber \\ 
&+{S^{(6)}}^{\prime} + \mathcal{H} S^{(6)} -  S^{(7)}
\end{align}

\subsection{Solution of the evolution equation for the second-order DM density contrast}

In this section we study the behavior of Eq.\ (\ref{EQ:seconddelta}). It is clear that the homogeneous part of this equation is equal to Eq.\ (\ref{EQ:lineardelta}). 
Thus, using Green's method, and Eqs.\ (\ref{EQ:growdec2}), we can find an analytical (in its integral form) solution for the evolution of the second-order DM density perturbations. 
Using Eqs.\ (\ref{EQ:linearvelocity}), (\ref{EQ:linearpi}), (\ref{EQ:linearphi}), (\ref{EQ:linear6}) and (\ref{EQ:lineardelta}), in the Poisson gauge the Fourier transform of the source term Eq.\ (\ref{EQ:sourceg}) becomes

\begin{align}
S^{(\delta)}&(a,\vec{k})=\int d^3 k_1d^3 k_2\delta^{(3)}(\vec{k}-\vec{k}_1-\vec{k}_2)\mathcal{K}(a,\vec{k}_1,\vec{k}_2) \delta^{(1)}(a,\vec{k}_1) \delta^{(1)}(a,\vec{k}_2)\,.
\end{align}
Here, the symmetrized kernel $\mathcal{K}(a,\vec{k}_1,\vec{k}_2)$ reads
\begin{align}\label{kernel}
\mathcal{K}(a,\vec{k}_1,\vec{k}_2)&\equiv \g_1(a) +\g_2(a) \,\frac{\left(\vec{k}_1\cdot\vec{k}_2\right)\left({k_1}^2+{k_2}^2\right)}{{k_1}^2 {k_2}^2} +\g_3(a) \frac{{\left(\vec{k}_1\cdot\vec{k}_2\right)}^2}{{k_1}^2 {k_2}^2} \nonumber\\
&+\g_4(a) \frac{a^2\mathcal{H}^2\left(\vec{k}_1\cdot\vec{k}_2\right)}{{k_1}^2 {k_2}^2} + \g_5(a) \frac{a^2\mathcal{H}^2}{k^2}+ \g_6(a)\left(\frac{a^2\mathcal{H}^2}{{k_1}^2}+\frac{a^2\mathcal{H}^2}{{k_2}^2}\right)\nonumber\\
& + \g_7(a)\frac{a^2\mathcal{H}^2\left({k_1}^4+{k_2}^4\right)}{k^2 {k_1}^2 {k_2}^2}+ \g_8(a)\frac{a^4\mathcal{H}^4}{{k_1}^2 {k_2}^2}\,,
\end{align}
where the background functions $\gamma_1(a)$, $\gamma_2(a)$ and $\gamma_3(a)$ are shown in the next section, while the other $\gamma_i(a)$ are listed in Appendix \ref{SEC:AppendixC}. Finally, using Green's method with the homogeneous solutions, Eq.\ (\ref{EQ:growdec2}), we can find the evolution of the second-order density fluctuations
\begin{align}
\label{delta2}
\delta^{(2)}(a,\vec{k})&= D_+(a)\delta^{(2)}(\vec{k})-D_+(a)\int_{a_m}^a da'\frac{D_-(a') S^{(\delta)}(a',\vec{k})}{{a'}^2 \mathcal{H}^2(a') W(a')} \nonumber\\
&+D_-(a)\int_{a_m}^a da'\frac{D_+(a') S^{(\delta)}(a',\vec{k})}{{a'}^2 \mathcal{H}^2(a') W(a')}\,,
\end{align}
where $W$ is the Wronskian
\begin{align}\label{wronskian}
W(a)\equiv D_+(a) {D_-}'(a) - D_-(a) {D_+}'(a)=-\frac{5 \mathcal{H}_0}{2 a^2 \mathcal{H}(a)}\,,
\end{align}
$a_m$ is some initial time deep inside the matter dominated era and $\delta^{(2)}(\vec{k})$ is the initial second-order DM perturbation. It is interesting to see that in this relation there is no explicit dependence on the coefficients $c_i$.

\section{Dark Matter power spectrum and bispectrum}\label{SEC:bispectrum}

To describe the DM distribution of the universe the first statistical interesting quantity is the power-spectrum
\begin{align}
\left\langle\delta (a,\vec{k}_1)\delta (a,\vec{k}_2)\right\rangle\equiv {\left(2\pi\right)}^3\delta^{(3)}(\vec{k}_1+\vec{k}_2) P(a,k_1)\,,
\end{align}
where $\delta^{(3)}(\vec{k}_1+\vec{k}_2)$ is the three dimensional Dirac delta function and $\langle\ldots\rangle$ indicates ensemble averaging. Note that, under the assumption of spatial isotropy, 
the power-spectrum depends only on the absolute value of $\vec{k}_1$. By the Wick theorem, for Gaussian distributed fluctuations the power-spectrum contains all the information about the DM distribution. 
The linear power-spectrum, calculated using first-order equations, reads
\begin{align}
P(a,k)\propto {\left| D_+(a)\right|}^2 T^2(k) {\left(\frac{k}{\mathcal{H}_0}\right)}^{n_s}\,,
\end{align}
where $n_s$ is the scalar spectral index of primordial fluctuations and $T(k)$ is the transfer function (for which we use for simplicity the fit provided in \cite{Bardeen:1985tr}). In the following computations we will take $n_s=0.96$~\cite{Hinshaw:2012fq}. The second statistic of interest is the bispectrum, defined by
\begin{align}
\left\langle\delta (a,\vec{k}_1)\delta (a,\vec{k}_2)\delta (a,\vec{k}_3)\right\rangle\equiv {\left(2\pi\right)}^3\delta^{(3)}(\vec{k}_1+\vec{k}_2+\vec{k}_3) B(a,k_1,k_2,k_3)\,,
\end{align}
where the Dirac delta function imposes that only closed triangle configurations are to be considered. 
Since we are interested in studying the contribution generated by gravitational instability at late times in the Galileon theory, we impose Gaussian initial conditions. 
It is convenient to use the reduced bispectrum \cite{Groth:1977gj}, defined by
\begin{align}\label{EQ:reducedbispectrum}
Q(a,k_1,k_2,k_3)&\equiv \frac{B(a,k_1,k_2,k_3)}{P(a,k_1) P(a,k_2) +cyc.}\,,
\end{align}
which has the property that it remove most of the scale dependence to lowest-order (tree-level) in non-linear perturbation theory. 
Using the results of the previous sections we can write the density contrast perturbation as
\footnote{Notice that in the following we neglect the contribution proportional to the initial second-order DM perturbation $\delta^{(2)}(\vec{k})$ in Eq.~(\ref{delta2}). $\delta^{(2)}(\vec{k})$ contains both a possible primordial 
NG, and a non-primordial contribution, see, e.g.~\cite{Bartolo:2005xa,Bartolo:2006fj}. However the non-primordial term gives a negligible contribution to our final results on the scales of the quasi-static regime.
}
\begin{align}
\delta (&a,\vec{k})\equiv\delta^{(1)} (a,\vec{k})+\frac{1}{2}\delta^{(2)} (a,\vec{k})=D_+(a)\delta^{(1)} (\vec{k})\nonumber\\
&+\int d^3q_1\int d^3q_2\delta^{(3)}(\vec{k}-\vec{q}_1-\vec{q}_2) F(a,\vec{q}_1,\vec{q}_2) \delta^{(1)}(a,\vec{q}_1) \delta^{(1)}(a,\vec{q}_2)\,,
\end{align}
where
\begin{align}\label{EQ:kernelF}
F&(a,\vec{q}_1,\vec{q}_2)=\int_{a_m}^a da' \frac{{D_+}^2(a')\left[D_-(a) D_+(a')-D_+(a) D_-(a')\right]}{2 {a'}^2 \mathcal{H}^2(a') W(a') {D_+}^2(a)} \mathcal{K}_{SH}(a',\vec{q}_1,\vec{q}_2)\nonumber\\
&= \frac{a\sqrt{\mathcal{H}(a)}}{\g(a)} \int_{a_m}^a da' \left(\int_{a'}^a \frac{da''}{\g^2(a'')}\right) \frac{\g^3(a')}{{a'}^3 \sqrt{\mathcal{H}(a')}} \frac{\mathcal{K}_{SH}(a',\vec{q}_1,\vec{q}_2)}{2\mathcal{H}^2(a')} \,.
\end{align}
The kernel $\mathcal{K}_{SH}(a',\vec{q}_1,\vec{q}_2)$ is the leading order of Eq.\ (\ref{kernel}) taking into account that we are working on scales much smaller than the horizon (${k_i}^2\gg a^2\mathcal{H}$). 
This kernel can be recast in a more convenient form as
\begin{align}\label{kernel2}
\frac{\mathcal{K}_{SH}(a,\vec{q}_1,\vec{q}_2)}{2 \mathcal{H}^2(a)}=\g_1(a) +\g_2(a) \frac{\left(\vec{k}_1\cdot\vec{k}_2\right) \left({k_1}^2+{k_2}^2\right)}{{k_1}^2 {k_2}^2} +\g_3(a) \frac{{\left(\vec{k}_1\cdot\vec{k}_2\right)}^2}{{k_1}^2 {k_2}^2}\,,
\end{align}
where
\begin{align}\label{growthrate}
\g_1(a)&\equiv f^2(a) + \frac{\rho_m a^2}{2 \Mpl^2 \mathcal{H}^2} -  \frac{c_3^2 a^2 \mathcal{H}^2 {\pi^\prime}^4 \rho_m}{4 c_2 M^6 \Mpl^4\a} + \frac{c_3^4 a^2 \mathcal{H}^4{\pi^\prime}^6 \rho_m^2}{8 c_2^3 M^{12} \Mpl^6\a^3}\nonumber\\
\g_2(a)&\equiv f^2(a) + \frac{\rho_m a^2}{4 \Mpl^2 \mathcal{H}^2} - \frac{c_3^2 a^2 \mathcal{H}^2 {\pi^\prime}^4 \rho_m}{8 c_2 M^6 \Mpl^4\a}\nonumber\\
\g_3(a)&\equiv f^2(a) - \frac{c_3^4 a^2 \mathcal{H}^4 {\pi^\prime}^6 \rho_m^2}{8 c_2^3 M^{12} \Mpl^6\a^3}\nonumber\\
f(a)&=1+\frac{3 p_\pi}{4\left(\rho_m+\rho_\pi\right)}+\frac{1}{2} a h(a)\,.
\end{align}

Here we introduce $h(a)=g(a)+7/(2 a)$, where $g(a)$ is the solution of Eq.\ (\ref{EQ:g}), to parametrize the contribution of the accelerated expansion on the growth rate. Eq. (\ref{kernel2}) is one of the main results of our paper. It reduces to the usual form of the Newtonian kernel in the limit of an Einstein-de Sitter (EdS) universe \cite{Peebles1980,Bernardeau:2001qr}. It shows that the different contributions to the bispectrum have the same scale dependence as in EdS and $\Lcdm$, while they are modulated by time 
dependent coefficients that depend on the particular Galileon model. Looking at Eq.\ (\ref{kernel2}) we can recognize three kind of modifications w.r.t.\ the $\lcdm$ kernel. The first is due to the different evolution of the growth rate w.r.t.\ $\Lcdm$ and, as stated before, should produce deviations in the bispectrum that can reach $\simeq 100\%$. The second comes from the different evolution of the background, while the third is related to the parameters $c_2$ and $c_3$.

The reduced bispectrum, Eq. (\ref{EQ:reducedbispectrum}), assumes the standard form
\begin{align}\label{EQ:reducedbispectrum2}
Q(a,k_1,k_2,k_3)&=\frac{2 F(a,\vec{k}_1,\vec{k}_2)P(a,k_1) P(a,k_2) +cyc.}{P(a,k_1) P(a,k_2) +cyc.}\,.
\end{align}

\begin{figure}[!ht] %  figure placement: here, top, bottom, or page
   %\centering
\includegraphics[width=0.495\textwidth]{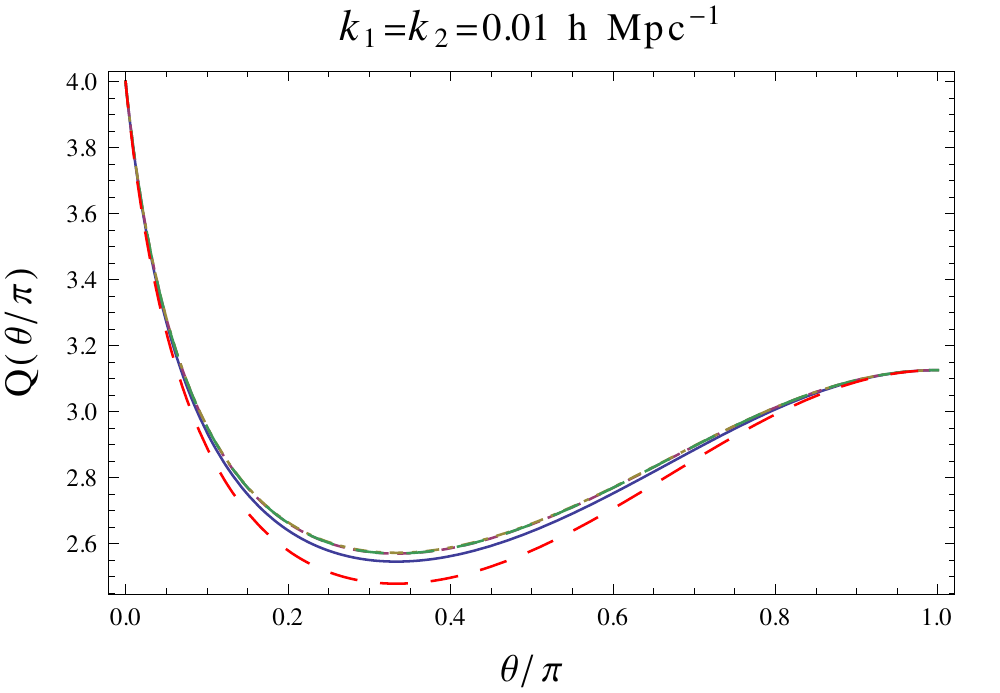} \includegraphics[width=0.505\textwidth]{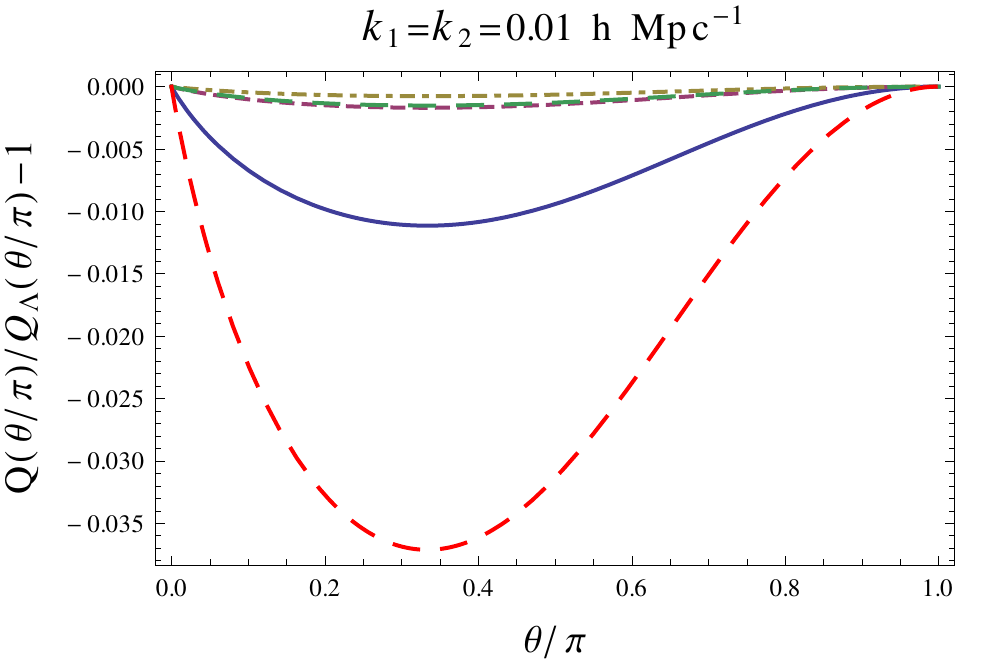}\\
\includegraphics[width=0.495\textwidth]{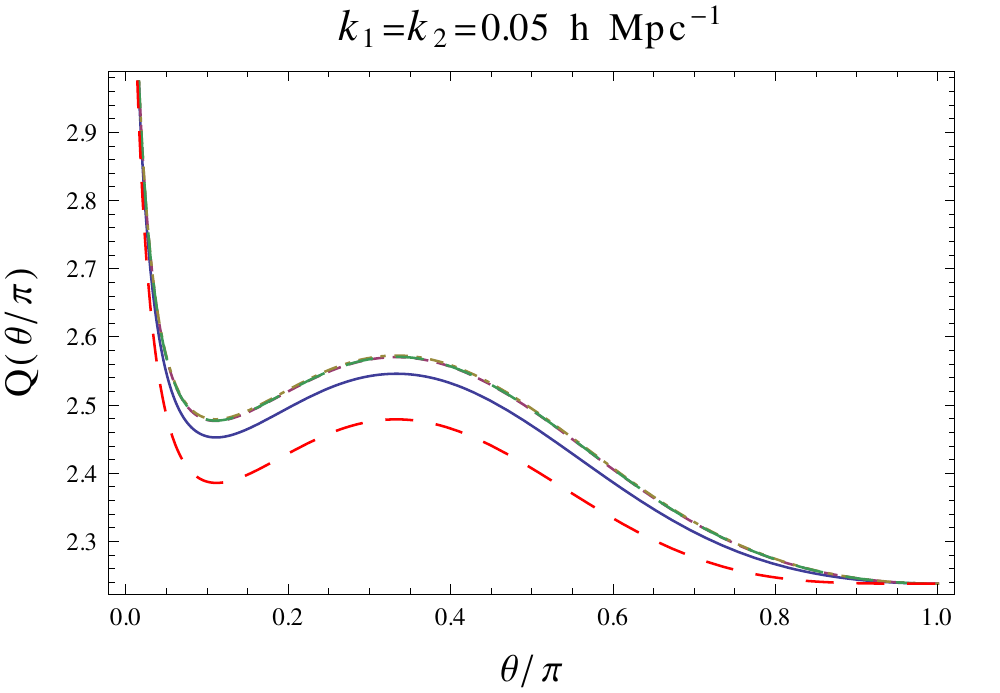} \includegraphics[width=0.505\textwidth]{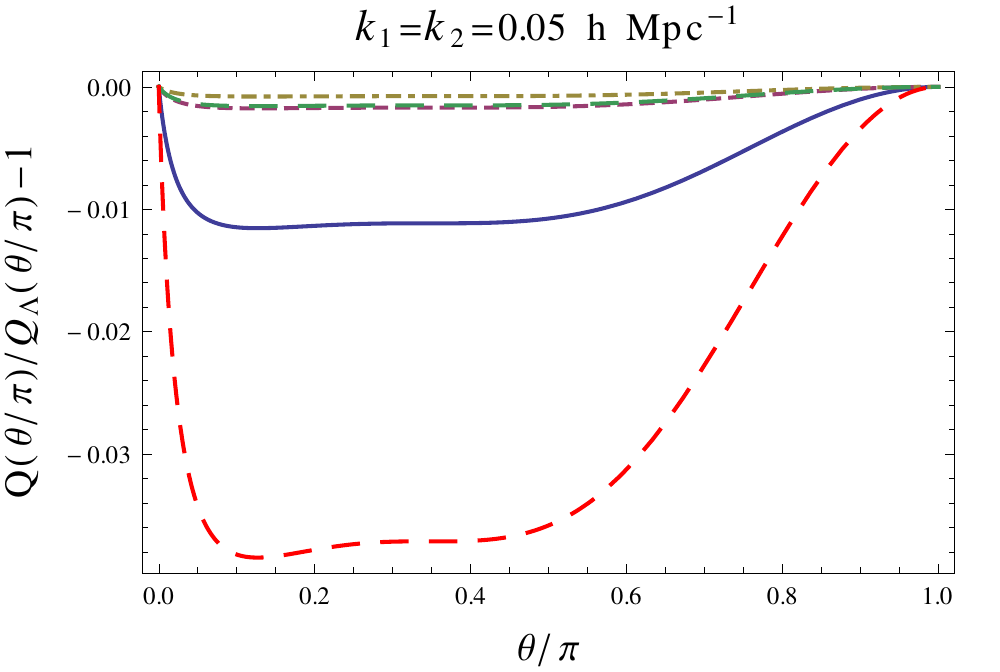}\\
\includegraphics[width=0.495\textwidth]{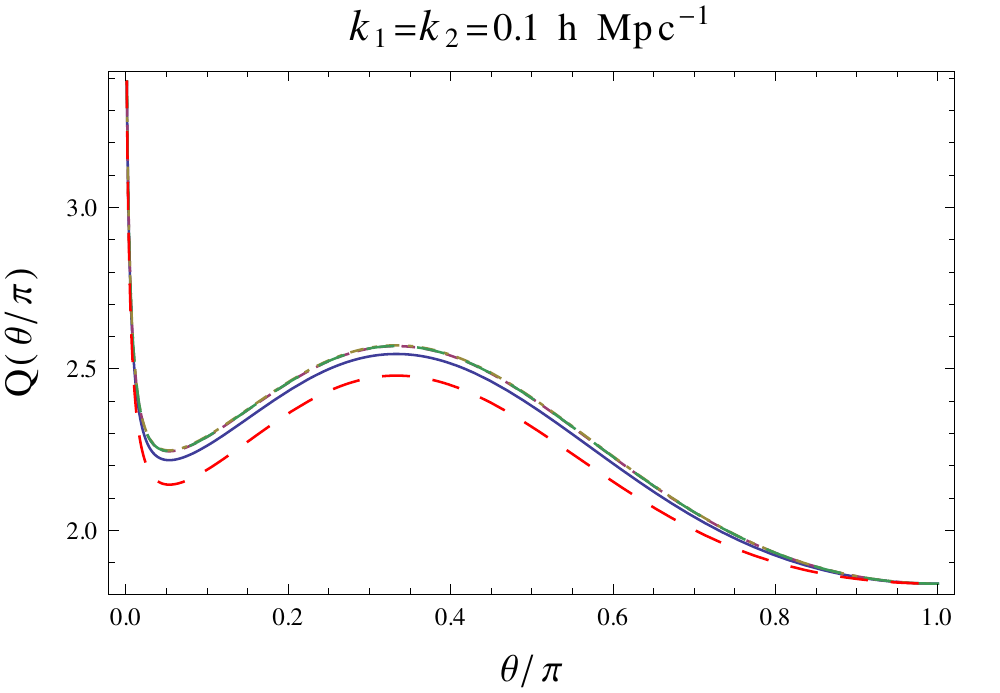} \includegraphics[width=0.505\textwidth]{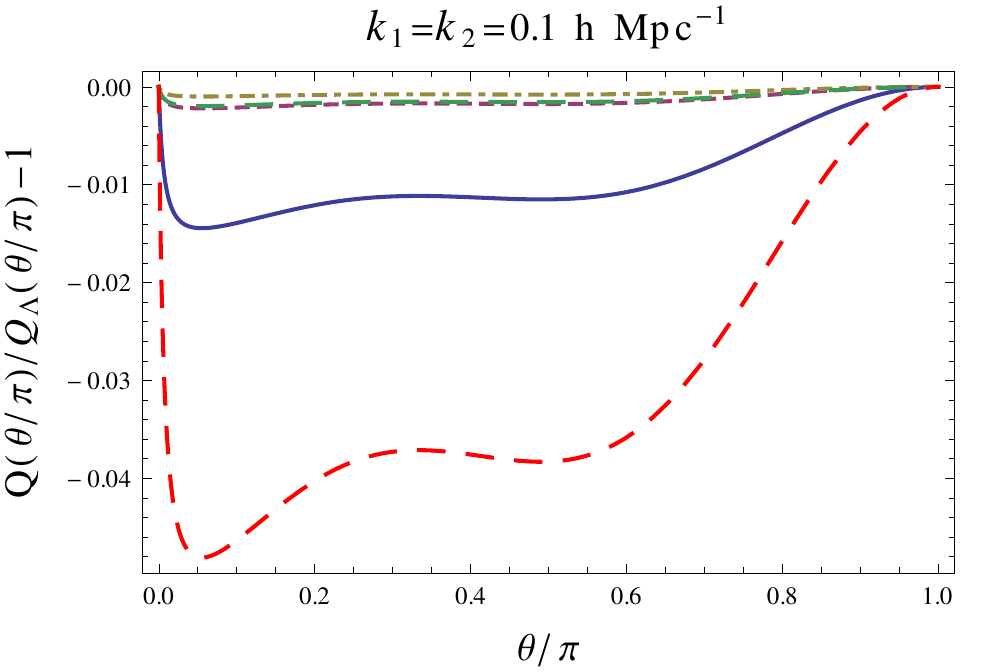}\\
\caption{In the left panels we plot the reduced bispectrum for some Galileon models as a function of the angle $\theta$, fixing $k_1=k_2$ at $a=1$. In the right panels we plot the relative deviations of the bispectrum of the Galileon w.r.t.\ the one of $\Lcdm$. The parameter values are: $c_1=1.6$, $c_2=0.04$, $c_3=10^{-3}$ (red line); $c_1=1.5$, $c_2=0.4$, $c_3=10^{-3}$ (blue line); $c_1=11$, $c_2=3.8$, $c_3=1$ (purple line); $c_1=6$, $c_2=3.6$, $c_3=1$ (yellow line); $c_1=10^{-4}$, $c_2=3.3$, $c_3=1$ (green line).}      \label{FIG:bispectrum1}
\end{figure}

\begin{figure}[!ht] %  figure placement: here, top, bottom, or page
   %\centering
\includegraphics[width=0.495\textwidth]{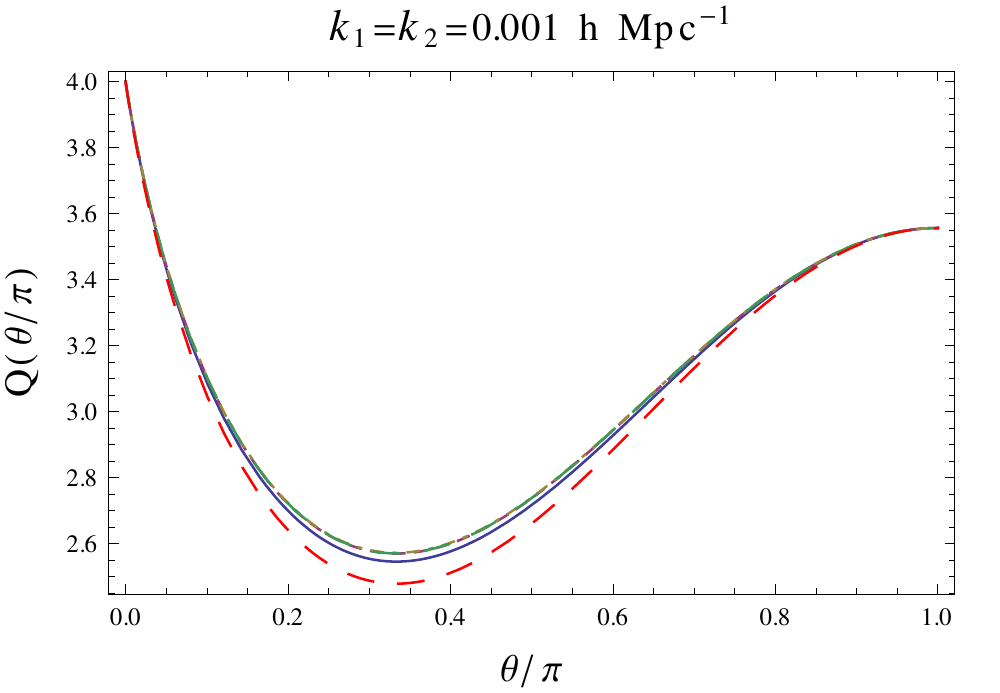} \includegraphics[width=0.505\textwidth]{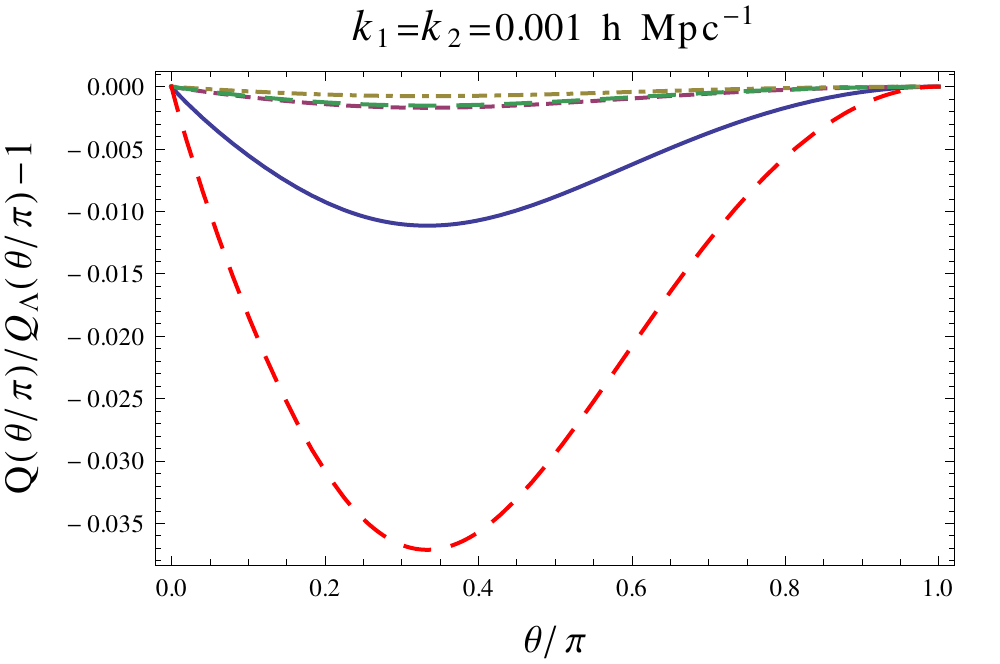}\\
\includegraphics[width=0.495\textwidth]{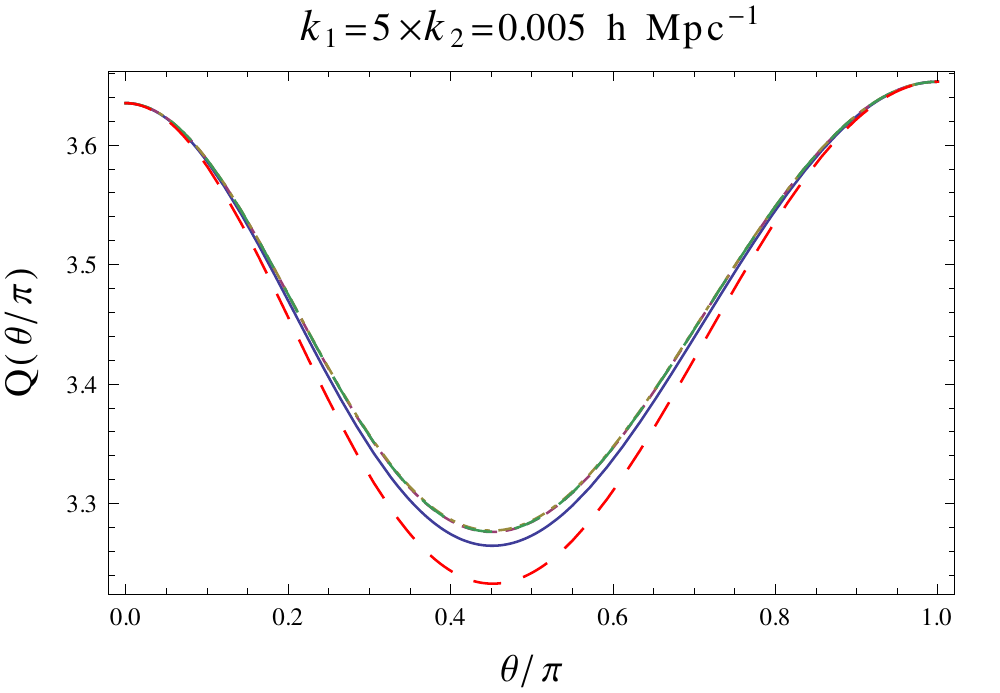} \includegraphics[width=0.505\textwidth]{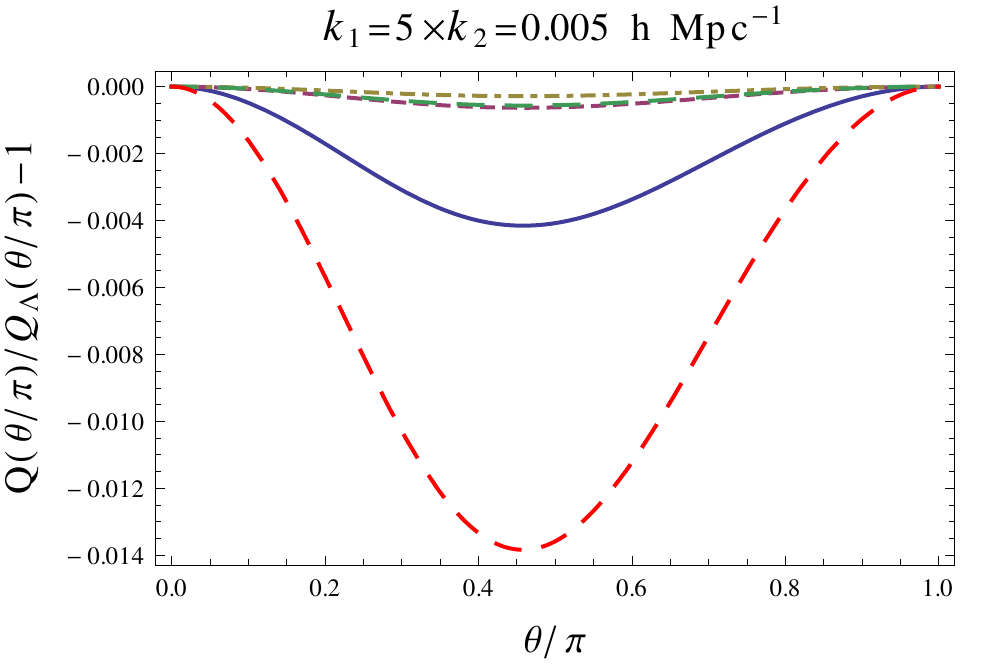}\\
\includegraphics[width=0.495\textwidth]{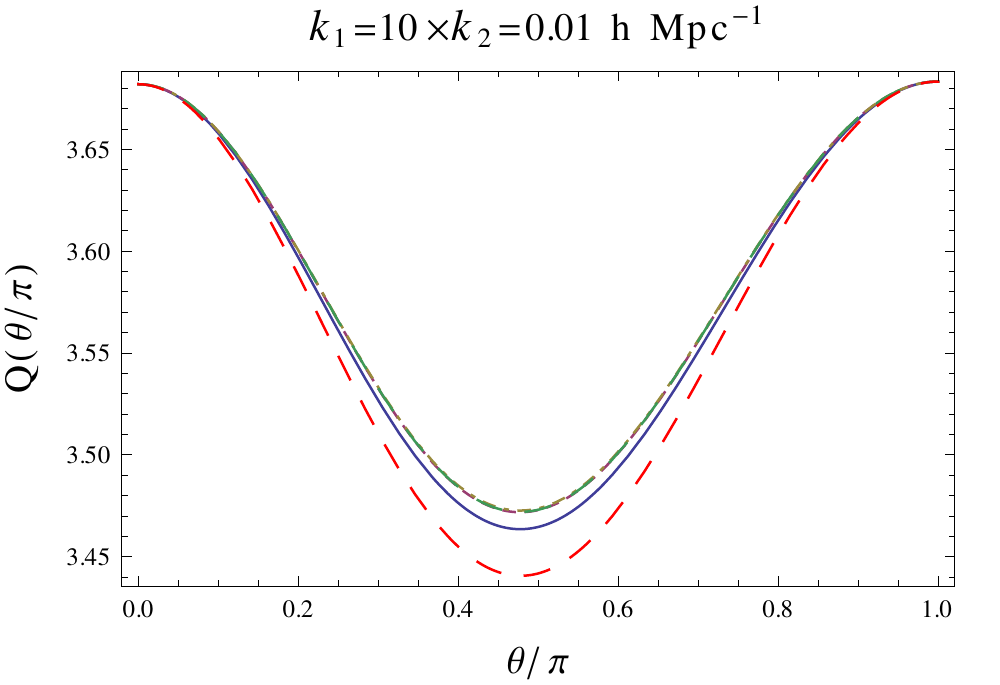} \includegraphics[width=0.505\textwidth]{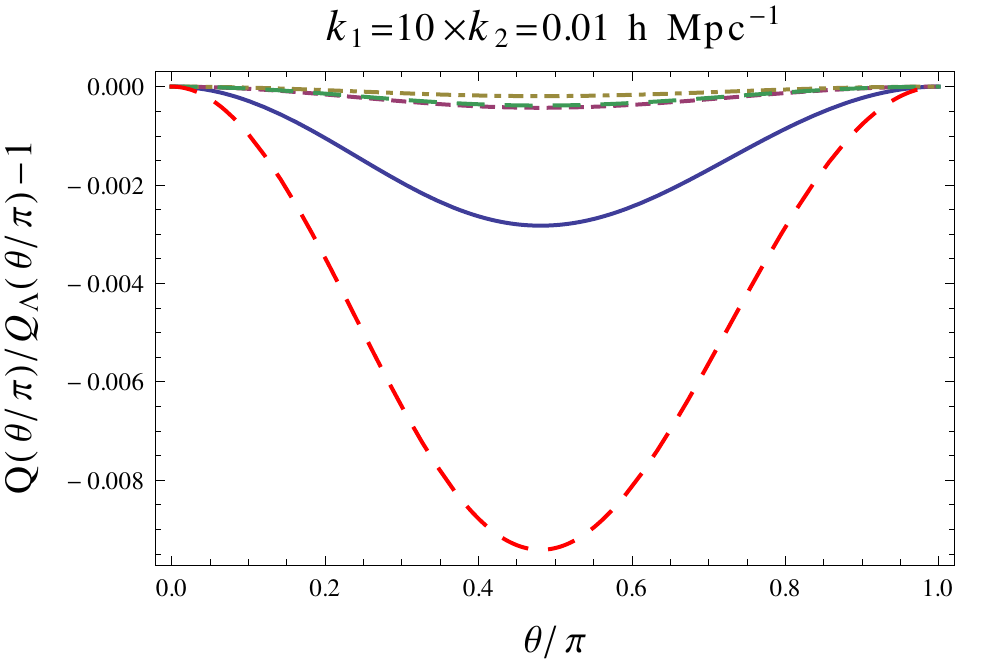}\\
\caption{The same as Fig.\ \ref{FIG:bispectrum1} fixing $k_1=const.\times k_2$ and $k_2=0.001 \,\rm{h\, {Mpc}^{-1}}$.} \label{FIG:bispectrum2}
\end{figure}

\begin{figure}[!ht] %  figure placement: here, top, bottom, or page
   %\centering
\includegraphics[width=0.495\textwidth]{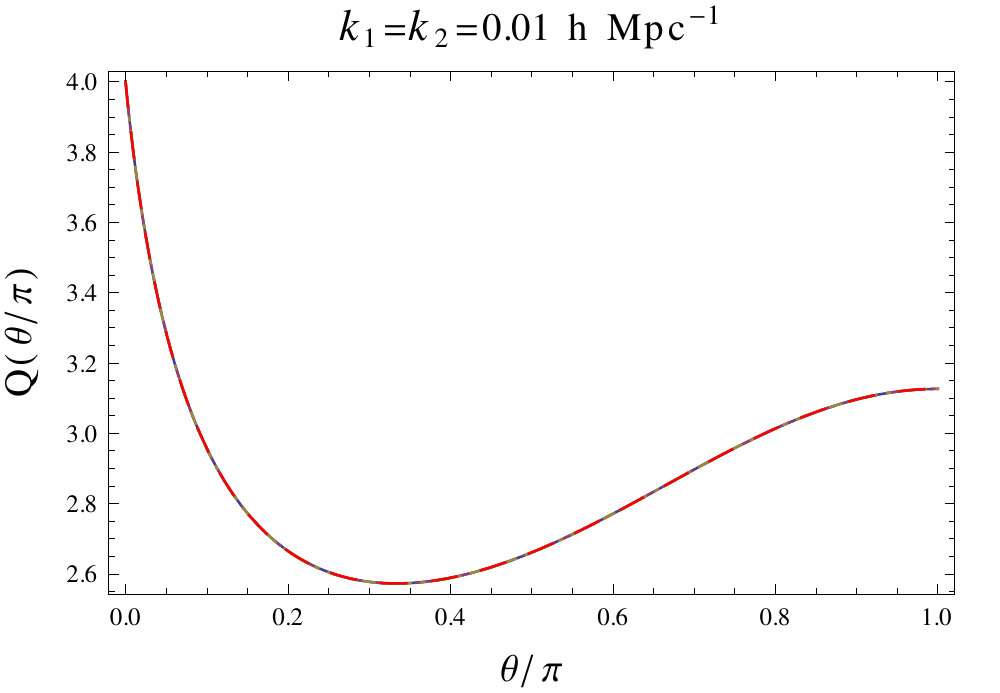} \includegraphics[width=0.505\textwidth]{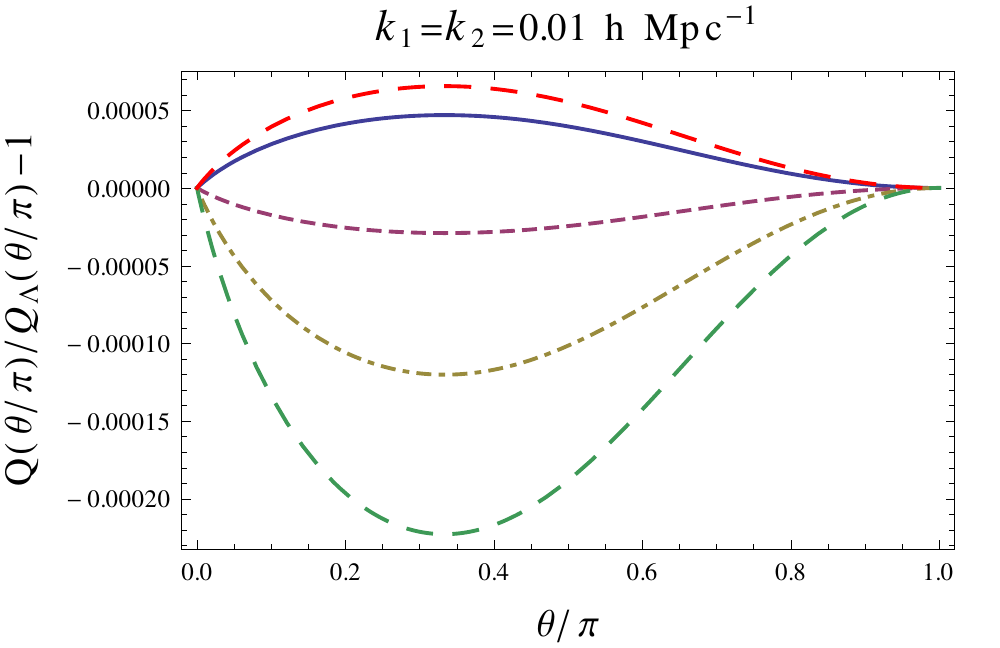}\\
\includegraphics[width=0.495\textwidth]{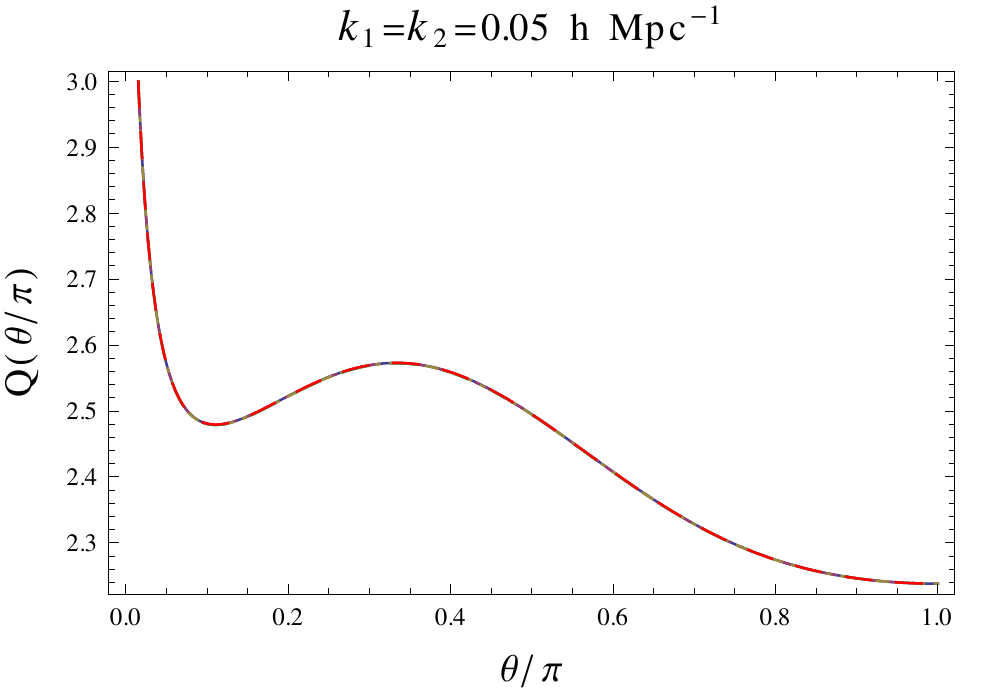} \includegraphics[width=0.505\textwidth]{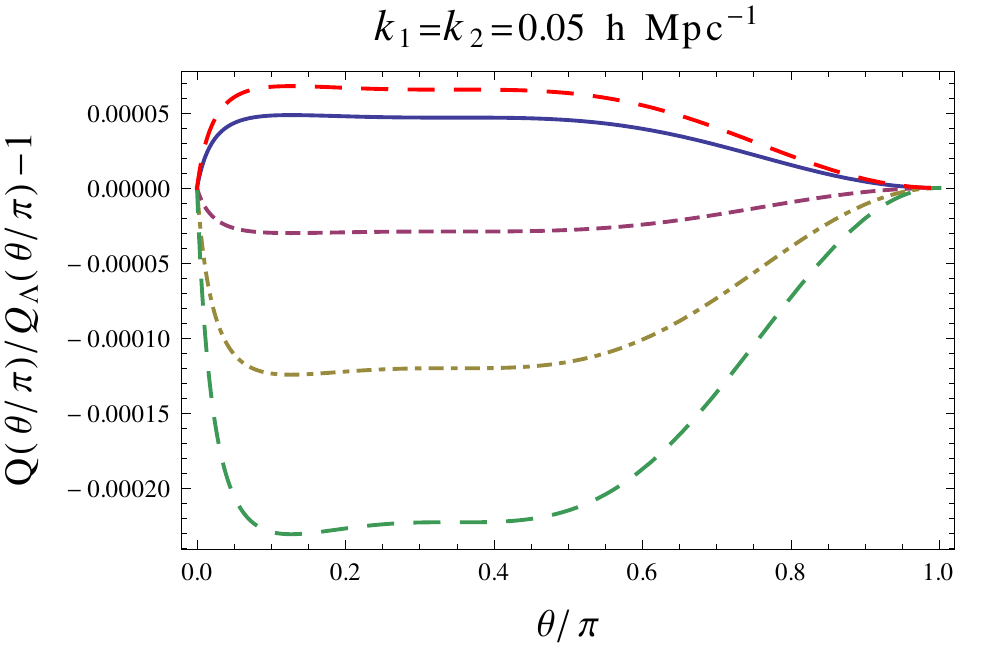}\\
\includegraphics[width=0.495\textwidth]{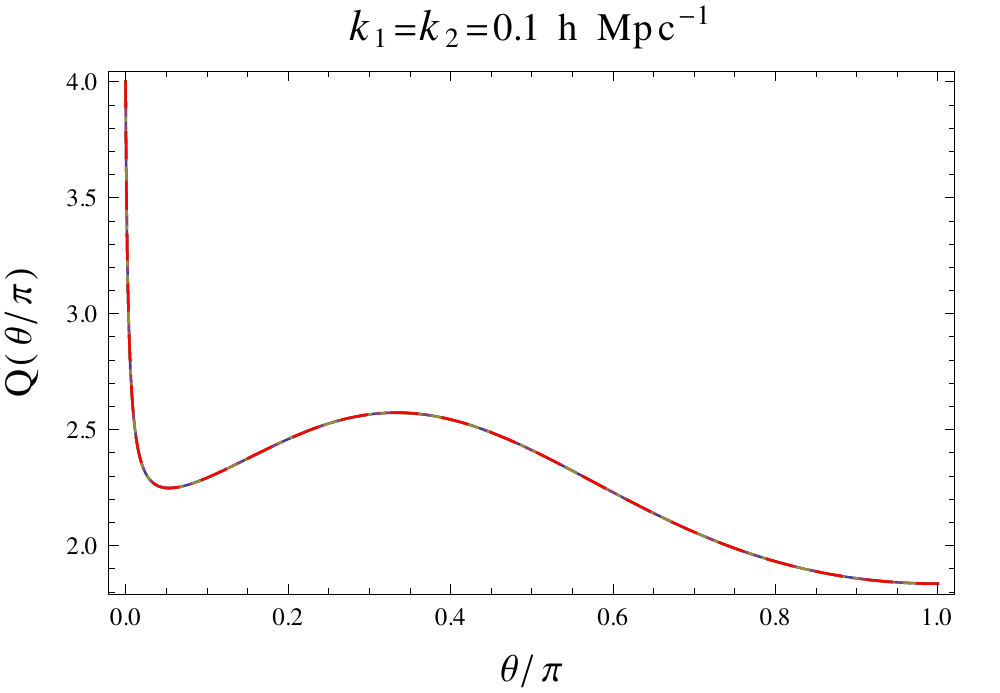} \includegraphics[width=0.505\textwidth]{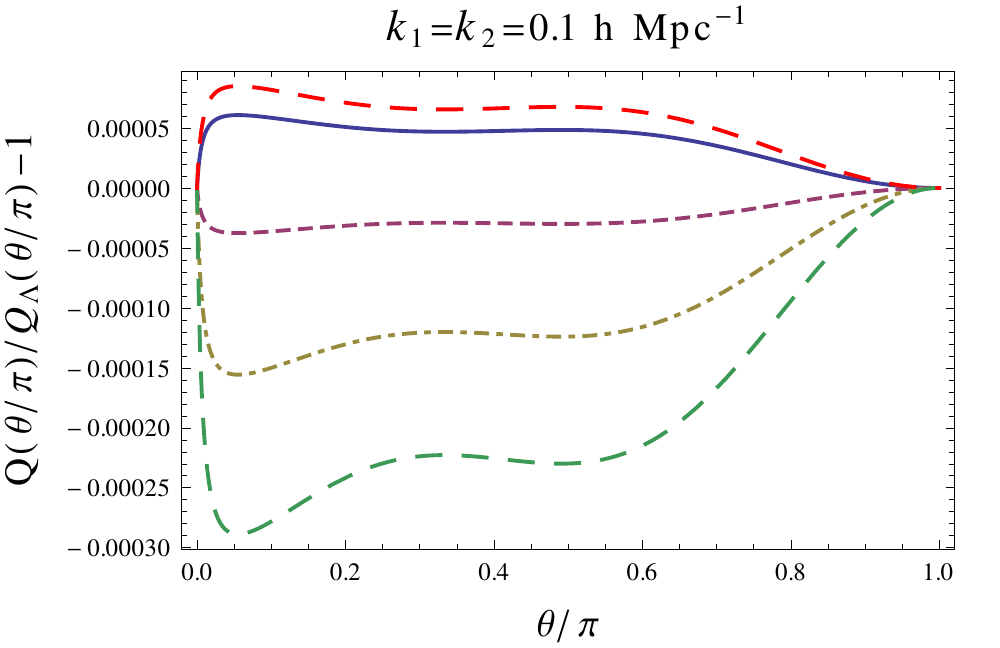}\\
\caption{In the left panels we plot the reduced bispectrum for some Galileon models as a function of the angle $\theta$, fixing $k_1=k_2$ at $a=0.6$. In the right panels we plot the relative deviations of the bispectrum of the Galileon w.r.t.\ the one of $\Lcdm$. The parameter values are: $c_1=1.6$, $c_2=0.04$, $c_3=10^{-3}$ (red line); $c_1=1.5$, $c_2=0.4$, $c_3=10^{-3}$ (blue line); $c_1=11$, $c_2=3.8$, $c_3=1$ (purple line); $c_1=6$, $c_2=3.6$, $c_3=1$ (yellow line); $c_1=10^{-4}$, $c_2=3.3$, $c_3=1$ (green line).}      \label{FIG:bispectrum3}
\end{figure}
\begin{figure}[!ht] %  figure placement: here, top, bottom, or page
   %\centering
\includegraphics[width=0.495\textwidth]{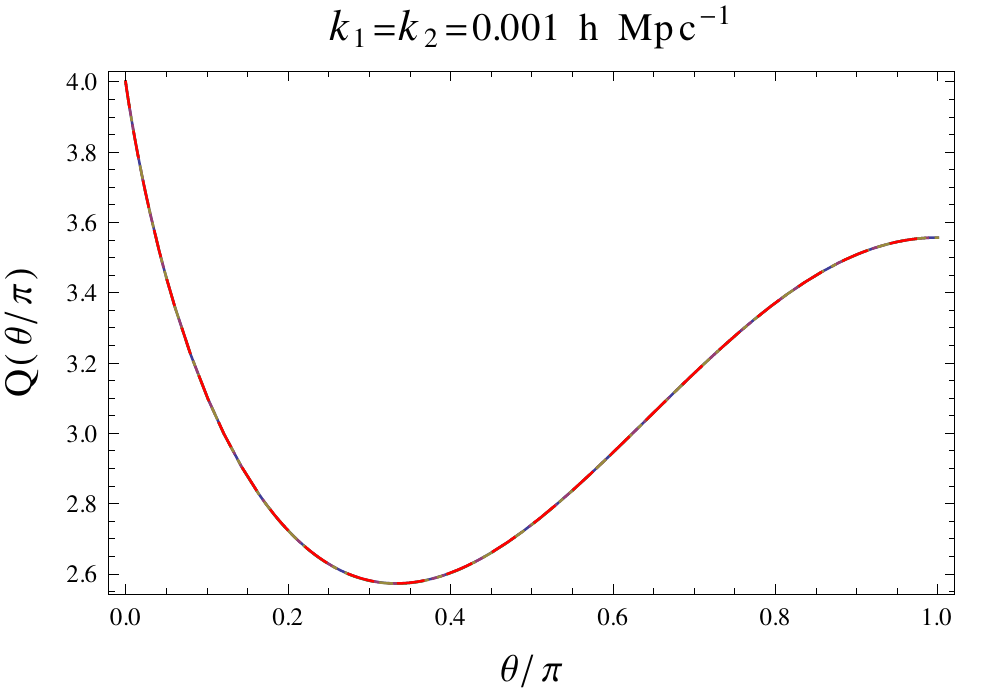} \includegraphics[width=0.505\textwidth]{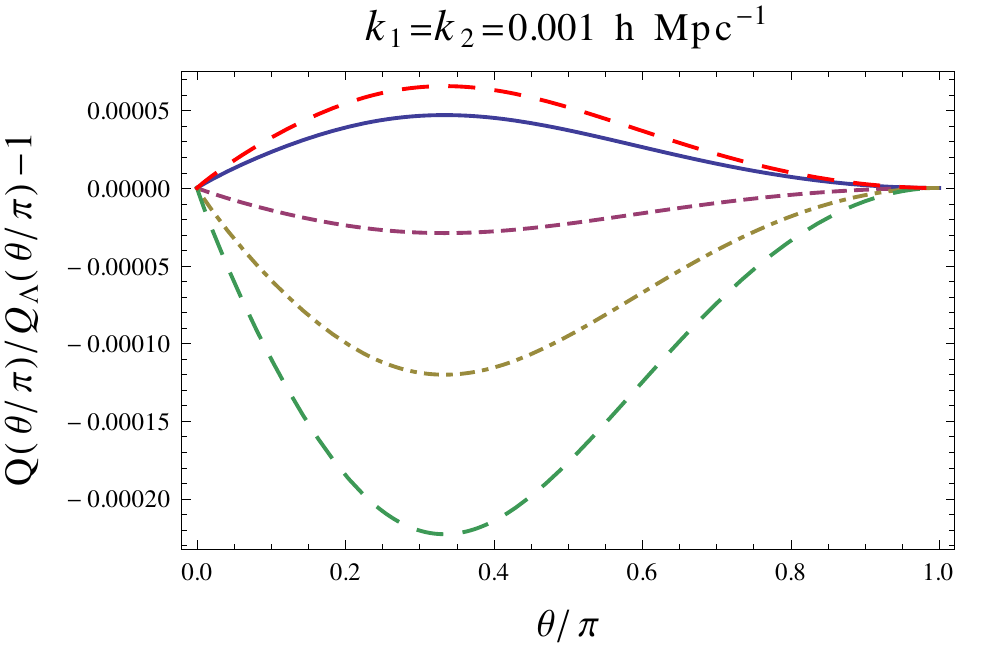}\\
\includegraphics[width=0.495\textwidth]{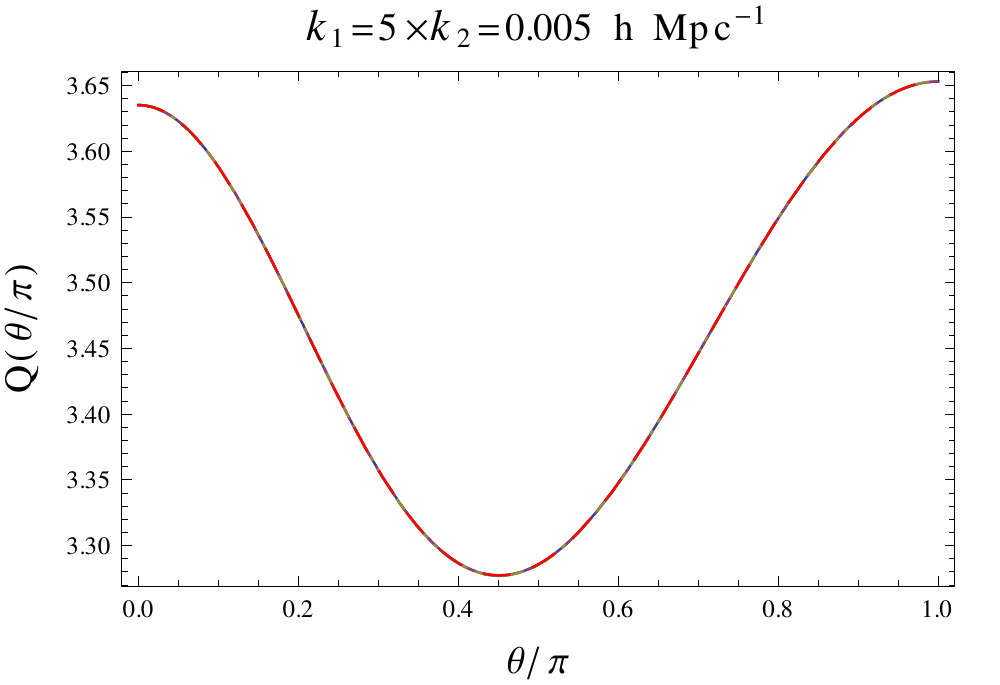} \includegraphics[width=0.505\textwidth]{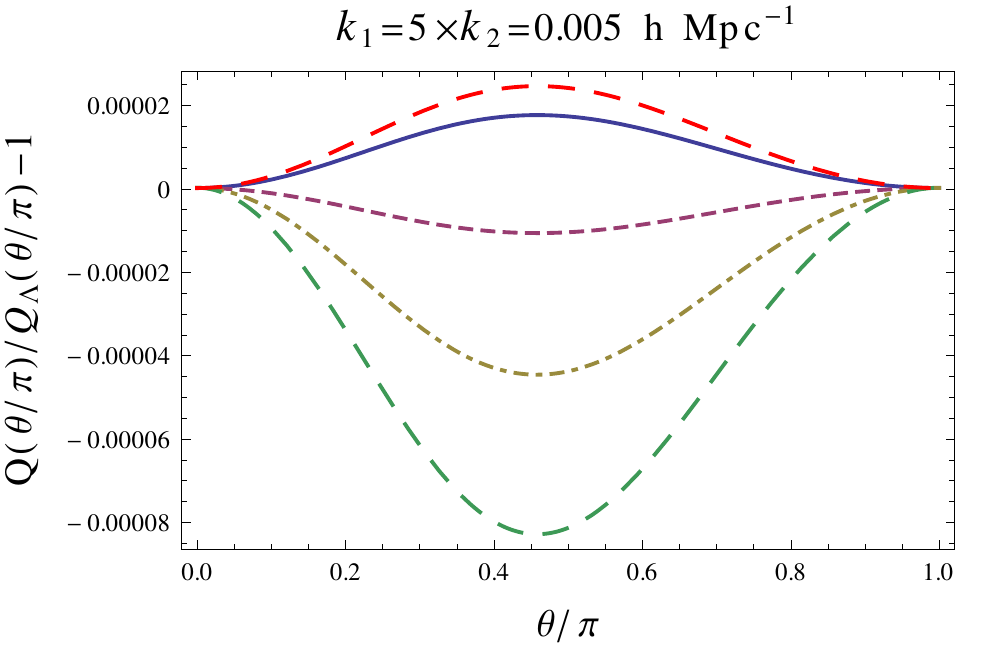}\\
\includegraphics[width=0.495\textwidth]{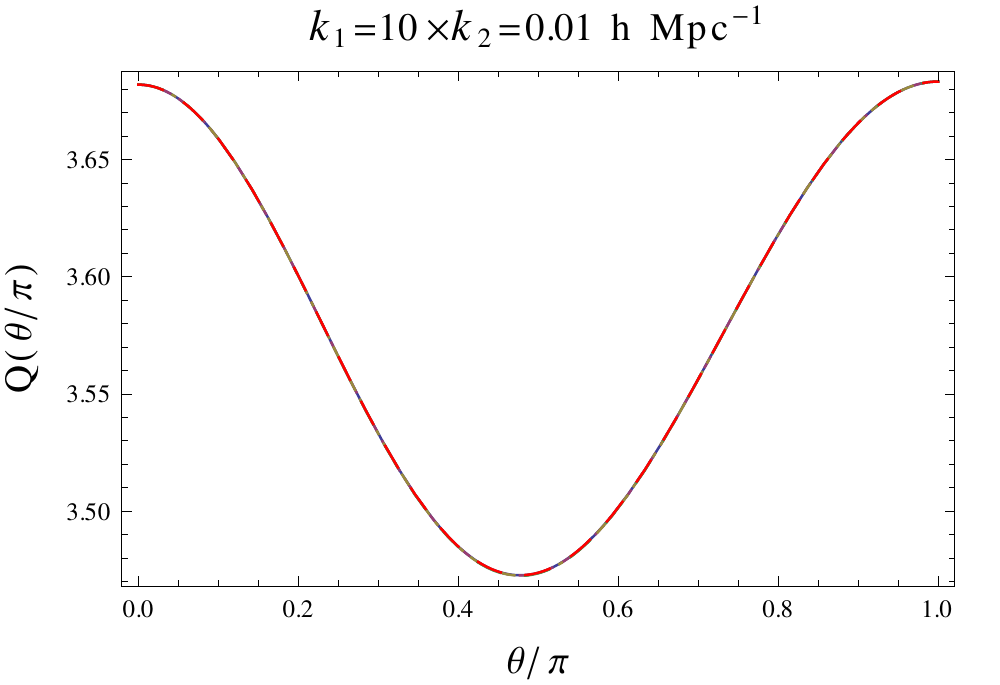} \includegraphics[width=0.505\textwidth]{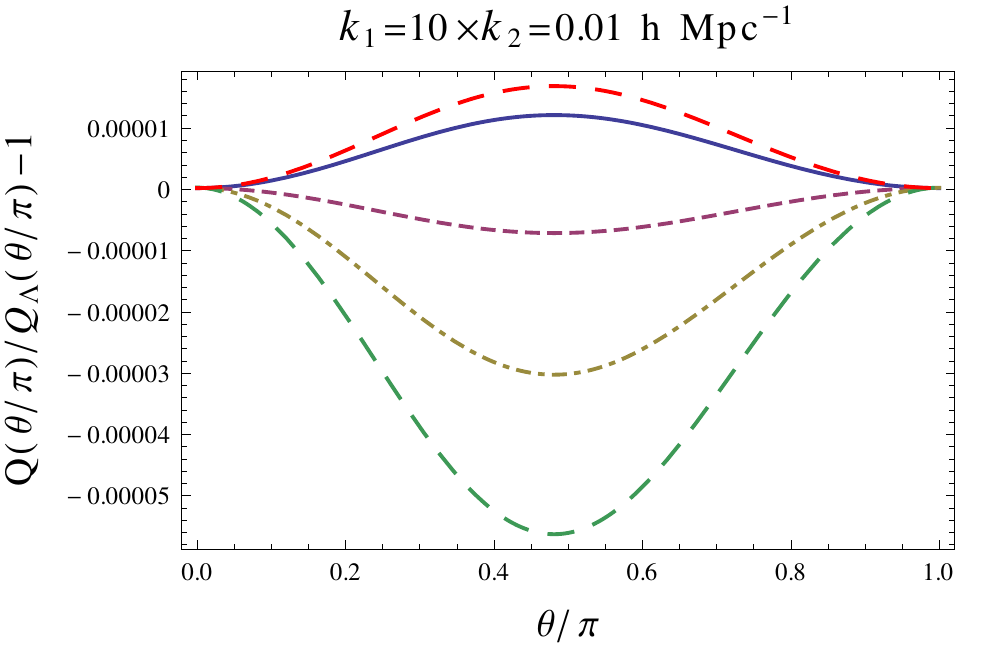}\\
\caption{The same as Fig.\ \ref{FIG:bispectrum3} fixing $k_1=const.\times k_2$ and $k_2=0.001 \,\rm{h\, {Mpc}^{-1}}$.} \label{FIG:bispectrum4}
\end{figure}

The scales at which our approximations can give valid results are $10^{-4} \,\rm{h\, {Mpc}^{-1}} \ll k \lesssim 10^{-1} \,\rm{h\, {Mpc}^{-1}}$. The first inequality follows from the sub-horizon approximation, while the second excludes the scales at which highly non-linear effects become non-negligible. In Figs.\ \ref{FIG:bispectrum1} and \ref{FIG:bispectrum3} we show the angular dependence of the reduced bispectrum  for different Galileon models, 
at $a=1$ and at $a=0.6$ respectively, fixing $k_1=k_2$, $\theta$ being the angle between $\vec{k}_1$ and $\vec{k}_2$ ($\vec{k}_1\cdot \vec{k}_2=k_1 k_2 \cos\theta$). In Fig.\ \ref{FIG:bispectrum2} and \ref{FIG:bispectrum4} we show the angular dependence of the reduced bispectrum, at $a=1$ and at $a=0.6$ respectively, fixing $k_1=const.\times k_2$ and $k_2=10^{-3} \,\rm{h\, {Mpc}^{-1}}$.

\begin{figure}[!ht] %  figure placement: here, top, bottom, or page
   %\centering
\begin{center}
\includegraphics[width=0.5\textwidth]{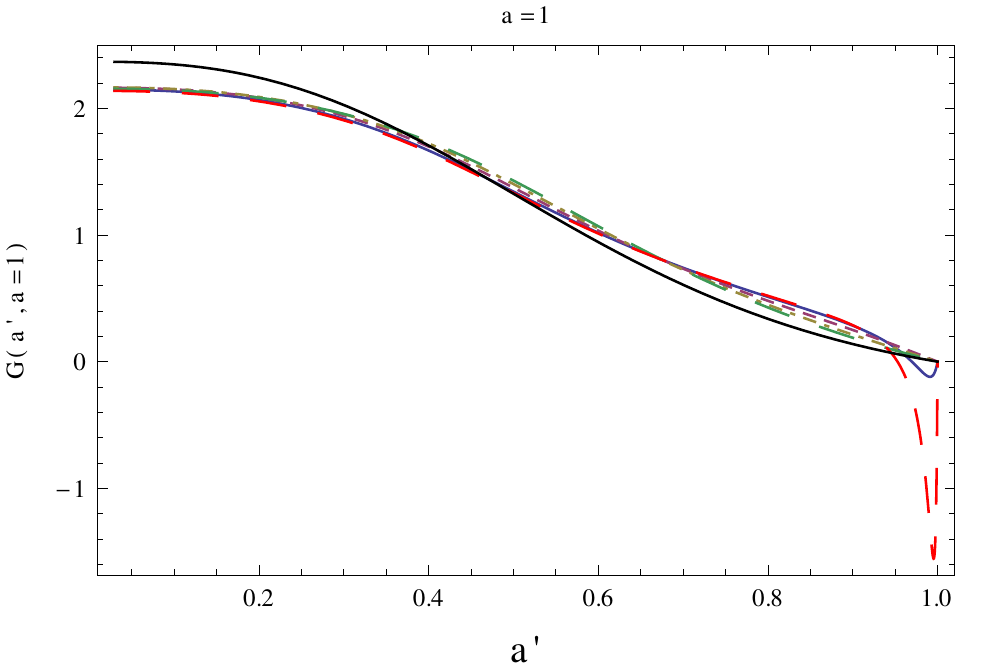}\includegraphics[width=0.5\textwidth]{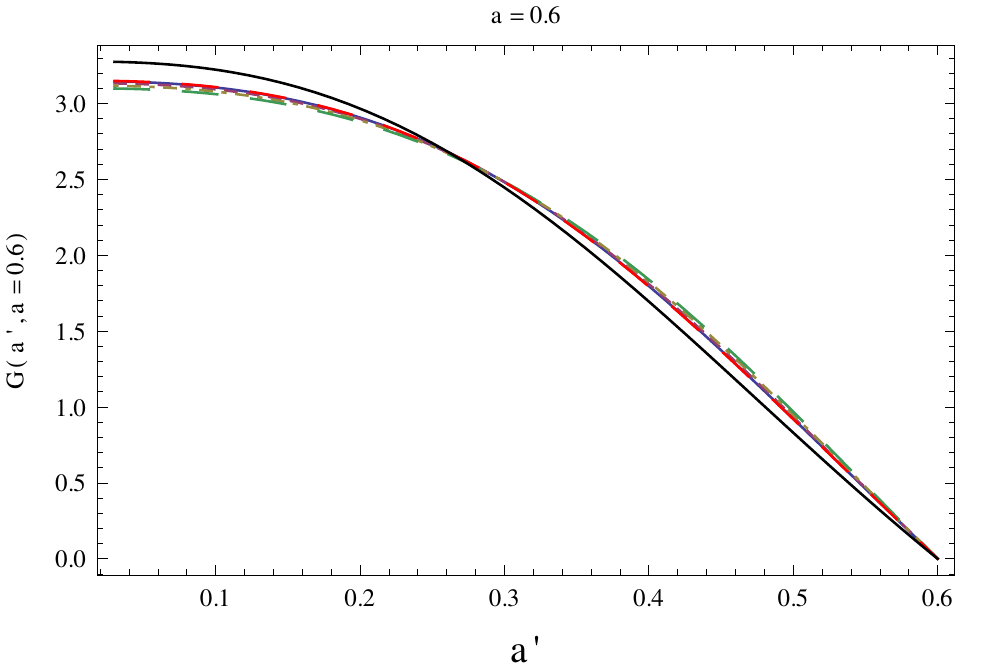}\\
\end{center}
\caption{The integrand of Eq.\ \ref{EQ:kernelF} for an equilateral configuration at $a=1$ (left panel) and $a=0.6$ (right panel). The models we plot are the same as in Fig.\ \ref{FIG:background}, while the black line represents the $\lcdm$} model. \label{FIG:Ga1}
\end{figure}

In Fig.\ \ref{FIG:Ga1} we show the evolution of
\begin{align}
 G(a^\prime,a)\equiv \frac{{D_+}^2(a')\left[D_-(a) D_+(a')-D_+(a) D_-(a')\right]}{{a'}^2 W(a') {D_+}^2(a)} \frac{\mathcal{K}_{SH}(a',\vec{q}_1,\vec{q}_2)}{2\mathcal{H}^2(a')}
\end{align}
for an equilateral configuration at $a=1$ (left panel) and $a=0.6$ (right panel). This configuration is useful to understand the behavior of the reduced bispectrum, Eq.\ \ref{EQ:reducedbispectrum2}, because it is totally independent of the power-spectrum, in fact $Q(a,k_1,k_1,k_1)=2 F(a,\vec{k}_1,\vec{k}_1)$. As one can see in the left panel of Fig.\ \ref{FIG:Ga1} the function $G(a=1,a^\prime)$ contains a compensation effect that reduces the deviations w.r.t.\ the $\lcdm$ model in the bispectrum, as shown in Figs.\ \ref{FIG:bispectrum1}, \ref{FIG:bispectrum2}, \ref{FIG:bispectrum3} and \ref{FIG:bispectrum4}. Let us notice that, for $a^\prime\lesssim 0.4$, the line of every Galileon model we consider lies below the $\lcdm$ line; viceversa, for $a^\prime\gtrsim 0.4$, except for the red and blue lines, for which we find the strongest deviations, the Galileon lines lie above the $\lcdm$ line (up to  the present epoch). Consequently, when we integrate $G(a=1,a^\prime)$, the deviations that we have 
obtained studying the background and the power-spectrum are 
attenuated considerably. Instead, when $w\gtrsim -0.85$ --  corresponding to the red and blue lines, see Fig.\ \ref{FIG:background} -- we see a minimum below the $\lcdm$ around $a^\prime\simeq 1$. This feature decreases the compensation effect and produces larger deviations in the dark matter bispectrum. This could be explained by the fact that the universe is not accelerating enough today and the evolution of the growth rate is strongly modified (see Fig.\ \ref{FIG:growthrate}). For these cases the deviations we find in the bispectrum are about $\simeq 5\%$. Instead, computing $B(a,k_1,k_2,k_3)$ before the acceleration of the universe, the compensation effect is conserved because the contribution of the Galileon is negligible and all models are indistinguishable (see for example the right panel of Fig.\ \ref{FIG:Ga1} and the tiny deviations seen in Figs.\ \ref{FIG:bispectrum3} and \ref{FIG:bispectrum4}).

\section{Conclusions}\label{SEC:Conclusions}

In this paper we have presented an analytic expression for the DM modified bispectrum in the context of the cubic covariant Galileon theory. We worked on sub-horizon scales at second order in the perturbations, to show the leading contribution in the weakly non-linear regime. We have first studied the background with the most general potential that preserves the Galilean shift symmetry in a flat space-time. The contribution of $c_3$ is crucial to drive the late-time cosmic acceleration, however the deviations w.r.t.\ the $\lcdm$ model are smaller if $c_1\sim c_3$. At the linear level we have studied the evolution of the DM perturbations finding semi-analytical expressions for the growing and the decaying modes. In Fig.\ \ref{FIG:growthrate} we plot the deviations of the Galileon growth rate w.r.t.\ the growth rate of the $\lcdm$ model.
For models in which the value of $c_3$ is negligible w.r.t.\ the value of $c_1$ the deviations are large (until about $100\%$), while, increasing $c_3$ the deviations decrease reaching $\simeq 10\%$.

Then, we have extended our analysis to second-order perturbations in order to calculate the DM bispectrum. Eq.\ (\ref{kernel2}), is one of our main results. It shows that the overall $\vec{k}$-dependence of the bispectrum is the same as in the $\lcdm$ model, with time dependent coefficients which depend on the particular Galileon model. 
We noted that, in general, there is a compensation effect (see Fig.\ \ref{FIG:Ga1}) in the integrand of Eq.\ (\ref{kernel2}), $G(a,a^\prime)$, that reduces the deviations w.r.t.\ the $\lcdm$ model in the bispectrum. 
This effect is conserved if we compute $B(a,k_1,k_2,k_3)$ before the accelerated phase of the universe (see the right panel of Fig.\ \ref{FIG:Ga1}), because the contribution of the Galileon is negligible and all models are indistinguishable. If the bispectrum is evaluated today and the model has $w\lesssim-0.85$, the compensation effect is preserved, giving deviations up to $\simeq 1\%$. Instead, we noted that this effect is less strong for those models which have 
$w\gtrsim-0.85$, allowing for larger deviations in the bispectrum up to $\simeq 5\%$. We argue that the Vainshtein mechanism can be a possible explanation for the overall suppression of the deviations w.r.t.\ the $\lcdm$ model in the DM bispectrum and we leave for future work further investigation of this aspect.

\acknowledgments{We thanks Bin Hu and Angelo Ricciardone for useful discussions. This research has been partially supported by the ASI/INAF Agreement I/072/09/0 for the Planck LFI Activity of Phase E2 and by the PRIN 2009 project ``La Ricerca di non-Gaussianit\`a Primordiale''. DB is supported by the South African Square Kilometre Array Project.}

\appendix

\section{Gauges in linear approximation}\label{SEC:AppendixB}

\subsection*{Poisson Gauge}

This gauge is very useful because in many cases the scalar metric perturbation $\psi$ can be interpreted as the Newtonian potential. It can be obtained suppressing the off-diagonal terms of the metric

\begin{align}
&\chi^{(1)}=0\nonumber\\
&\omega^{(1)}=0\,.
\end{align}

From Eq.\ (\ref{EQ:linear4}), we obtain the standard result
\beq\label{EQ:linearpsi}
\psi^{(1)}=\phi^{(1)}\,,
\eeq
while Eq.\ (\ref{EQ:linear7}) reads
\beq\label{EQ:linearvelocity}
\psi^{(1)} + v^{(1)\prime}=0\,.
\eeq

In sub-horizon approximation the field equation for the galileon, Eq.\ (\ref{EQ:linear5}) reads
\begin{align}\label{EQ:linearpi}
\left(c_2 - \frac{2 c_3 \ddpi}{M^3 a^2} - 2 \frac{c_3 \mathcal{H}\dpi}{M^3 a^2}\right) k^2 \pif - \frac{c_3 \dpi^2}{M^3 a^2} k^2 \psi^{(1)}=0\,.
\end{align}

Using also the time-time component of the Einstein equations, Eq.\ (\ref{EQ:linear1}), we obtain
\begin{align}\label{EQ:linearphi}
&\left(1 + \frac{c_3^2 \dpi^4}{2 c_2 M^6 \Mpl^2 a^4} - \frac{2 c_3 \ddpi}{c_2 M^3 a^2} - \frac{2 c_3 \mathcal{H}\dpi}{c_2 M^3 a^2}\right) k^2 \psi^{(1)} \nonumber\\
&+ \left(\frac{1}{2 \Mpl^2} - \frac{c_3 \ddpi}{c_2 M^3 \Mpl^2 a^2} - \frac{c_3 \mathcal{H}\dpi}{c_2 M^3 \Mpl^2 a^2}\right) a^2\rho_m\delta^{(1)}=0\,.
\end{align}

Substituting Eqs.\ (\ref{EQ:linearvelocity}) and then (\ref{EQ:linearpsi}) into the derivative of Eq.\ (\ref{EQ:linear6}), in sub-horizon approximation we obtain
\beq\label{EQ:lineardeltaprovv}
\delta^{(1)\prime\prime} + \mathcal{H} \delta^{(1)\prime} =- k^2 \psi^{(1)}\,.
\eeq

Using Eq.\ (\ref{EQ:linearphi}) to eliminate the metric perturbation $\phi$ in Eq.\ (\ref{EQ:lineardeltaprovv}), the result is
\beq\label{EQ:lineardeltaP}
\delta^{(1)\prime\prime} + \mathcal{H} \delta^{(1)\prime} = 4 \pi G\left(1-\frac{{c_3}^2 {\pi^\prime}^4}{2 c_2 M^6 \Mpl^2 a^4 \a}\right) a^2 \rho_m\delta^{(1)}\,.
\eeq

This equations studies the dynamics of the DM perturbation $\delta^{(1)}$, and it is the same equation obtained without choosing a gauge, Eq.\ (\ref{EQ:lineardelta}).

\subsection*{Spatially Flat Gauge}

The spatially flat gauge can be obtained by considering the spatial scalar fluctuations equal to zero

\begin{align}
&\phi^{(1)}=0\nonumber\\
&\chi^{(1)}=0\,.
\end{align}

In this gauge Eq.\ (\ref{EQ:linear7}) remains the same, while from Eq.\ (\ref{EQ:linear4}) and its derivative we can solve for $\psi^{(1)}$
\begin{align}
\psi^{(1)}= - \domf - 2 \mathcal{H} \om^{(1)}\,.
\end{align}

To separate the galileon perturbation we use Eq.\ (\ref{EQ:linear5}) in sub-horizon approximation

\begin{align}
\left(c_2 - \frac{2 c_3 \ddpi}{M^3 a^2} - \frac{2 c_3 \mathcal{H}\dpi}{M^3 a^2}\right) \pif + \left(c_2 \dpi - \frac{2 c_3 \dpi\ddpi}{M^3 a^2} - \frac{c_3 \mathcal{H}\dpi^2}{M^3 a^2}\right) \om^{(1)}=0\,.
\end{align}

Using the last equation in Eq.\ (\ref{EQ:linear1}), after a sub-horizon approximation, to eliminate the galileon field $\pif$ we obtain
\begin{align}
&\left(c_2 - \frac{2 c_3 \ddpi}{M^3 a^2} - \frac{2 c_3 \mathcal{H}\dpi}{M^3 a^2}\right) a^2\rho_m \delta^{(1)} \nonumber\\
&= \left(2 c_2 \Mpl^2 \mathcal{H} + \frac{c_3^2 \mathcal{H}\dpi^4}{M^6 a^4} - \frac{4 c_3 \Mpl^2 \mathcal{H}\ddpi}{M^3 a^2} - \frac{4 c_3 \Mpl^2 \mathcal{H}^2\dpi}{M^3 a^2}\right) k^2 \om^{(1)}\,.
\end{align}

Finally, to find the dynamics of $\delta^{(1)}$ we have to substitute $\omega^{(1)}$ from the last equation in the derivative of Eq.\ (\ref{EQ:linear6}). It is straightforward to show that also in this gauge the result is identical w.r.t.\ Eq.\ (\ref{EQ:lineardelta}).

\subsection*{Synchronous Gauge}

The synchronous gauge is a gauge that, at first order, leaves only the spatial scalar perturbations

\begin{align}
&\psi^{(1)}=0\nonumber\\
&\omega^{(1)}=0\,.
\end{align}

It is slightly different from the other gauges described, because it has a residual gauge freedom. From Eq.\ (\ref{EQ:linear7}) we find that the velocity $v^{(1)}$ must satisfy
\beq\label{EQ:velocityS}
v^{(1)\prime}+\mathcal{H}v^{(1)}=0\,.
\eeq

One can fix the residual gauge freedom imposing the additional condition $v^{(1)}=0$. However we do not need to fix it to decouple on sub-horizon scales the DM density fluctuation $\delta^{(1)}$. Taking the difference between Eq.\ (\ref{EQ:linear3}) and Eq.\ (\ref{EQ:linear1}), and performing a sub-horizon approximation the result is
\begin{align}\label{EQ:linearpsiS}
6\Mpl^2 \ddphif + \left(6\Mpl^2 \mathcal{H} + \frac{3 c_3 \dpi^3}{M^3 a^2}\right) \dphif = a^2\rho_m\delta^{(1)}+ \frac{c_3\dpi^2}{M^3 a^2} k^2 \pif\,.
\end{align}

In this gauge Eq.\ (\ref{EQ:linear5}) reads
\begin{align}\label{EQ:linearpiS}
\frac{3 c_3 \dpi^2}{M^3 a^2} \ddphif &+ \left(\frac{6 c_3\dpi\ddpi}{M^3 a^2} + \frac{9 c_3 \mathcal{H}\dpi^2}{M^3 a^2} - 3 c_2 \dpi\right) \dphif \nonumber\\
&= - \left(c_2 - \frac{2 c_3\ddpi}{M^3 a^2} - \frac{2 c_3 \mathcal{H}\dpi}{M^3 a^2}\right) k^2 \pif\,.
\end{align}

Combining Eqs.\ (\ref{EQ:linearpsiS}) and (\ref{EQ:linearpiS}) to eliminate the galileon field $\pif$ we obtain

\begin{align}
&\left(6 c_2 \Mpl^2 + \frac{3 c_3^2 \dpi^4}{M^6 a^4} - \frac{12 c_3 \Mpl^2 \left(\ddpi +\mathcal{H}\dpi\right)}{M^3 a^2}\right) \left[\ddphif + \mathcal{H}\dphif\right] \nonumber\\
&=\left(c_2 + \frac{2 c_3 \left(\ddpi +\mathcal{H}\dpi\right)}{M^3}\right) a^2\rho_m \delta^{(1)}\,.
\end{align}

It is now straightforward to use this equation, Eq.\ (\ref{EQ:velocityS}) and the derivative of Eq.\ (\ref{EQ:linear6}),

\begin{align}
\delta^{(1)\prime\prime} + \mathcal{H}\delta^{(1)\prime} = 3 \left(\ddphif + \mathcal{H} \dphif\right)\,,
\end{align}
to obtain Eq.\ (\ref{EQ:lineardelta}).

\section{Source terms for the second-order equations of motion} \label{SEC:AppendixA}

In the following we give the explicit expression in a general gauge of the source terms found in Sec.\ \ref{SEC:second}. They reads

\begin{align}
 S^{(1)} &\equiv c_2 {{\pi^{(1)}}^{\prime}}^{2} + 6 {\Mpl}^{2} {{\phi^{(1)}}^{\prime}}^{2} + \frac{1}{12} {\Mpl}^{2} {\nabla^2 {\chi^{(1)}}^{\prime}}^{2} + \frac{2}{3} {\Mpl}^{2} \mathcal{H} \nabla^2 {\chi^{(1)}}^{\prime} \nabla^2 \chi^{(1)} + \frac{1}{9} {\Mpl}^{2} \nabla^2 \nabla^2 \chi^{(1)} \nabla^2 \chi^{(1)} \nonumber \\ 
& + 4 {\Mpl}^{2} {\phi^{(1)}}^{\prime} \nabla^2 \omega^{(1)} -  \frac{1}{3} {\Mpl}^{2} \nabla^2 {\chi^{(1)}}^{\prime} \nabla^2 \omega^{(1)} -  \frac{4}{3} {\Mpl}^{2} \mathcal{H} \nabla^2 \chi^{(1)} \nabla^2 \omega^{(1)} + {\Mpl}^{2} {\nabla^2 \omega^{(1)}}^{2} \nonumber \\ 
&+ \frac{2}{3} {\Mpl}^{2} \nabla^2 \chi^{(1)} \nabla^2 \phi^{(1)} - 24 {\Mpl}^{2} {\phi^{(1)}}^{\prime} \mathcal{H} \phi^{(1)} + \frac{8}{3} {\Mpl}^{2} \nabla^2 \nabla^2 \chi^{(1)} \phi^{(1)} - 8 {\Mpl}^{2} \mathcal{H} \nabla^2 \omega^{(1)} \phi^{(1)} \nonumber \\ 
&+ 16 {\Mpl}^{2} \nabla^2 \phi^{(1)} \phi^{(1)}- 4 c_2 {\pi}^{\prime} {\pi^{(1)}}^{\prime} \psi^{(1)} + 24 {\Mpl}^{2} {\phi^{(1)}}^{\prime} \mathcal{H} \psi^{(1)} + 8 {\Mpl}^{2} \mathcal{H} \nabla^2 \omega^{(1)} \psi^{(1)} + 4 c_2 {{\pi}^{\prime}}^{2} {\psi^{(1)}}^{2} \nonumber \\ 
&+ 24 {\Mpl}^{2} {\mathcal{H}}^{2} {\psi^{(1)}}^{2} + \frac{18 c_3 {{\pi}^{\prime}}^{2} {\pi^{(1)}}^{\prime} {\phi^{(1)}}^{\prime}}{{M}^{3} {a}^{2}} -  \frac{18 c_3 {\pi}^{\prime} {{\pi^{(1)}}^{\prime}}^{2} \mathcal{H}}{{M}^{3} {a}^{2}} -  \frac{c_3 {{\pi}^{\prime}}^{3} \nabla^2 {\chi^{(1)}}^{\prime} \nabla^2 \chi^{(1)}}{3 {M}^{3} {a}^{2}} + \frac{4 c_3 {\pi}^{\prime} {\pi^{(1)}}^{\prime} \nabla^2 \pi^{(1)}}{{M}^{3} {a}^{2}} \nonumber \\ 
& + \frac{2 c_3 {{\pi}^{\prime}}^{2} \nabla^2 \chi^{(1)} \nabla^2 \pi^{(1)}}{3 {M}^{3} {a}^{2}} + \frac{6 c_3 {{\pi}^{\prime}}^{2} {\pi^{(1)}}^{\prime} \nabla^2 \omega^{(1)}}{{M}^{3} {a}^{2}} + \frac{2 c_3 {{\pi}^{\prime}}^{3} \nabla^2 \chi^{(1)} \nabla^2 \omega^{(1)}}{3 {M}^{3} {a}^{2}} \nonumber \\ 
& + \frac{12 c_3 {{\pi}^{\prime}}^{3} {\phi^{(1)}}^{\prime} \phi^{(1)}}{{M}^{3} {a}^{2}} + \frac{4 c_3 {{\pi}^{\prime}}^{2} \nabla^2 \pi^{(1)} \phi^{(1)}}{{M}^{3} {a}^{2}} + \frac{4 c_3 {{\pi}^{\prime}}^{3} \nabla^2 \omega^{(1)} \phi^{(1)}}{{M}^{3} {a}^{2}} -  \frac{24 c_3 {{\pi}^{\prime}}^{3} {\phi^{(1)}}^{\prime} \psi^{(1)}}{{M}^{3} {a}^{2}} \nonumber \\ 
& + \frac{72 c_3 {{\pi}^{\prime}}^{2} {\pi^{(1)}}^{\prime} \mathcal{H} \psi^{(1)}}{{M}^{3} {a}^{2}} -  \frac{4 c_3 {{\pi}^{\prime}}^{2} \nabla^2 \pi^{(1)} \psi^{(1)}}{{M}^{3} {a}^{2}} -  \frac{8 c_3 {{\pi}^{\prime}}^{3} \nabla^2 \omega^{(1)} \psi^{(1)}}{{M}^{3} {a}^{2}} \nonumber \\ 
& -  \frac{72 c_3 {{\pi}^{\prime}}^{3} \mathcal{H} {\psi^{(1)}}^{2}}{{M}^{3} {a}^{2}} + \frac{2 c_3 {{\pi}^{\prime}}^{2} \partial_{i}\omega^{(1)} \partial^{i}{\pi^{(1)}}^{\prime}}{{M}^{3} {a}^{2}} + 4 {\Mpl}^{2} \partial_{i}\omega^{(1)} \partial^{i}{\phi^{(1)}}^{\prime} \nonumber \\ 
& + \frac{2}{3} {\Mpl}^{2} \partial_{i}\omega^{(1)} \partial^{i}\nabla^2 {\chi^{(1)}}^{\prime} -  \frac{5}{12} {\Mpl}^{2} \partial_{i}\nabla^2 \chi^{(1)} \partial^{i}\nabla^2 \chi^{(1)} -  \frac{4 c_3 {{\pi}^{\prime}}^{2} \partial_{i}\pi^{(1)} \partial^{i}\nabla^2 \chi^{(1)}}{3 {M}^{3} {a}^{2}} \nonumber \\ 
& + \frac{8}{3} {\Mpl}^{2} \mathcal{H} \partial_{i}\omega^{(1)} \partial^{i}\nabla^2 \chi^{(1)} -  \frac{4 c_3 {{\pi}^{\prime}}^{3} \partial_{i}\omega^{(1)} \partial^{i}\nabla^2 \chi^{(1)}}{3 {M}^{3} {a}^{2}} + c_2 \partial_{i}\pi^{(1)} \partial^{i}\pi^{(1)} \nonumber \\ 
& + \frac{2 c_3 {\pi}^{\prime} \mathcal{H} \partial_{i}\pi^{(1)} \partial^{i}\pi^{(1)}}{{M}^{3} {a}^{2}} + \frac{12 c_3 {{\pi}^{\prime}}^{2} \mathcal{H} \partial_{i}\pi^{(1)} \partial^{i}\omega^{(1)}}{{M}^{3} {a}^{2}} -  c_2 {{\pi}^{\prime}}^{2} \partial_{i}\omega^{(1)} \partial^{i}\omega^{(1)} \nonumber \\ 
& - 6 {\Mpl}^{2} {\mathcal{H}}^{2} \partial_{i}\omega^{(1)} \partial^{i}\omega^{(1)} + \frac{12 c_3 {{\pi}^{\prime}}^{3} \mathcal{H} \partial_{i}\omega^{(1)} \partial^{i}\omega^{(1)}}{{M}^{3} {a}^{2}} -  \frac{2 c_3 {{\pi}^{\prime}}^{2} \partial_{i}\pi^{(1)} \partial^{i}\phi^{(1)}}{{M}^{3} {a}^{2}} \nonumber \\ 
& + 4 {\Mpl}^{2} \mathcal{H} \partial_{i}\omega^{(1)} \partial^{i}\phi^{(1)} -  \frac{2 c_3 {{\pi}^{\prime}}^{3} \partial_{i}\omega^{(1)} \partial^{i}\phi^{(1)}}{{M}^{3} {a}^{2}} + 6 {\Mpl}^{2} \partial_{i}\phi^{(1)} \partial^{i}\phi^{(1)} + 4 {\Mpl}^{2} \mathcal{H} \partial_{i}\omega^{(1)} \partial^{i}\psi^{(1)} \nonumber \\ 
& -  \frac{2 c_3 {{\pi}^{\prime}}^{3} \partial_{i}\omega^{(1)} \partial^{i}\psi^{(1)}}{{M}^{3} {a}^{2}} - 2 \rho_m {a}^{2} \partial_{i}\omega^{(1)} \partial^{i}v^{(1)} - 2 \rho_m {a}^{2} \partial_{i}v^{(1)} \partial^{i}v^{(1)} -  \frac{1}{4} {\Mpl}^{2} \partial_{j}\partial^{i}{\chi^{(1)}}^{\prime} \partial^{j}\partial_{i}{\chi^{(1)}}^{\prime} \nonumber \\ 
& - 2 {\Mpl}^{2} \mathcal{H} \partial_{j}\partial^{i}\chi^{(1)} \partial^{j}\partial_{i}{\chi^{(1)}}^{\prime} + \frac{c_3 {{\pi}^{\prime}}^{3} \partial_{j}\partial^{i}\chi^{(1)} \partial^{j}\partial_{i}{\chi^{(1)}}^{\prime}}{{M}^{3} {a}^{2}} + {\Mpl}^{2} \partial_{j}\partial^{i}\omega^{(1)} \partial^{j}\partial_{i}{\chi^{(1)}}^{\prime} \nonumber \\ 
& -  \frac{1}{3} {\Mpl}^{2} \partial_{j}\partial^{i}\chi^{(1)} \partial^{j}\partial_{i}\nabla^2 \chi^{(1)} -  \frac{2 c_3 {{\pi}^{\prime}}^{2} \partial_{j}\partial^{i}\chi^{(1)} \partial^{j}\partial_{i}\pi^{(1)}}{{M}^{3} {a}^{2}} + 4 {\Mpl}^{2} \mathcal{H} \partial_{j}\partial^{i}\chi^{(1)} \partial^{j}\partial_{i}\omega^{(1)} \nonumber \\ 
& -  \frac{2 c_3 {{\pi}^{\prime}}^{3} \partial_{j}\partial^{i}\chi^{(1)} \partial^{j}\partial_{i}\omega^{(1)}}{{M}^{3} {a}^{2}} -  {\Mpl}^{2} \partial_{j}\partial^{i}\omega^{(1)} \partial^{j}\partial_{i}\omega^{(1)} - 2 {\Mpl}^{2} \partial_{j}\partial^{i}\chi^{(1)} \partial^{j}\partial_{i}\phi^{(1)} \nonumber \\ 
& + \frac{1}{4} {\Mpl}^{2} \partial_{k}\partial^{j}\partial^{i}\chi^{(1)} \partial^{k}\partial_{j}\partial_{i}\chi^{(1)}
\end{align}

\begin{align}
 S^{(2)} &\equiv\frac{4}{3} {\Mpl}^{2} {\phi^{(1)}}^{\prime} \nabla^2 \nabla^2 \chi^{(1)} + \frac{1}{18} {\Mpl}^{2} \nabla^2 {\chi^{(1)}}^{\prime} \nabla^2 \nabla^2 \chi^{(1)} -  \frac{2}{3} {\Mpl}^{2} \nabla^2 {\phi^{(1)}}^{\prime} \nabla^2 \chi^{(1)} -  \frac{1}{9} {\Mpl}^{2} \nabla^2 \nabla^2 {\chi^{(1)}}^{\prime} \nabla^2 \chi^{(1)} \nonumber \\ 
& + 2 c_2 {\pi^{(1)}}^{\prime} \nabla^2 \pi^{(1)} -  \frac{1}{3} {\Mpl}^{2} \nabla^2 \nabla^2 \chi^{(1)} \nabla^2 \omega^{(1)} + 8 {\Mpl}^{2} {\phi^{(1)}}^{\prime} \nabla^2 \phi^{(1)} + \frac{1}{3} {\Mpl}^{2} \nabla^2 {\chi^{(1)}}^{\prime} \nabla^2 \phi^{(1)} \nonumber \\ 
&- 2 {\Mpl}^{2} \nabla^2 \omega^{(1)} \nabla^2 \phi^{(1)} - 4 {\Mpl}^{2} {\phi^{(1)}}^{\prime} \nabla^2 \psi^{(1)} + \frac{1}{3} {\Mpl}^{2} \nabla^2 {\chi^{(1)}}^{\prime} \nabla^2 \psi^{(1)} - 2 {\Mpl}^{2} \nabla^2 \omega^{(1)} \nabla^2 \psi^{(1)} \nonumber \\ 
&+ 8 {\Mpl}^{2} \nabla^2 {\phi^{(1)}}^{\prime} \phi^{(1)} + \frac{4}{3} {\Mpl}^{2} \nabla^2 \nabla^2 {\chi^{(1)}}^{\prime} \phi^{(1)} - 8 {\Mpl}^{2} \nabla^2 {\phi^{(1)}}^{\prime} \psi^{(1)} -  \frac{4}{3} {\Mpl}^{2} \nabla^2 \nabla^2 {\chi^{(1)}}^{\prime} \psi^{(1)} \nonumber \\ 
&- 4 c_2 {\pi}^{\prime} \nabla^2 \pi^{(1)} \psi^{(1)}- 16 {\Mpl}^{2} \mathcal{H} \nabla^2 \psi^{(1)} \psi^{(1)} + \frac{4 c_3 {\pi}^{\prime} {\pi^{(1)}}^{\prime} \nabla^2 {\pi^{(1)}}^{\prime}}{{M}^{3} {a}^{2}} + \frac{6 c_3 {{\pi}^{\prime}}^{2} {\phi^{(1)}}^{\prime} \nabla^2 \pi^{(1)}}{{M}^{3} {a}^{2}} \nonumber \\ 
&-  \frac{12 c_3 {\pi}^{\prime} {\pi^{(1)}}^{\prime} \mathcal{H} \nabla^2 \pi^{(1)}}{{M}^{3} {a}^{2}} + \frac{2 c_3 {\pi}^{\prime} {\nabla^2 \pi^{(1)}}^{2}}{{M}^{3} {a}^{2}} + \frac{2 c_3 {{\pi}^{\prime}}^{2} \nabla^2 \pi^{(1)} \nabla^2 \omega^{(1)}}{{M}^{3} {a}^{2}} -  \frac{6 c_3 {{\pi}^{\prime}}^{2} {\pi^{(1)}}^{\prime} \nabla^2 \psi^{(1)}}{{M}^{3} {a}^{2}} \nonumber \\ 
&-  \frac{8 c_3 {{\pi}^{\prime}}^{2} \nabla^2 {\pi^{(1)}}^{\prime} \psi^{(1)}}{{M}^{3} {a}^{2}} + \frac{24 c_3 {{\pi}^{\prime}}^{2} \mathcal{H} \nabla^2 \pi^{(1)} \psi^{(1)}}{{M}^{3} {a}^{2}} + \frac{12 c_3 {{\pi}^{\prime}}^{3} \nabla^2 \psi^{(1)} \psi^{(1)}}{{M}^{3} {a}^{2}} -  \frac{2}{3} \rho_m \nabla^2 \chi^{(1)} \nabla^2 v^{(1)} {a}^{2} \nonumber \\ 
&+ 2 \rho_m \nabla^2 \omega^{(1)} \delta^{(1)} {a}^{2} + 2 \rho_m \nabla^2 v^{(1)} \delta^{(1)} {a}^{2} - 4 \rho_m \nabla^2 v^{(1)} \phi^{(1)} {a}^{2} - 4 \rho_m \nabla^2 \omega^{(1)} \psi^{(1)} {a}^{2} \nonumber \\ 
&- 2 \rho_m \nabla^2 v^{(1)} \psi^{(1)} {a}^{2} + \frac{4 c_3 {\pi}^{\prime} \partial_{i}{\pi^{(1)}}^{\prime} \partial^{i}{\pi^{(1)}}^{\prime}}{{M}^{3} {a}^{2}} + 2 c_2 \partial_{i}\pi^{(1)} \partial^{i}{\pi^{(1)}}^{\prime} -  \frac{12 c_3 {\pi}^{\prime} \mathcal{H} \partial_{i}\pi^{(1)} \partial^{i}{\pi^{(1)}}^{\prime}}{{M}^{3} {a}^{2}} \nonumber \\ 
& + \frac{6 c_3 {{\pi}^{\prime}}^{2} \partial_{i}\pi^{(1)} \partial^{i}{\phi^{(1)}}^{\prime}}{{M}^{3} {a}^{2}} + 16 {\Mpl}^{2} \partial_{i}\phi^{(1)} \partial^{i}{\phi^{(1)}}^{\prime} - 12 {\Mpl}^{2} \partial_{i}\psi^{(1)} \partial^{i}{\phi^{(1)}}^{\prime} -  \frac{7}{18} {\Mpl}^{2} \partial_{i}\nabla^2 \chi^{(1)} \partial^{i}\nabla^2 {\chi^{(1)}}^{\prime} \nonumber \\ 
& + \frac{2}{3} {\Mpl}^{2} \partial_{i}\phi^{(1)} \partial^{i}\nabla^2 {\chi^{(1)}}^{\prime} - 2 {\Mpl}^{2} \partial_{i}\psi^{(1)} \partial^{i}\nabla^2 {\chi^{(1)}}^{\prime} -  \frac{2}{3} {\Mpl}^{2} \partial_{i}\omega^{(1)} \partial^{i}\nabla^2 \nabla^2 \chi^{(1)} \nonumber \\ 
& + \frac{8}{3} {\Mpl}^{2} \partial_{i}{\phi^{(1)}}^{\prime} \partial^{i}\nabla^2 \chi^{(1)} + \frac{4}{3} \rho_m {a}^{2} \partial_{i}v^{(1)} \partial^{i}\nabla^2 \chi^{(1)} -  \frac{2 c_3 {{\pi}^{\prime}}^{2} \partial_{i}\omega^{(1)} \partial^{i}\nabla^2 \pi^{(1)}}{{M}^{3} {a}^{2}} \nonumber \\ 
& + 4 {\Mpl}^{2} \mathcal{H} \partial_{i}\omega^{(1)} \partial^{i}\nabla^2 \omega^{(1)} -  \frac{2 c_3 {{\pi}^{\prime}}^{3} \partial_{i}\omega^{(1)} \partial^{i}\nabla^2 \omega^{(1)}}{{M}^{3} {a}^{2}} - 4 {\Mpl}^{2} \partial_{i}\omega^{(1)} \partial^{i}\nabla^2 \phi^{(1)} \nonumber \\ 
& + 2 \rho_m {a}^{2} \partial_{i}\delta^{(1)} \partial^{i}\omega^{(1)} - 4 c_2 {\pi}^{\prime} \partial_{i}\pi^{(1)} \partial^{i}\psi^{(1)} + \frac{24 c_3 {{\pi}^{\prime}}^{2} \mathcal{H} \partial_{i}\pi^{(1)} \partial^{i}\psi^{(1)}}{{M}^{3} {a}^{2}} - 4 \rho_m {a}^{2} \partial_{i}\omega^{(1)} \partial^{i}\psi^{(1)} \nonumber \\ 
& - 16 {\Mpl}^{2} \mathcal{H} \partial_{i}\psi^{(1)} \partial^{i}\psi^{(1)} + \frac{12 c_3 {{\pi}^{\prime}}^{3} \partial_{i}\psi^{(1)} \partial^{i}\psi^{(1)}}{{M}^{3} {a}^{2}} + 2 \rho_m {a}^{2} \partial_{i}\delta^{(1)} \partial^{i}v^{(1)} - 4 \rho_m {a}^{2} \partial_{i}\phi^{(1)} \partial^{i}v^{(1)} \nonumber \\ 
& - 2 \rho_m {a}^{2} \partial_{i}\psi^{(1)} \partial^{i}v^{(1)} -  {\Mpl}^{2} \partial_{j}\partial^{i}\phi^{(1)} \partial^{j}\partial_{i}{\chi^{(1)}}^{\prime} -  {\Mpl}^{2} \partial_{j}\partial^{i}\psi^{(1)} \partial^{j}\partial_{i}{\chi^{(1)}}^{\prime} + 2 {\Mpl}^{2} \partial_{j}\partial^{i}\chi^{(1)} \partial^{j}\partial_{i}{\phi^{(1)}}^{\prime} \nonumber \\ 
& + \frac{1}{3} {\Mpl}^{2} \partial_{j}\partial^{i}\chi^{(1)} \partial^{j}\partial_{i}\nabla^2 {\chi^{(1)}}^{\prime} -  \frac{1}{6} {\Mpl}^{2} \partial_{j}\partial^{i}{\chi^{(1)}}^{\prime} \partial^{j}\partial_{i}\nabla^2 \chi^{(1)} -  \frac{1}{3} {\Mpl}^{2} \partial_{j}\partial^{i}\omega^{(1)} \partial^{j}\partial_{i}\nabla^2 \chi^{(1)} \nonumber \\ 
& -  \frac{2 c_3 {\pi}^{\prime} \partial_{j}\partial^{i}\pi^{(1)} \partial^{j}\partial_{i}\pi^{(1)}}{{M}^{3} {a}^{2}} -  \frac{4 c_3 {{\pi}^{\prime}}^{2} \partial_{j}\partial^{i}\pi^{(1)} \partial^{j}\partial_{i}\omega^{(1)}}{{M}^{3} {a}^{2}} + 4 {\Mpl}^{2} \mathcal{H} \partial_{j}\partial^{i}\omega^{(1)} \partial^{j}\partial_{i}\omega^{(1)} \nonumber \\ 
& -  \frac{2 c_3 {{\pi}^{\prime}}^{3} \partial_{j}\partial^{i}\omega^{(1)} \partial^{j}\partial_{i}\omega^{(1)}}{{M}^{3} {a}^{2}} - 2 {\Mpl}^{2} \partial_{j}\partial^{i}\omega^{(1)} \partial^{j}\partial_{i}\phi^{(1)} + 2 {\Mpl}^{2} \partial_{j}\partial^{i}\omega^{(1)} \partial^{j}\partial_{i}\psi^{(1)} \nonumber \\ 
& + 2 \rho_m {a}^{2} \partial_{j}\partial^{i}\chi^{(1)} \partial^{j}\partial_{i}v^{(1)} + \frac{1}{2} {\Mpl}^{2} \partial_{k}\partial^{j}\partial^{i}\chi^{(1)} \partial^{k}\partial_{j}\partial_{i}{\chi^{(1)}}^{\prime}-  \frac{14 c_3 {{\pi}^{\prime}}^{2} \partial_{i}\psi^{(1)} \partial^{i}{\pi^{(1)}}^{\prime}}{{M}^{3} {a}^{2}}
\end{align}

\begin{align}
 S^{(3)} &\equiv- \frac{12 c_3 {\pi^{(1)}}^{\prime\prime} {\pi}^{\prime} {\pi^{(1)}}^{\prime}}{{M}^{3} {a}^{4}} -  \frac{6 c_3 {\pi}^{\prime\prime} {{\pi^{(1)}}^{\prime}}^{2}}{{M}^{3} {a}^{4}} + \frac{18 c_3 {{\pi}^{\prime}}^{2} {\pi^{(1)}}^{\prime} {\psi^{(1)}}^{\prime}}{{M}^{3} {a}^{4}} \nonumber \\ 
& + \frac{18 c_3 {\pi}^{\prime} {{\pi^{(1)}}^{\prime}}^{2} \mathcal{H}}{{M}^{3} {a}^{4}} + \frac{24 c_3 {\pi^{(1)}}^{\prime\prime} {{\pi}^{\prime}}^{2} \psi^{(1)}}{{M}^{3} {a}^{4}} + \frac{48 c_3 {\pi}^{\prime\prime} {\pi}^{\prime} {\pi^{(1)}}^{\prime} \psi^{(1)}}{{M}^{3} {a}^{4}} \nonumber \\ 
& -  \frac{36 c_3 {{\pi}^{\prime}}^{3} {\psi^{(1)}}^{\prime} \psi^{(1)}}{{M}^{3} {a}^{4}} -  \frac{72 c_3 {{\pi}^{\prime}}^{2} {\pi^{(1)}}^{\prime} \mathcal{H} \psi^{(1)}}{{M}^{3} {a}^{4}} -  \frac{72 c_3 {\pi}^{\prime\prime} {{\pi}^{\prime}}^{2} {\psi^{(1)}}^{2}}{{M}^{3} {a}^{4}} \nonumber \\ 
& + \frac{72 c_3 {{\pi}^{\prime}}^{3} \mathcal{H} {\psi^{(1)}}^{2}}{{M}^{3} {a}^{4}} -  \frac{3 c_2 {{\pi^{(1)}}^{\prime}}^{2}}{{a}^{2}} -  \frac{6 {\Mpl}^{2} {{\phi^{(1)}}^{\prime}}^{2}}{{a}^{2}} + \frac{12 {\Mpl}^{2} {\phi^{(1)}}^{\prime} {\psi^{(1)}}^{\prime}}{{a}^{2}} + \frac{5 {\Mpl}^{2} {\nabla^2 {\chi^{(1)}}^{\prime}}^{2}}{12 {a}^{2}} \nonumber \\ 
& + \frac{2 {\Mpl}^{2} \nabla^2 {\chi^{(1)}}^{\prime\prime} \nabla^2 \chi^{(1)}}{3 {a}^{2}} + \frac{4 {\Mpl}^{2} \mathcal{H} \nabla^2 {\chi^{(1)}}^{\prime} \nabla^2 \chi^{(1)}}{3 {a}^{2}} -  \frac{4 {\Mpl}^{2} \nabla^2 {\omega^{(1)}}^{\prime} \nabla^2 \chi^{(1)}}{3 {a}^{2}} \nonumber \\ 
& + \frac{{\Mpl}^{2} \nabla^2 \nabla^2 \chi^{(1)} \nabla^2 \chi^{(1)}}{9 {a}^{2}} + \frac{4 {\Mpl}^{2} {\phi^{(1)}}^{\prime} \nabla^2 \omega^{(1)}}{{a}^{2}} + \frac{4 {\Mpl}^{2} {\psi^{(1)}}^{\prime} \nabla^2 \omega^{(1)}}{{a}^{2}} -  \frac{{\Mpl}^{2} \nabla^2 {\chi^{(1)}}^{\prime} \nabla^2 \omega^{(1)}}{3 {a}^{2}} \nonumber \\ 
& -  \frac{8 {\Mpl}^{2} \mathcal{H} \nabla^2 \chi^{(1)} \nabla^2 \omega^{(1)}}{3 {a}^{2}} + \frac{{\Mpl}^{2} {\nabla^2 \omega^{(1)}}^{2}}{{a}^{2}} + \frac{2 {\Mpl}^{2} \nabla^2 \chi^{(1)} \nabla^2 \phi^{(1)}}{3 {a}^{2}} -  \frac{4 {\Mpl}^{2} \nabla^2 \chi^{(1)} \nabla^2 \psi^{(1)}}{3 {a}^{2}} \nonumber \\ 
& -  \frac{24 {\Mpl}^{2} {\phi^{(1)}}^{\prime\prime} \phi^{(1)}}{{a}^{2}} -  \frac{48 {\Mpl}^{2} {\phi^{(1)}}^{\prime} \mathcal{H} \phi^{(1)}}{{a}^{2}} -  \frac{8 {\Mpl}^{2} \nabla^2 {\omega^{(1)}}^{\prime} \phi^{(1)}}{{a}^{2}} + \frac{8 {\Mpl}^{2} \nabla^2 \nabla^2 \chi^{(1)} \phi^{(1)}}{3 {a}^{2}} \nonumber \\ 
& -  \frac{16 {\Mpl}^{2} \mathcal{H} \nabla^2 \omega^{(1)} \phi^{(1)}}{{a}^{2}} + \frac{16 {\Mpl}^{2} \nabla^2 \phi^{(1)} \phi^{(1)}}{{a}^{2}} -  \frac{8 {\Mpl}^{2} \nabla^2 \psi^{(1)} \phi^{(1)}}{{a}^{2}} + \frac{24 {\Mpl}^{2} {\phi^{(1)}}^{\prime\prime} \psi^{(1)}}{{a}^{2}} \nonumber \\ 
& + \frac{12 c_2 {\pi}^{\prime} {\pi^{(1)}}^{\prime} \psi^{(1)}}{{a}^{2}} + \frac{48 {\Mpl}^{2} {\phi^{(1)}}^{\prime} \mathcal{H} \psi^{(1)}}{{a}^{2}} + \frac{48 {\Mpl}^{2} {\psi^{(1)}}^{\prime} \mathcal{H} \psi^{(1)}}{{a}^{2}} + \frac{8 {\Mpl}^{2} \nabla^2 {\omega^{(1)}}^{\prime} \psi^{(1)}}{{a}^{2}} \nonumber \\ 
& + \frac{16 {\Mpl}^{2} \mathcal{H} \nabla^2 \omega^{(1)} \psi^{(1)}}{{a}^{2}} + \frac{8 {\Mpl}^{2} \nabla^2 \psi^{(1)} \psi^{(1)}}{{a}^{2}} -  \frac{12 c_2 {{\pi}^{\prime}}^{2} {\psi^{(1)}}^{2}}{{a}^{2}} + \frac{48 {\Mpl}^{2} {\mathcal{H}}^{\prime} {\psi^{(1)}}^{2}}{{a}^{2}} \nonumber \\ 
& + \frac{24 {\Mpl}^{2} {\mathcal{H}}^{2} {\psi^{(1)}}^{2}}{{a}^{2}} + \frac{8 c_3 {\pi}^{\prime} \partial_{i}\pi^{(1)} \partial^{i}{\pi^{(1)}}^{\prime}}{{M}^{3} {a}^{4}} + \frac{10 c_3 {{\pi}^{\prime}}^{2} \partial_{i}\omega^{(1)} \partial^{i}{\pi^{(1)}}^{\prime}}{{M}^{3} {a}^{4}} \nonumber \\ 
& + \frac{6 c_3 {{\pi}^{\prime}}^{2} \partial_{i}\pi^{(1)} \partial^{i}{\omega^{(1)}}^{\prime}}{{M}^{3} {a}^{4}} + \frac{6 c_3 {{\pi}^{\prime}}^{3} \partial_{i}\omega^{(1)} \partial^{i}{\omega^{(1)}}^{\prime}}{{M}^{3} {a}^{4}} -  \frac{12 {\Mpl}^{2} \mathcal{H} \partial_{i}\omega^{(1)} \partial^{i}{\omega^{(1)}}^{\prime}}{{a}^{2}} \nonumber \\ 
& + \frac{4 {\Mpl}^{2} \partial_{i}\phi^{(1)} \partial^{i}{\omega^{(1)}}^{\prime}}{{a}^{2}} + \frac{12 {\Mpl}^{2} \partial_{i}\omega^{(1)} \partial^{i}{\phi^{(1)}}^{\prime}}{{a}^{2}} + \frac{2 {\Mpl}^{2} \partial_{i}\omega^{(1)} \partial^{i}\nabla^2 {\chi^{(1)}}^{\prime}}{{a}^{2}} \nonumber \\ 
& + \frac{8 {\Mpl}^{2} \partial_{i}{\omega^{(1)}}^{\prime} \partial^{i}\nabla^2 \chi^{(1)}}{3 {a}^{2}} -  \frac{5 {\Mpl}^{2} \partial_{i}\nabla^2 \chi^{(1)} \partial^{i}\nabla^2 \chi^{(1)}}{12 {a}^{2}} + \frac{16 {\Mpl}^{2} \mathcal{H} \partial_{i}\omega^{(1)} \partial^{i}\nabla^2 \chi^{(1)}}{3 {a}^{2}} \nonumber \\ 
& + \frac{8 {\Mpl}^{2} \partial_{i}\psi^{(1)} \partial^{i}\nabla^2 \chi^{(1)}}{3 {a}^{2}} + \frac{2 c_3 {\pi}^{\prime\prime} \partial_{i}\pi^{(1)} \partial^{i}\pi^{(1)}}{{M}^{3} {a}^{4}} -  \frac{2 c_3 {\pi}^{\prime} \mathcal{H} \partial_{i}\pi^{(1)} \partial^{i}\pi^{(1)}}{{M}^{3} {a}^{4}} \nonumber \\ 
& + \frac{c_2 \partial_{i}\pi^{(1)} \partial^{i}\pi^{(1)}}{{a}^{2}} + \frac{12 c_3 {\pi}^{\prime\prime} {\pi}^{\prime} \partial_{i}\pi^{(1)} \partial^{i}\omega^{(1)}}{{M}^{3} {a}^{4}} -  \frac{12 c_3 {{\pi}^{\prime}}^{2} \mathcal{H} \partial_{i}\pi^{(1)} \partial^{i}\omega^{(1)}}{{M}^{3} {a}^{4}} \nonumber \\ 
& + \frac{4 c_2 {\pi}^{\prime} \partial_{i}\pi^{(1)} \partial^{i}\omega^{(1)}}{{a}^{2}} + \frac{12 c_3 {\pi}^{\prime\prime} {{\pi}^{\prime}}^{2} \partial_{i}\omega^{(1)} \partial^{i}\omega^{(1)}}{{M}^{3} {a}^{4}} -  \frac{12 c_3 {{\pi}^{\prime}}^{3} \mathcal{H} \partial_{i}\omega^{(1)} \partial^{i}\omega^{(1)}}{{M}^{3} {a}^{4}} \nonumber \\ 
& + \frac{3 c_2 {{\pi}^{\prime}}^{2} \partial_{i}\omega^{(1)} \partial^{i}\omega^{(1)}}{{a}^{2}} -  \frac{12 {\Mpl}^{2} {\mathcal{H}}^{\prime} \partial_{i}\omega^{(1)} \partial^{i}\omega^{(1)}}{{a}^{2}} -  \frac{6 {\Mpl}^{2} {\mathcal{H}}^{2} \partial_{i}\omega^{(1)} \partial^{i}\omega^{(1)}}{{a}^{2}} \nonumber \\ 
& + \frac{8 {\Mpl}^{2} \mathcal{H} \partial_{i}\omega^{(1)} \partial^{i}\phi^{(1)}}{{a}^{2}} + \frac{6 {\Mpl}^{2} \partial_{i}\phi^{(1)} \partial^{i}\phi^{(1)}}{{a}^{2}} -  \frac{2 c_3 {{\pi}^{\prime}}^{2} \partial_{i}\pi^{(1)} \partial^{i}\psi^{(1)}}{{M}^{3} {a}^{4}} \nonumber \\ 
& -  \frac{4 c_3 {{\pi}^{\prime}}^{3} \partial_{i}\omega^{(1)} \partial^{i}\psi^{(1)}}{{M}^{3} {a}^{4}} + \frac{8 {\Mpl}^{2} \mathcal{H} \partial_{i}\omega^{(1)} \partial^{i}\psi^{(1)}}{{a}^{2}} + \frac{4 {\Mpl}^{2} \partial_{i}\phi^{(1)} \partial^{i}\psi^{(1)}}{{a}^{2}} \nonumber \\ 
& + \frac{4 {\Mpl}^{2} \partial_{i}\psi^{(1)} \partial^{i}\psi^{(1)}}{{a}^{2}} + 2 \rho_m \partial_{i}\omega^{(1)} \partial^{i}v^{(1)} + 2 \rho_m \partial_{i}v^{(1)} \partial^{i}v^{(1)} -  \frac{2 {\Mpl}^{2} \partial_{j}\partial^{i}\chi^{(1)} \partial^{j}\partial_{i}{\chi^{(1)}}^{\prime\prime}}{{a}^{2}} \nonumber \\ 
& -  \frac{5 {\Mpl}^{2} \partial_{j}\partial^{i}{\chi^{(1)}}^{\prime} \partial^{j}\partial_{i}{\chi^{(1)}}^{\prime}}{4 {a}^{2}} -  \frac{4 {\Mpl}^{2} \mathcal{H} \partial_{j}\partial^{i}\chi^{(1)} \partial^{j}\partial_{i}{\chi^{(1)}}^{\prime}}{{a}^{2}} \nonumber \\ 
& + \frac{{\Mpl}^{2} \partial_{j}\partial^{i}\omega^{(1)} \partial^{j}\partial_{i}{\chi^{(1)}}^{\prime}}{{a}^{2}} + \frac{4 {\Mpl}^{2} \partial_{j}\partial^{i}\chi^{(1)} \partial^{j}\partial_{i}{\omega^{(1)}}^{\prime}}{{a}^{2}} -  \frac{{\Mpl}^{2} \partial_{j}\partial^{i}\chi^{(1)} \partial^{j}\partial_{i}\nabla^2 \chi^{(1)}}{3 {a}^{2}} \nonumber \\ 
& + \frac{8 {\Mpl}^{2} \mathcal{H} \partial_{j}\partial^{i}\chi^{(1)} \partial^{j}\partial_{i}\omega^{(1)}}{{a}^{2}} -  \frac{{\Mpl}^{2} \partial_{j}\partial^{i}\omega^{(1)} \partial^{j}\partial_{i}\omega^{(1)}}{{a}^{2}} -  \frac{2 {\Mpl}^{2} \partial_{j}\partial^{i}\chi^{(1)} \partial^{j}\partial_{i}\phi^{(1)}}{{a}^{2}} \nonumber \\ 
& + \frac{4 {\Mpl}^{2} \partial_{j}\partial^{i}\chi^{(1)} \partial^{j}\partial_{i}\psi^{(1)}}{{a}^{2}} + \frac{{\Mpl}^{2} \partial_{k}\partial^{j}\partial^{i}\chi^{(1)} \partial^{k}\partial_{j}\partial_{i}\chi^{(1)}}{4 {a}^{2}}
\end{align}

\begin{align}
S^{(4)} &\equiv\frac{1}{2} \nabla^2 {\chi^{(1)}}^{\prime} \nabla^2 {\phi^{(1)}}^{\prime} -  \frac{1}{2} \nabla^2 {\chi^{(1)}}^{\prime} \nabla^2 {\psi^{(1)}}^{\prime} -  {\phi^{(1)}}^{\prime} \nabla^2 \nabla^2 {\chi^{(1)}}^{\prime} + {\psi^{(1)}}^{\prime} \nabla^2 \nabla^2 {\chi^{(1)}}^{\prime} \nonumber \\ 
& -  \frac{1}{3} \nabla^2 {\chi^{(1)}}^{\prime} \nabla^2 \nabla^2 {\chi^{(1)}}^{\prime} - 2 {\phi^{(1)}}^{\prime\prime} \nabla^2 \nabla^2 \chi^{(1)} - 4 {\phi^{(1)}}^{\prime} \mathcal{H} \nabla^2 \nabla^2 \chi^{(1)} -  \frac{1}{6} \nabla^2 {\chi^{(1)}}^{\prime\prime} \nabla^2 \nabla^2 \chi^{(1)} \nonumber \\ 
& -  \frac{1}{3} \mathcal{H} \nabla^2 {\chi^{(1)}}^{\prime} \nabla^2 \nabla^2 \chi^{(1)} + \frac{2}{3} \nabla^2 {\omega^{(1)}}^{\prime} \nabla^2 \nabla^2 \chi^{(1)} + \frac{1}{36} {\nabla^2 \nabla^2 \chi^{(1)}}^{2} - 2 {\phi^{(1)}}^{\prime} \nabla^2 \nabla^2 \omega^{(1)} - 2 {\psi^{(1)}}^{\prime} \nabla^2 \nabla^2 \omega^{(1)} \nonumber \\ 
& + \frac{1}{6} \nabla^2 {\chi^{(1)}}^{\prime} \nabla^2 \nabla^2 \omega^{(1)} + \nabla^2 {\phi^{(1)}}^{\prime\prime} \nabla^2 \chi^{(1)} + 2 \mathcal{H} \nabla^2 {\phi^{(1)}}^{\prime} \nabla^2 \chi^{(1)} -  \frac{1}{6} \nabla^2 \nabla^2 {\chi^{(1)}}^{\prime\prime} \nabla^2 \chi^{(1)} \nonumber \\ 
& -  \frac{1}{3} \mathcal{H} \nabla^2 \nabla^2 {\chi^{(1)}}^{\prime} \nabla^2 \chi^{(1)} + \frac{2}{3} \nabla^2 \nabla^2 {\omega^{(1)}}^{\prime} \nabla^2 \chi^{(1)} -  \frac{1}{18} \nabla^2 \nabla^2 \nabla^2 \chi^{(1)} \nabla^2 \chi^{(1)} + \frac{4}{3} \mathcal{H} \nabla^2 \nabla^2 \omega^{(1)} \nabla^2 \chi^{(1)} \nonumber \\ 
& -  \frac{1}{3} \nabla^2 \nabla^2 \phi^{(1)} \nabla^2 \chi^{(1)} + \frac{2}{3} \nabla^2 \nabla^2 \psi^{(1)} \nabla^2 \chi^{(1)} -  \frac{3 c_2 {\nabla^2 \pi^{(1)}}^{2}}{{\Mpl}^{2}} + \nabla^2 {\phi^{(1)}}^{\prime} \nabla^2 \omega^{(1)} + \nabla^2 {\psi^{(1)}}^{\prime} \nabla^2 \omega^{(1)} \nonumber \\ 
& + \frac{2}{3} \nabla^2 \nabla^2 {\chi^{(1)}}^{\prime} \nabla^2 \omega^{(1)} + \frac{4}{3} \mathcal{H} \nabla^2 \nabla^2 \chi^{(1)} \nabla^2 \omega^{(1)} -  \nabla^2 \nabla^2 \omega^{(1)} \nabla^2 \omega^{(1)} -  \frac{3 c_2 {\pi}^{\prime} \nabla^2 \pi^{(1)} \nabla^2 \omega^{(1)}}{{\Mpl}^{2}} \nonumber \\ 
& + \nabla^2 {\chi^{(1)}}^{\prime\prime} \nabla^2 \phi^{(1)} + 2 \mathcal{H} \nabla^2 {\chi^{(1)}}^{\prime} \nabla^2 \phi^{(1)} + 4 \nabla^2 {\omega^{(1)}}^{\prime} \nabla^2 \phi^{(1)} -  \frac{2}{3} \nabla^2 \nabla^2 \chi^{(1)} \nabla^2 \phi^{(1)} + 8 \mathcal{H} \nabla^2 \omega^{(1)} \nabla^2 \phi^{(1)} \nonumber \\ 
& - 5 {\nabla^2 \phi^{(1)}}^{2} -  \nabla^2 {\chi^{(1)}}^{\prime\prime} \nabla^2 \psi^{(1)} - 2 \mathcal{H} \nabla^2 {\chi^{(1)}}^{\prime} \nabla^2 \psi^{(1)} + 2 \nabla^2 {\omega^{(1)}}^{\prime} \nabla^2 \psi^{(1)} + \frac{2}{3} \nabla^2 \nabla^2 \chi^{(1)} \nabla^2 \psi^{(1)} \nonumber \\ 
& - 2 \mathcal{H} \nabla^2 \omega^{(1)} \nabla^2 \psi^{(1)} + 4 \nabla^2 \phi^{(1)} \nabla^2 \psi^{(1)} -  {\nabla^2 \psi^{(1)}}^{2} - 2 \nabla^2 \nabla^2 {\chi^{(1)}}^{\prime\prime} \phi^{(1)} - 4 \mathcal{H} \nabla^2 \nabla^2 {\chi^{(1)}}^{\prime} \phi^{(1)} \nonumber \\ 
&+ 4 \nabla^2 \nabla^2 {\omega^{(1)}}^{\prime} \phi^{(1)} -  \frac{4}{3} \nabla^2 \nabla^2 \nabla^2 \chi^{(1)} \phi^{(1)} + 8 \mathcal{H} \nabla^2 \nabla^2 \omega^{(1)} \phi^{(1)} - 8 \nabla^2 \nabla^2 \phi^{(1)} \phi^{(1)} + 4 \nabla^2 \nabla^2 \psi^{(1)} \phi^{(1)} \nonumber \\ 
&+ 2 \nabla^2 \nabla^2 {\chi^{(1)}}^{\prime\prime} \psi^{(1)} + 4 \mathcal{H} \nabla^2 \nabla^2 {\chi^{(1)}}^{\prime} \psi^{(1)} - 4 \nabla^2 \nabla^2 {\omega^{(1)}}^{\prime} \psi^{(1)} - 8 \mathcal{H} \nabla^2 \nabla^2 \omega^{(1)} \psi^{(1)} - 4 \nabla^2 \nabla^2 \psi^{(1)} \psi^{(1)} \nonumber \\ 
&-  \frac{6 c_3 {\pi}^{\prime} \nabla^2 {\pi^{(1)}}^{\prime} \nabla^2 \pi^{(1)}}{{M}^{3} {\Mpl}^{2} {a}^{2}} + \frac{3 c_3 {\pi}^{\prime\prime} {\nabla^2 \pi^{(1)}}^{2}}{{M}^{3} {\Mpl}^{2} {a}^{2}} + \frac{6 c_3 {\pi}^{\prime} \mathcal{H} {\nabla^2 \pi^{(1)}}^{2}}{{M}^{3} {\Mpl}^{2} {a}^{2}} -  \frac{3 c_3 {{\pi}^{\prime}}^{2} \nabla^2 {\pi^{(1)}}^{\prime} \nabla^2 \omega^{(1)}}{{M}^{3} {\Mpl}^{2} {a}^{2}} \nonumber \\ 
& + \frac{9 c_3 {{\pi}^{\prime}}^{2} \mathcal{H} \nabla^2 \pi^{(1)} \nabla^2 \omega^{(1)}}{{M}^{3} {\Mpl}^{2} {a}^{2}} + \frac{6 c_3 {{\pi}^{\prime}}^{2} \nabla^2 \pi^{(1)} \nabla^2 \psi^{(1)}}{{M}^{3} {\Mpl}^{2} {a}^{2}} + \frac{3 c_3 {{\pi}^{\prime}}^{3} \nabla^2 \omega^{(1)} \nabla^2 \psi^{(1)}}{{M}^{3} {\Mpl}^{2} {a}^{2}} \nonumber \\ 
& + \frac{3 \rho_m \nabla^2 \omega^{(1)} \nabla^2 v^{(1)} {a}^{2}}{{\Mpl}^{2}} + \frac{3 \rho_m {\nabla^2 v^{(1)}}^{2} {a}^{2}}{{\Mpl}^{2}} - 4 \partial_{i}\nabla^2 \chi^{(1)} \partial^{i}{\phi^{(1)}}^{\prime\prime} + \frac{1}{6} \partial_{i}\nabla^2 \chi^{(1)} \partial^{i}\nabla^2 {\chi^{(1)}}^{\prime\prime} \nonumber \\ 
& - 4 \partial_{i}\phi^{(1)} \partial^{i}\nabla^2 {\chi^{(1)}}^{\prime\prime} + 4 \partial_{i}\psi^{(1)} \partial^{i}\nabla^2 {\chi^{(1)}}^{\prime\prime} - 2 \partial_{i}{\phi^{(1)}}^{\prime} \partial^{i}\nabla^2 {\chi^{(1)}}^{\prime} + 2 \partial_{i}{\psi^{(1)}}^{\prime} \partial^{i}\nabla^2 {\chi^{(1)}}^{\prime} \nonumber \\ 
& + \frac{1}{6} \partial_{i}\nabla^2 {\chi^{(1)}}^{\prime} \partial^{i}\nabla^2 {\chi^{(1)}}^{\prime} + \frac{1}{3} \mathcal{H} \partial_{i}\nabla^2 \chi^{(1)} \partial^{i}\nabla^2 {\chi^{(1)}}^{\prime} + \frac{1}{3} \partial_{i}\nabla^2 \omega^{(1)} \partial^{i}\nabla^2 {\chi^{(1)}}^{\prime} \nonumber \\ 
& - 8 \mathcal{H} \partial_{i}\phi^{(1)} \partial^{i}\nabla^2 {\chi^{(1)}}^{\prime} + 8 \mathcal{H} \partial_{i}\psi^{(1)} \partial^{i}\nabla^2 {\chi^{(1)}}^{\prime} -  \frac{4 c_3 {\pi}^{\prime} \partial_{i}\pi^{(1)} \partial^{i}\nabla^2 {\pi^{(1)}}^{\prime}}{{M}^{3} {\Mpl}^{2} {a}^{2}} \nonumber \\ 
& -  \frac{2 c_3 {{\pi}^{\prime}}^{2} \partial_{i}\omega^{(1)} \partial^{i}\nabla^2 {\pi^{(1)}}^{\prime}}{{M}^{3} {\Mpl}^{2} {a}^{2}} -  \frac{1}{2} \partial_{i}\nabla^2 \chi^{(1)} \partial^{i}\nabla^2 {\omega^{(1)}}^{\prime} + 12 \partial_{i}\phi^{(1)} \partial^{i}\nabla^2 {\omega^{(1)}}^{\prime} \nonumber \\ 
& - 8 \partial_{i}\psi^{(1)} \partial^{i}\nabla^2 {\omega^{(1)}}^{\prime} -  \frac{1}{3} \partial_{i}{\omega^{(1)}}^{\prime} \partial^{i}\nabla^2 \nabla^2 \chi^{(1)} + \frac{7}{18} \partial_{i}\nabla^2 \chi^{(1)} \partial^{i}\nabla^2 \nabla^2 \chi^{(1)} -  \frac{2}{3} \mathcal{H} \partial_{i}\omega^{(1)} \partial^{i}\nabla^2 \nabla^2 \chi^{(1)} \nonumber \\ 
& -  \frac{11}{3} \partial_{i}\phi^{(1)} \partial^{i}\nabla^2 \nabla^2 \chi^{(1)} -  \frac{1}{3} \partial_{i}\psi^{(1)} \partial^{i}\nabla^2 \nabla^2 \chi^{(1)} - 8 \mathcal{H} \partial_{i}{\phi^{(1)}}^{\prime} \partial^{i}\nabla^2 \chi^{(1)} \nonumber \\ 
& -  \frac{4 c_3 {\pi}^{\prime} \partial_{i}{\pi^{(1)}}^{\prime} \partial^{i}\nabla^2 \pi^{(1)}}{{M}^{3} {\Mpl}^{2} {a}^{2}} -  \frac{4 c_2 \partial_{i}\pi^{(1)} \partial^{i}\nabla^2 \pi^{(1)}}{{\Mpl}^{2}} + \frac{4 c_3 {\pi}^{\prime\prime} \partial_{i}\pi^{(1)} \partial^{i}\nabla^2 \pi^{(1)}}{{M}^{3} {\Mpl}^{2} {a}^{2}} \nonumber \\ 
& + \frac{8 c_3 {\pi}^{\prime} \mathcal{H} \partial_{i}\pi^{(1)} \partial^{i}\nabla^2 \pi^{(1)}}{{M}^{3} {\Mpl}^{2} {a}^{2}} -  \frac{2 c_2 {\pi}^{\prime} \partial_{i}\omega^{(1)} \partial^{i}\nabla^2 \pi^{(1)}}{{\Mpl}^{2}} + \frac{6 c_3 {{\pi}^{\prime}}^{2} \mathcal{H} \partial_{i}\omega^{(1)} \partial^{i}\nabla^2 \pi^{(1)}}{{M}^{3} {\Mpl}^{2} {a}^{2}} \nonumber \\ 
& + \frac{4 c_3 {{\pi}^{\prime}}^{2} \partial_{i}\psi^{(1)} \partial^{i}\nabla^2 \pi^{(1)}}{{M}^{3} {\Mpl}^{2} {a}^{2}} -  \frac{2 c_3 {{\pi}^{\prime}}^{2} \partial_{i}{\pi^{(1)}}^{\prime} \partial^{i}\nabla^2 \omega^{(1)}}{{M}^{3} {\Mpl}^{2} {a}^{2}} - 4 \partial_{i}{\phi^{(1)}}^{\prime} \partial^{i}\nabla^2 \omega^{(1)} \nonumber \\ 
& - 4 \partial_{i}{\psi^{(1)}}^{\prime} \partial^{i}\nabla^2 \omega^{(1)} -  \mathcal{H} \partial_{i}\nabla^2 \chi^{(1)} \partial^{i}\nabla^2 \omega^{(1)} -  \partial_{i}\nabla^2 \omega^{(1)} \partial^{i}\nabla^2 \omega^{(1)} -  \frac{2 c_2 {\pi}^{\prime} \partial_{i}\pi^{(1)} \partial^{i}\nabla^2 \omega^{(1)}}{{\Mpl}^{2}} \nonumber \\ 
& + \frac{6 c_3 {{\pi}^{\prime}}^{2} \mathcal{H} \partial_{i}\pi^{(1)} \partial^{i}\nabla^2 \omega^{(1)}}{{M}^{3} {\Mpl}^{2} {a}^{2}} + 24 \mathcal{H} \partial_{i}\phi^{(1)} \partial^{i}\nabla^2 \omega^{(1)} - 20 \mathcal{H} \partial_{i}\psi^{(1)} \partial^{i}\nabla^2 \omega^{(1)} \nonumber \\ 
& + \frac{2 c_3 {{\pi}^{\prime}}^{3} \partial_{i}\psi^{(1)} \partial^{i}\nabla^2 \omega^{(1)}}{{M}^{3} {\Mpl}^{2} {a}^{2}} + \frac{2 \rho_m {a}^{2} \partial_{i}v^{(1)} \partial^{i}\nabla^2 \omega^{(1)}}{{\Mpl}^{2}} + 4 \partial_{i}{\omega^{(1)}}^{\prime} \partial^{i}\nabla^2 \phi^{(1)} \nonumber \\ 
& -  \frac{1}{6} \partial_{i}\nabla^2 \chi^{(1)} \partial^{i}\nabla^2 \phi^{(1)} + 8 \mathcal{H} \partial_{i}\omega^{(1)} \partial^{i}\nabla^2 \phi^{(1)} - 28 \partial_{i}\phi^{(1)} \partial^{i}\nabla^2 \phi^{(1)} + 4 \partial_{i}\psi^{(1)} \partial^{i}\nabla^2 \phi^{(1)} \nonumber \\ 
& -  \frac{1}{2} \partial_{i}\nabla^2 \chi^{(1)} \partial^{i}\nabla^2 \psi^{(1)} + \frac{4 c_3 {{\pi}^{\prime}}^{2} \partial_{i}\pi^{(1)} \partial^{i}\nabla^2 \psi^{(1)}}{{M}^{3} {\Mpl}^{2} {a}^{2}} - 4 \mathcal{H} \partial_{i}\omega^{(1)} \partial^{i}\nabla^2 \psi^{(1)} \nonumber \\ 
& + \frac{2 c_3 {{\pi}^{\prime}}^{3} \partial_{i}\omega^{(1)} \partial^{i}\nabla^2 \psi^{(1)}}{{M}^{3} {\Mpl}^{2} {a}^{2}} + 12 \partial_{i}\phi^{(1)} \partial^{i}\nabla^2 \psi^{(1)} - 12 \partial_{i}\psi^{(1)} \partial^{i}\nabla^2 \psi^{(1)} + \frac{2 \rho_m {a}^{2} \partial_{i}\omega^{(1)} \partial^{i}\nabla^2 v^{(1)}}{{\Mpl}^{2}} \nonumber \\ 
& + \frac{4 \rho_m {a}^{2} \partial_{i}v^{(1)} \partial^{i}\nabla^2 v^{(1)}}{{\Mpl}^{2}} - 3 \partial_{j}\partial^{i}\phi^{(1)} \partial^{j}\partial_{i}{\chi^{(1)}}^{\prime\prime} + 3 \partial_{j}\partial^{i}\psi^{(1)} \partial^{j}\partial_{i}{\chi^{(1)}}^{\prime\prime} \nonumber \\ 
& - 3 \partial_{j}\partial^{i}\chi^{(1)} \partial^{j}\partial_{i}{\phi^{(1)}}^{\prime\prime} - 6 \mathcal{H} \partial_{j}\partial^{i}\phi^{(1)} \partial^{j}\partial_{i}{\chi^{(1)}}^{\prime} + 6 \mathcal{H} \partial_{j}\partial^{i}\psi^{(1)} \partial^{j}\partial_{i}{\chi^{(1)}}^{\prime} \nonumber \\ 
& -  \frac{2 c_3 {\pi}^{\prime} \partial_{j}\partial^{i}\pi^{(1)} \partial^{j}\partial_{i}{\pi^{(1)}}^{\prime}}{{M}^{3} {\Mpl}^{2} {a}^{2}} -  \frac{c_3 {{\pi}^{\prime}}^{2} \partial_{j}\partial^{i}\omega^{(1)} \partial^{j}\partial_{i}{\pi^{(1)}}^{\prime}}{{M}^{3} {\Mpl}^{2} {a}^{2}} \nonumber \\ 
& + 8 \partial_{j}\partial^{i}\phi^{(1)} \partial^{j}\partial_{i}{\omega^{(1)}}^{\prime} - 6 \partial_{j}\partial^{i}\psi^{(1)} \partial^{j}\partial_{i}{\omega^{(1)}}^{\prime} -  \frac{3}{2} \partial_{j}\partial^{i}{\chi^{(1)}}^{\prime} \partial^{j}\partial_{i}{\phi^{(1)}}^{\prime} \nonumber \\ 
& - 6 \mathcal{H} \partial_{j}\partial^{i}\chi^{(1)} \partial^{j}\partial_{i}{\phi^{(1)}}^{\prime} - 3 \partial_{j}\partial^{i}\omega^{(1)} \partial^{j}\partial_{i}{\phi^{(1)}}^{\prime} + \frac{3}{2} \partial_{j}\partial^{i}{\chi^{(1)}}^{\prime} \partial^{j}\partial_{i}{\psi^{(1)}}^{\prime} \nonumber \\ 
& - 3 \partial_{j}\partial^{i}\omega^{(1)} \partial^{j}\partial_{i}{\psi^{(1)}}^{\prime} + \frac{1}{2} \partial_{j}\partial^{i}\chi^{(1)} \partial^{j}\partial_{i}\nabla^2 {\chi^{(1)}}^{\prime\prime} + \partial_{j}\partial^{i}{\chi^{(1)}}^{\prime} \partial^{j}\partial_{i}\nabla^2 {\chi^{(1)}}^{\prime} \nonumber \\ 
& + \mathcal{H} \partial_{j}\partial^{i}\chi^{(1)} \partial^{j}\partial_{i}\nabla^2 {\chi^{(1)}}^{\prime} -  \partial_{j}\partial^{i}\omega^{(1)} \partial^{j}\partial_{i}\nabla^2 {\chi^{(1)}}^{\prime} - 2 \partial_{j}\partial^{i}\chi^{(1)} \partial^{j}\partial_{i}\nabla^2 {\omega^{(1)}}^{\prime} \nonumber \\ 
& + \frac{1}{6} \partial_{j}\partial^{i}\chi^{(1)} \partial^{j}\partial_{i}\nabla^2 \nabla^2 \chi^{(1)} + \frac{1}{2} \partial_{j}\partial^{i}{\chi^{(1)}}^{\prime\prime} \partial^{j}\partial_{i}\nabla^2 \chi^{(1)} + \mathcal{H} \partial_{j}\partial^{i}{\chi^{(1)}}^{\prime} \partial^{j}\partial_{i}\nabla^2 \chi^{(1)} \nonumber \\ 
& -  \frac{8}{3} \partial_{j}\partial^{i}{\omega^{(1)}}^{\prime} \partial^{j}\partial_{i}\nabla^2 \chi^{(1)} + \frac{1}{3} \partial_{j}\partial^{i}\nabla^2 \chi^{(1)} \partial^{j}\partial_{i}\nabla^2 \chi^{(1)} -  \frac{16}{3} \mathcal{H} \partial_{j}\partial^{i}\omega^{(1)} \partial^{j}\partial_{i}\nabla^2 \chi^{(1)} \nonumber \\ 
& - 2 \partial_{j}\partial^{i}\phi^{(1)} \partial^{j}\partial_{i}\nabla^2 \chi^{(1)} -  \frac{8}{3} \partial_{j}\partial^{i}\psi^{(1)} \partial^{j}\partial_{i}\nabla^2 \chi^{(1)} -  \frac{1}{2} \partial_{j}\partial^{i}{\chi^{(1)}}^{\prime} \partial^{j}\partial_{i}\nabla^2 \omega^{(1)} \nonumber \\ 
& - 4 \mathcal{H} \partial_{j}\partial^{i}\chi^{(1)} \partial^{j}\partial_{i}\nabla^2 \omega^{(1)} + \partial_{j}\partial^{i}\omega^{(1)} \partial^{j}\partial_{i}\nabla^2 \omega^{(1)} + \partial_{j}\partial^{i}\chi^{(1)} \partial^{j}\partial_{i}\nabla^2 \phi^{(1)} \nonumber \\ 
& - 2 \partial_{j}\partial^{i}\chi^{(1)} \partial^{j}\partial_{i}\nabla^2 \psi^{(1)} -  \frac{c_2 \partial_{j}\partial^{i}\pi^{(1)} \partial^{j}\partial_{i}\pi^{(1)}}{{\Mpl}^{2}} + \frac{c_3 {\pi}^{\prime\prime} \partial_{j}\partial^{i}\pi^{(1)} \partial^{j}\partial_{i}\pi^{(1)}}{{M}^{3} {\Mpl}^{2} {a}^{2}} \nonumber \\ 
& + \frac{2 c_3 {\pi}^{\prime} \mathcal{H} \partial_{j}\partial^{i}\pi^{(1)} \partial^{j}\partial_{i}\pi^{(1)}}{{M}^{3} {\Mpl}^{2} {a}^{2}} -  \frac{c_2 {\pi}^{\prime} \partial_{j}\partial^{i}\pi^{(1)} \partial^{j}\partial_{i}\omega^{(1)}}{{\Mpl}^{2}} \nonumber \\ 
& + \frac{3 c_3 {{\pi}^{\prime}}^{2} \mathcal{H} \partial_{j}\partial^{i}\pi^{(1)} \partial^{j}\partial_{i}\omega^{(1)}}{{M}^{3} {\Mpl}^{2} {a}^{2}} + 16 \mathcal{H} \partial_{j}\partial^{i}\omega^{(1)} \partial^{j}\partial_{i}\phi^{(1)} - 15 \partial_{j}\partial^{i}\phi^{(1)} \partial^{j}\partial_{i}\phi^{(1)} \nonumber \\ 
& + \frac{2 c_3 {{\pi}^{\prime}}^{2} \partial_{j}\partial^{i}\pi^{(1)} \partial^{j}\partial_{i}\psi^{(1)}}{{M}^{3} {\Mpl}^{2} {a}^{2}} - 14 \mathcal{H} \partial_{j}\partial^{i}\omega^{(1)} \partial^{j}\partial_{i}\psi^{(1)} + \frac{c_3 {{\pi}^{\prime}}^{3} \partial_{j}\partial^{i}\omega^{(1)} \partial^{j}\partial_{i}\psi^{(1)}}{{M}^{3} {\Mpl}^{2} {a}^{2}} \nonumber \\ 
& + 8 \partial_{j}\partial^{i}\phi^{(1)} \partial^{j}\partial_{i}\psi^{(1)} - 7 \partial_{j}\partial^{i}\psi^{(1)} \partial^{j}\partial_{i}\psi^{(1)} + \frac{\rho_m {a}^{2} \partial_{j}\partial^{i}\omega^{(1)} \partial^{j}\partial_{i}v^{(1)}}{{\Mpl}^{2}} \nonumber \\ 
& + \frac{\rho_m {a}^{2} \partial_{j}\partial^{i}v^{(1)} \partial^{j}\partial_{i}v^{(1)}}{{\Mpl}^{2}} + \frac{1}{2} \partial_{k}\partial^{j}\partial^{i}\chi^{(1)} \partial^{k}\partial_{j}\partial_{i}{\chi^{(1)}}^{\prime\prime} + \frac{1}{2} \partial_{k}\partial^{j}\partial^{i}{\chi^{(1)}}^{\prime} \partial^{k}\partial_{j}\partial_{i}{\chi^{(1)}}^{\prime} \nonumber \\ 
& + \mathcal{H} \partial_{k}\partial^{j}\partial^{i}\chi^{(1)} \partial^{k}\partial_{j}\partial_{i}{\chi^{(1)}}^{\prime} -  \partial_{k}\partial^{j}\partial^{i}\omega^{(1)} \partial^{k}\partial_{j}\partial_{i}{\chi^{(1)}}^{\prime} -  \frac{5}{2} \partial_{k}\partial^{j}\partial^{i}\chi^{(1)} \partial^{k}\partial_{j}\partial_{i}{\omega^{(1)}}^{\prime} \nonumber \\ 
& -  \frac{1}{6} \partial_{k}\partial^{j}\partial^{i}\chi^{(1)} \partial^{k}\partial_{j}\partial_{i}\nabla^2 \chi^{(1)} - 5 \mathcal{H} \partial_{k}\partial^{j}\partial^{i}\chi^{(1)} \partial^{k}\partial_{j}\partial_{i}\omega^{(1)} + \partial_{k}\partial^{j}\partial^{i}\omega^{(1)} \partial^{k}\partial_{j}\partial_{i}\omega^{(1)} \nonumber \\ 
& + \frac{1}{2} \partial_{k}\partial^{j}\partial^{i}\chi^{(1)} \partial^{k}\partial_{j}\partial_{i}\phi^{(1)} -  \frac{5}{2} \partial_{k}\partial^{j}\partial^{i}\chi^{(1)} \partial^{k}\partial_{j}\partial_{i}\psi^{(1)} -  \frac{1}{4} \partial_{l}\partial^{k}\partial^{j}\partial^{i}\chi^{(1)} \partial^{l}\partial_{k}\partial_{j}\partial_{i}\chi^{(1)}
\end{align}

\begin{align}
S^{(5)} &\equiv {\pi^{(1)}}^{\prime} \psi^{(1)} (8 c_2 \mathcal{H} -  \frac{48 c_3 {\pi}^{\prime} {\mathcal{H}}^{\prime}}{{M}^{3} {a}^{2}} -  \frac{48 c_3 {\pi}^{\prime\prime} \mathcal{H}}{{M}^{3} {a}^{2}}) + {\pi^{(1)}}^{\prime\prime} \psi^{(1)} (4 c_2 -  \frac{48 c_3 {\pi}^{\prime} \mathcal{H}}{{M}^{3} {a}^{2}}) \nonumber \\ 
& + {\pi^{(1)}}^{\prime} {\psi^{(1)}}^{\prime} (2 c_2 -  \frac{36 c_3 {\pi}^{\prime} \mathcal{H}}{{M}^{3} {a}^{2}}) + {\pi^{(1)}}^{\prime} {\phi^{(1)}}^{\prime} (6 c_2 -  \frac{12 c_3 {\pi}^{\prime\prime}}{{M}^{3} {a}^{2}} -  \frac{36 c_3 {\pi}^{\prime} \mathcal{H}}{{M}^{3} {a}^{2}}) \nonumber \\ 
& + {\pi^{(1)}}^{\prime} \nabla^2 \omega^{(1)} (2 c_2 -  \frac{4 c_3 {\pi}^{\prime\prime}}{{M}^{3} {a}^{2}} -  \frac{12 c_3 {\pi}^{\prime} \mathcal{H}}{{M}^{3} {a}^{2}}) + \nabla^2 \pi^{(1)} \phi^{(1)} (4 c_2 -  \frac{8 c_3 {\pi}^{\prime\prime}}{{M}^{3} {a}^{2}} -  \frac{8 c_3 {\pi}^{\prime} \mathcal{H}}{{M}^{3} {a}^{2}}) \nonumber \\ 
& + \nabla^2 \chi^{(1)} \nabla^2 \pi^{(1)} (\frac{2}{3} c_2 -  \frac{4 c_3 {\pi}^{\prime\prime}}{3 {M}^{3} {a}^{2}} -  \frac{4 c_3 {\pi}^{\prime} \mathcal{H}}{3 {M}^{3} {a}^{2}}) + \nabla^2 \pi^{(1)} \psi^{(1)} (\frac{8 c_3 {\pi}^{\prime\prime}}{{M}^{3} {a}^{2}} + \frac{8 c_3 {\pi}^{\prime} \mathcal{H}}{{M}^{3} {a}^{2}}) \nonumber \\ 
& + {\psi^{(1)}}^{2} (-8 c_2 {\pi}^{\prime\prime} - 16 c_2 {\pi}^{\prime} \mathcal{H} + \frac{72 c_3 {{\pi}^{\prime}}^{2} {\mathcal{H}}^{\prime}}{{M}^{3} {a}^{2}} + \frac{144 c_3 {\pi}^{\prime\prime} {\pi}^{\prime} \mathcal{H}}{{M}^{3} {a}^{2}}) + {\phi^{(1)}}^{\prime} \phi^{(1)} (12 c_2 {\pi}^{\prime} \nonumber \\ 
& -  \frac{24 c_3 {\pi}^{\prime\prime} {\pi}^{\prime}}{{M}^{3} {a}^{2}} -  \frac{36 c_3 {{\pi}^{\prime}}^{2} \mathcal{H}}{{M}^{3} {a}^{2}}) + \nabla^2 \omega^{(1)} \phi^{(1)} (4 c_2 {\pi}^{\prime} -  \frac{8 c_3 {\pi}^{\prime\prime} {\pi}^{\prime}}{{M}^{3} {a}^{2}} -  \frac{12 c_3 {{\pi}^{\prime}}^{2} \mathcal{H}}{{M}^{3} {a}^{2}}) \nonumber \\ 
& + \nabla^2 \chi^{(1)} \nabla^2 \omega^{(1)} (\frac{2}{3} c_2 {\pi}^{\prime} -  \frac{4 c_3 {\pi}^{\prime\prime} {\pi}^{\prime}}{3 {M}^{3} {a}^{2}} -  \frac{2 c_3 {{\pi}^{\prime}}^{2} \mathcal{H}}{{M}^{3} {a}^{2}}) + \nabla^2 {\chi^{(1)}}^{\prime} \nabla^2 \chi^{(1)} (- \frac{1}{3} c_2 {\pi}^{\prime} \nonumber \\ 
& + \frac{2 c_3 {\pi}^{\prime\prime} {\pi}^{\prime}}{3 {M}^{3} {a}^{2}} + \frac{c_3 {{\pi}^{\prime}}^{2} \mathcal{H}}{{M}^{3} {a}^{2}}) + \nabla^2 \omega^{(1)} \psi^{(1)} (-4 c_2 {\pi}^{\prime} + \frac{16 c_3 {\pi}^{\prime\prime} {\pi}^{\prime}}{{M}^{3} {a}^{2}} + \frac{24 c_3 {{\pi}^{\prime}}^{2} \mathcal{H}}{{M}^{3} {a}^{2}}) \nonumber \\ 
& + {\phi^{(1)}}^{\prime} \psi^{(1)} (-12 c_2 {\pi}^{\prime} + \frac{48 c_3 {\pi}^{\prime\prime} {\pi}^{\prime}}{{M}^{3} {a}^{2}} + \frac{72 c_3 {{\pi}^{\prime}}^{2} \mathcal{H}}{{M}^{3} {a}^{2}}) + {\psi^{(1)}}^{\prime} \psi^{(1)} (-8 c_2 {\pi}^{\prime} + \frac{108 c_3 {{\pi}^{\prime}}^{2} \mathcal{H}}{{M}^{3} {a}^{2}}) \nonumber \\ 
& -  \frac{12 c_3 {\phi^{(1)}}^{\prime\prime} {\pi}^{\prime} {\pi^{(1)}}^{\prime}}{{M}^{3} {a}^{2}} + \frac{6 c_3 {\mathcal{H}}^{\prime} {{\pi^{(1)}}^{\prime}}^{2}}{{M}^{3} {a}^{2}} -  \frac{12 c_3 {\pi^{(1)}}^{\prime\prime} {\pi}^{\prime} {\phi^{(1)}}^{\prime}}{{M}^{3} {a}^{2}} -  \frac{4 c_3 {\pi}^{\prime} \partial_{j}\partial^{i}\pi^{(1)} \partial^{j}\partial_{i}\omega^{(1)}}{{M}^{3} {a}^{2}}\nonumber \\ 
& -  \frac{2 c_3 {{\pi}^{\prime}}^{2} \partial_{j}\partial^{i}\omega^{(1)} \partial^{j}\partial_{i}\omega^{(1)}}{{M}^{3} {a}^{2}} + \frac{6 c_3 {{\pi}^{\prime}}^{2} {{\phi^{(1)}}^{\prime}}^{2}}{{M}^{3} {a}^{2}} + \frac{18 c_3 {{\pi}^{\prime}}^{2} {\phi^{(1)}}^{\prime} {\psi^{(1)}}^{\prime}}{{M}^{3} {a}^{2}} + \frac{12 c_3 {\pi^{(1)}}^{\prime\prime} {\pi^{(1)}}^{\prime} \mathcal{H}}{{M}^{3} {a}^{2}} \nonumber \\ 
& + \frac{c_3 {{\pi}^{\prime}}^{2} {\nabla^2 {\chi^{(1)}}^{\prime}}^{2}}{3 {M}^{3} {a}^{2}} -  \frac{4 c_3 {\pi}^{\prime} {\pi^{(1)}}^{\prime} \nabla^2 {\omega^{(1)}}^{\prime}}{{M}^{3} {a}^{2}} + \frac{c_3 {{\pi}^{\prime}}^{2} \nabla^2 {\chi^{(1)}}^{\prime\prime} \nabla^2 \chi^{(1)}}{3 {M}^{3} {a}^{2}} \nonumber \\ 
& -  \frac{2 c_3 {{\pi}^{\prime}}^{2} \nabla^2 {\omega^{(1)}}^{\prime} \nabla^2 \chi^{(1)}}{3 {M}^{3} {a}^{2}} -  \frac{4 c_3 {\pi^{(1)}}^{\prime\prime} \nabla^2 \pi^{(1)}}{{M}^{3} {a}^{2}} + \frac{8 c_3 {\pi}^{\prime} {\phi^{(1)}}^{\prime} \nabla^2 \pi^{(1)}}{{M}^{3} {a}^{2}} \nonumber \\ 
& + \frac{4 c_3 {\pi}^{\prime} {\psi^{(1)}}^{\prime} \nabla^2 \pi^{(1)}}{{M}^{3} {a}^{2}} -  \frac{4 c_3 {\pi^{(1)}}^{\prime} \mathcal{H} \nabla^2 \pi^{(1)}}{{M}^{3} {a}^{2}} -  \frac{2 c_3 {\pi}^{\prime} \nabla^2 {\chi^{(1)}}^{\prime} \nabla^2 \pi^{(1)}}{3 {M}^{3} {a}^{2}} + \frac{2 c_3 {\nabla^2 \pi^{(1)}}^{2}}{{M}^{3} {a}^{2}} \nonumber \\ 
& -  \frac{4 c_3 {\pi^{(1)}}^{\prime\prime} {\pi}^{\prime} \nabla^2 \omega^{(1)}}{{M}^{3} {a}^{2}} + \frac{8 c_3 {{\pi}^{\prime}}^{2} {\phi^{(1)}}^{\prime} \nabla^2 \omega^{(1)}}{{M}^{3} {a}^{2}} + \frac{6 c_3 {{\pi}^{\prime}}^{2} {\psi^{(1)}}^{\prime} \nabla^2 \omega^{(1)}}{{M}^{3} {a}^{2}} \nonumber \\ 
& -  \frac{2 c_3 {{\pi}^{\prime}}^{2} \nabla^2 {\chi^{(1)}}^{\prime} \nabla^2 \omega^{(1)}}{3 {M}^{3} {a}^{2}} + \frac{4 c_3 {\pi}^{\prime} \nabla^2 \pi^{(1)} \nabla^2 \omega^{(1)}}{{M}^{3} {a}^{2}} + \frac{2 c_3 {{\pi}^{\prime}}^{2} {\nabla^2 \omega^{(1)}}^{2}}{{M}^{3} {a}^{2}} \nonumber \\ 
& -  \frac{4 c_3 {\pi}^{\prime} {\pi^{(1)}}^{\prime} \nabla^2 \psi^{(1)}}{{M}^{3} {a}^{2}} -  \frac{2 c_3 {{\pi}^{\prime}}^{2} \nabla^2 \chi^{(1)} \nabla^2 \psi^{(1)}}{3 {M}^{3} {a}^{2}} -  \frac{12 c_3 {\phi^{(1)}}^{\prime\prime} {{\pi}^{\prime}}^{2} \phi^{(1)}}{{M}^{3} {a}^{2}} \nonumber \\ 
& -  \frac{4 c_3 {{\pi}^{\prime}}^{2} \nabla^2 {\omega^{(1)}}^{\prime} \phi^{(1)}}{{M}^{3} {a}^{2}} -  \frac{4 c_3 {{\pi}^{\prime}}^{2} \nabla^2 \psi^{(1)} \phi^{(1)}}{{M}^{3} {a}^{2}} + \frac{24 c_3 {\phi^{(1)}}^{\prime\prime} {{\pi}^{\prime}}^{2} \psi^{(1)}}{{M}^{3} {a}^{2}} \nonumber \\ 
& + \frac{8 c_3 {{\pi}^{\prime}}^{2} \nabla^2 {\omega^{(1)}}^{\prime} \psi^{(1)}}{{M}^{3} {a}^{2}} + \frac{8 c_3 {{\pi}^{\prime}}^{2} \nabla^2 \psi^{(1)} \psi^{(1)}}{{M}^{3} {a}^{2}} + \frac{4 c_3 \partial_{i}{\pi^{(1)}}^{\prime} \partial^{i}{\pi^{(1)}}^{\prime}}{{M}^{3} {a}^{2}} \nonumber \\ 
& -  \frac{8 c_3 \mathcal{H} \partial_{i}\pi^{(1)} \partial^{i}{\pi^{(1)}}^{\prime}}{{M}^{3} {a}^{2}} + 4 c_2 \partial_{i}\omega^{(1)} \partial^{i}{\pi^{(1)}}^{\prime} -  \frac{24 c_3 {\pi}^{\prime} \mathcal{H} \partial_{i}\omega^{(1)} \partial^{i}{\pi^{(1)}}^{\prime}}{{M}^{3} {a}^{2}} \nonumber \\ 
& -  \frac{8 c_3 {\pi}^{\prime} \partial_{i}\psi^{(1)} \partial^{i}{\pi^{(1)}}^{\prime}}{{M}^{3} {a}^{2}} + 2 c_2 \partial_{i}\pi^{(1)} \partial^{i}{\omega^{(1)}}^{\prime} -  \frac{12 c_3 {\pi}^{\prime} \mathcal{H} \partial_{i}\pi^{(1)} \partial^{i}{\omega^{(1)}}^{\prime}}{{M}^{3} {a}^{2}} \nonumber \\ 
& + 2 c_2 {\pi}^{\prime} \partial_{i}\omega^{(1)} \partial^{i}{\omega^{(1)}}^{\prime} -  \frac{18 c_3 {{\pi}^{\prime}}^{2} \mathcal{H} \partial_{i}\omega^{(1)} \partial^{i}{\omega^{(1)}}^{\prime}}{{M}^{3} {a}^{2}} + \frac{2 c_3 {{\pi}^{\prime}}^{2} \partial_{i}\phi^{(1)} \partial^{i}{\omega^{(1)}}^{\prime}}{{M}^{3} {a}^{2}} \nonumber \\ 
& + \frac{8 c_3 {\pi}^{\prime} \partial_{i}\pi^{(1)} \partial^{i}{\phi^{(1)}}^{\prime}}{{M}^{3} {a}^{2}} + \frac{8 c_3 {{\pi}^{\prime}}^{2} \partial_{i}\omega^{(1)} \partial^{i}{\phi^{(1)}}^{\prime}}{{M}^{3} {a}^{2}} + \frac{4 c_3 {\pi}^{\prime} \partial_{i}\pi^{(1)} \partial^{i}\nabla^2 {\chi^{(1)}}^{\prime}}{3 {M}^{3} {a}^{2}} \nonumber \\ 
& + \frac{4 c_3 {{\pi}^{\prime}}^{2} \partial_{i}\omega^{(1)} \partial^{i}\nabla^2 {\chi^{(1)}}^{\prime}}{3 {M}^{3} {a}^{2}} + \frac{4 c_3 {{\pi}^{\prime}}^{2} \partial_{i}{\omega^{(1)}}^{\prime} \partial^{i}\nabla^2 \chi^{(1)}}{3 {M}^{3} {a}^{2}} -  \frac{4}{3} c_2 \partial_{i}\pi^{(1)} \partial^{i}\nabla^2 \chi^{(1)} \nonumber \\ 
& + \frac{8 c_3 {\pi}^{\prime\prime} \partial_{i}\pi^{(1)} \partial^{i}\nabla^2 \chi^{(1)}}{3 {M}^{3} {a}^{2}} + \frac{8 c_3 {\pi}^{\prime} \mathcal{H} \partial_{i}\pi^{(1)} \partial^{i}\nabla^2 \chi^{(1)}}{3 {M}^{3} {a}^{2}} -  \frac{4}{3} c_2 {\pi}^{\prime} \partial_{i}\omega^{(1)} \partial^{i}\nabla^2 \chi^{(1)} \nonumber \\ 
& + \frac{8 c_3 {\pi}^{\prime\prime} {\pi}^{\prime} \partial_{i}\omega^{(1)} \partial^{i}\nabla^2 \chi^{(1)}}{3 {M}^{3} {a}^{2}} + \frac{4 c_3 {{\pi}^{\prime}}^{2} \mathcal{H} \partial_{i}\omega^{(1)} \partial^{i}\nabla^2 \chi^{(1)}}{{M}^{3} {a}^{2}} \nonumber \\ 
& + \frac{4 c_3 {{\pi}^{\prime}}^{2} \partial_{i}\psi^{(1)} \partial^{i}\nabla^2 \chi^{(1)}}{3 {M}^{3} {a}^{2}} -  \frac{2 c_3 {\mathcal{H}}^{\prime} \partial_{i}\pi^{(1)} \partial^{i}\pi^{(1)}}{{M}^{3} {a}^{2}} + 4 c_2 \mathcal{H} \partial_{i}\pi^{(1)} \partial^{i}\omega^{(1)} \nonumber \\ 
& -  \frac{12 c_3 {\pi}^{\prime} {\mathcal{H}}^{\prime} \partial_{i}\pi^{(1)} \partial^{i}\omega^{(1)}}{{M}^{3} {a}^{2}} -  \frac{12 c_3 {\pi}^{\prime\prime} \mathcal{H} \partial_{i}\pi^{(1)} \partial^{i}\omega^{(1)}}{{M}^{3} {a}^{2}} + 2 c_2 {\pi}^{\prime\prime} \partial_{i}\omega^{(1)} \partial^{i}\omega^{(1)} \nonumber \\ 
& + 4 c_2 {\pi}^{\prime} \mathcal{H} \partial_{i}\omega^{(1)} \partial^{i}\omega^{(1)} -  \frac{12 c_3 {{\pi}^{\prime}}^{2} {\mathcal{H}}^{\prime} \partial_{i}\omega^{(1)} \partial^{i}\omega^{(1)}}{{M}^{3} {a}^{2}} -  \frac{24 c_3 {\pi}^{\prime\prime} {\pi}^{\prime} \mathcal{H} \partial_{i}\omega^{(1)} \partial^{i}\omega^{(1)}}{{M}^{3} {a}^{2}} \nonumber \\ 
& - 2 c_2 \partial_{i}\pi^{(1)} \partial^{i}\phi^{(1)} + \frac{4 c_3 {\pi}^{\prime\prime} \partial_{i}\pi^{(1)} \partial^{i}\phi^{(1)}}{{M}^{3} {a}^{2}} + \frac{4 c_3 {\pi}^{\prime} \mathcal{H} \partial_{i}\pi^{(1)} \partial^{i}\phi^{(1)}}{{M}^{3} {a}^{2}} - 2 c_2 {\pi}^{\prime} \partial_{i}\omega^{(1)} \partial^{i}\phi^{(1)} \nonumber \\ 
& + \frac{4 c_3 {\pi}^{\prime\prime} {\pi}^{\prime} \partial_{i}\omega^{(1)} \partial^{i}\phi^{(1)}}{{M}^{3} {a}^{2}} + \frac{6 c_3 {{\pi}^{\prime}}^{2} \mathcal{H} \partial_{i}\omega^{(1)} \partial^{i}\phi^{(1)}}{{M}^{3} {a}^{2}} + 2 c_2 \partial_{i}\pi^{(1)} \partial^{i}\psi^{(1)} \nonumber \\ 
& + \frac{4 c_3 {\pi}^{\prime} \mathcal{H} \partial_{i}\pi^{(1)} \partial^{i}\psi^{(1)}}{{M}^{3} {a}^{2}} - 2 c_2 {\pi}^{\prime} \partial_{i}\omega^{(1)} \partial^{i}\psi^{(1)} + \frac{18 c_3 {{\pi}^{\prime}}^{2} \mathcal{H} \partial_{i}\omega^{(1)} \partial^{i}\psi^{(1)}}{{M}^{3} {a}^{2}} \nonumber \\ 
& + \frac{2 c_3 {{\pi}^{\prime}}^{2} \partial_{i}\phi^{(1)} \partial^{i}\psi^{(1)}}{{M}^{3} {a}^{2}} + \frac{6 c_3 {{\pi}^{\prime}}^{2} \partial_{i}\psi^{(1)} \partial^{i}\psi^{(1)}}{{M}^{3} {a}^{2}} -  \frac{c_3 {{\pi}^{\prime}}^{2} \partial_{j}\partial^{i}\chi^{(1)} \partial^{j}\partial_{i}{\chi^{(1)}}^{\prime\prime}}{{M}^{3} {a}^{2}} \nonumber \\ 
& -  \frac{c_3 {{\pi}^{\prime}}^{2} \partial_{j}\partial^{i}{\chi^{(1)}}^{\prime} \partial^{j}\partial_{i}{\chi^{(1)}}^{\prime}}{{M}^{3} {a}^{2}} + c_2 {\pi}^{\prime} \partial_{j}\partial^{i}\chi^{(1)} \partial^{j}\partial_{i}{\chi^{(1)}}^{\prime} -  \frac{2 c_3 {\pi}^{\prime\prime} {\pi}^{\prime} \partial_{j}\partial^{i}\chi^{(1)} \partial^{j}\partial_{i}{\chi^{(1)}}^{\prime}}{{M}^{3} {a}^{2}}\nonumber \\ 
& -  \frac{3 c_3 {{\pi}^{\prime}}^{2} \mathcal{H} \partial_{j}\partial^{i}\chi^{(1)} \partial^{j}\partial_{i}{\chi^{(1)}}^{\prime}}{{M}^{3} {a}^{2}} + \frac{2 c_3 {\pi}^{\prime} \partial_{j}\partial^{i}\pi^{(1)} \partial^{j}\partial_{i}{\chi^{(1)}}^{\prime}}{{M}^{3} {a}^{2}} + \frac{2 c_3 {{\pi}^{\prime}}^{2} \partial_{j}\partial^{i}\omega^{(1)} \partial^{j}\partial_{i}{\chi^{(1)}}^{\prime}}{{M}^{3} {a}^{2}} \nonumber \\ 
& + \frac{2 c_3 {{\pi}^{\prime}}^{2} \partial_{j}\partial^{i}\chi^{(1)} \partial^{j}\partial_{i}{\omega^{(1)}}^{\prime}}{{M}^{3} {a}^{2}} - 2 c_2 \partial_{j}\partial^{i}\chi^{(1)} \partial^{j}\partial_{i}\pi^{(1)} + \frac{4 c_3 {\pi}^{\prime\prime} \partial_{j}\partial^{i}\chi^{(1)} \partial^{j}\partial_{i}\pi^{(1)}}{{M}^{3} {a}^{2}} \nonumber \\ 
& + \frac{4 c_3 {\pi}^{\prime} \mathcal{H} \partial_{j}\partial^{i}\chi^{(1)} \partial^{j}\partial_{i}\pi^{(1)}}{{M}^{3} {a}^{2}} -  \frac{2 c_3 \partial_{j}\partial^{i}\pi^{(1)} \partial^{j}\partial_{i}\pi^{(1)}}{{M}^{3} {a}^{2}} - 2 c_2 {\pi}^{\prime} \partial_{j}\partial^{i}\chi^{(1)} \partial^{j}\partial_{i}\omega^{(1)} \nonumber \\ 
& + \frac{4 c_3 {\pi}^{\prime\prime} {\pi}^{\prime} \partial_{j}\partial^{i}\chi^{(1)} \partial^{j}\partial_{i}\omega^{(1)}}{{M}^{3} {a}^{2}} + \frac{6 c_3 {{\pi}^{\prime}}^{2} \mathcal{H} \partial_{j}\partial^{i}\chi^{(1)} \partial^{j}\partial_{i}\omega^{(1)}}{{M}^{3} {a}^{2}} + \frac{2 c_3 {{\pi}^{\prime}}^{2} \partial_{j}\partial^{i}\chi^{(1)} \partial^{j}\partial_{i}\psi^{(1)}}{{M}^{3} {a}^{2}}
\end{align}

\begin{align}
S^{(6)} &\equiv- \frac{1}{3} \nabla^2 {\chi^{(1)}}^{\prime} \nabla^2 \chi^{(1)} + 6 {\phi^{(1)}}^{\prime} \delta^{(1)} - 2 \nabla^2 v^{(1)} \delta^{(1)} + 12 {\phi^{(1)}}^{\prime} \phi^{(1)} - 2 \nabla^2 v^{(1)} \psi^{(1)} - 2 \partial_{i}\omega^{(1)} \partial^{i}{\omega^{(1)}}^{\prime} \nonumber \\ 
& - 2 \partial_{i}\omega^{(1)} \partial^{i}{v^{(1)}}^{\prime} - 4 \partial_{i}v^{(1)} \partial^{i}{v^{(1)}}^{\prime} - 2 \partial_{i}\delta^{(1)} \partial^{i}v^{(1)} - 2 \mathcal{H} \partial_{i}\omega^{(1)} \partial^{i}v^{(1)} + 6 \partial_{i}\phi^{(1)} \partial^{i}v^{(1)} \nonumber \\ 
& - 2 \mathcal{H} \partial_{i}v^{(1)} \partial^{i}v^{(1)} + \partial_{j}\partial^{i}\chi^{(1)} \partial^{j}\partial_{i}{\chi^{(1)}}^{\prime}- 4 \partial_{i}v^{(1)} \partial^{i}{\omega^{(1)}}^{\prime}- 4 \partial_{i}\psi^{(1)} \partial^{i}v^{(1)}
\end{align}

\begin{align}
S^{(7)} &\equiv\frac{2}{3} \nabla^2 {v^{(1)}}^{\prime} \nabla^2 \chi^{(1)} - 2 {\delta^{(1)}}^{\prime} \nabla^2 \omega^{(1)} + 6 {\phi^{(1)}}^{\prime} \nabla^2 \omega^{(1)} + 2 {\psi^{(1)}}^{\prime} \nabla^2 \omega^{(1)} - 2 {\delta^{(1)}}^{\prime} \nabla^2 v^{(1)} + 10 {\phi^{(1)}}^{\prime} \nabla^2 v^{(1)} \nonumber \\ 
& + \frac{2}{3} \nabla^2 {\chi^{(1)}}^{\prime} \nabla^2 v^{(1)} + \frac{2}{3} \mathcal{H} \nabla^2 \chi^{(1)} \nabla^2 v^{(1)} - 2 \nabla^2 \omega^{(1)} \nabla^2 v^{(1)} - 2 {\nabla^2 v^{(1)}}^{2} - 2 \nabla^2 {\omega^{(1)}}^{\prime} \delta^{(1)} - 2 \nabla^2 {v^{(1)}}^{\prime} \delta^{(1)} \nonumber \\ 
& - 2 \mathcal{H} \nabla^2 \omega^{(1)} \delta^{(1)} - 2 \nabla^2 \psi^{(1)} \delta^{(1)} - 2 \mathcal{H} \nabla^2 v^{(1)} \delta^{(1)} + 4 \nabla^2 {v^{(1)}}^{\prime} \phi^{(1)} + 4 \mathcal{H} \nabla^2 v^{(1)} \phi^{(1)} + 4 \nabla^2 {\omega^{(1)}}^{\prime} \psi^{(1)} \nonumber \\ 
& + 4 \mathcal{H} \nabla^2 \omega^{(1)} \psi^{(1)} + 4 \nabla^2 \psi^{(1)} \psi^{(1)} + 2 \mathcal{H} \nabla^2 v^{(1)} \psi^{(1)} - 2 \partial_{i}\omega^{(1)} \partial^{i}{\delta^{(1)}}^{\prime} - 2 \partial_{i}v^{(1)} \partial^{i}{\delta^{(1)}}^{\prime} - 2 \partial_{i}\delta^{(1)} \partial^{i}{\omega^{(1)}}^{\prime} \nonumber \\ 
& + 4 \partial_{i}\psi^{(1)} \partial^{i}{\omega^{(1)}}^{\prime} + 6 \partial_{i}\omega^{(1)} \partial^{i}{\phi^{(1)}}^{\prime} + 10 \partial_{i}v^{(1)} \partial^{i}{\phi^{(1)}}^{\prime} + 2 \partial_{i}\omega^{(1)} \partial^{i}{\psi^{(1)}}^{\prime} - 2 \partial_{i}\delta^{(1)} \partial^{i}{v^{(1)}}^{\prime} + 2 \nabla^2 {v^{(1)}}^{\prime} \psi^{(1)}\nonumber \\ 
& + 4 \partial_{i}\phi^{(1)} \partial^{i}{v^{(1)}}^{\prime} + 2 \partial_{i}\psi^{(1)} \partial^{i}{v^{(1)}}^{\prime} -  \frac{4}{3} \partial_{i}v^{(1)} \partial^{i}\nabla^2 {\chi^{(1)}}^{\prime} -  \frac{4}{3} \partial_{i}{v^{(1)}}^{\prime} \partial^{i}\nabla^2 \chi^{(1)} - 2 \partial_{j}\partial^{i}v^{(1)} \partial^{j}\partial_{i}v^{(1)}\nonumber \\ 
& -  \frac{4}{3} \mathcal{H} \partial_{i}v^{(1)} \partial^{i}\nabla^2 \chi^{(1)} - 2 \partial_{i}\omega^{(1)} \partial^{i}\nabla^2 v^{(1)} - 4 \partial_{i}v^{(1)} \partial^{i}\nabla^2 v^{(1)} - 2 \mathcal{H} \partial_{i}\delta^{(1)} \partial^{i}\omega^{(1)} - 2 \partial_{i}\delta^{(1)} \partial^{i}\psi^{(1)} \nonumber \\ 
& + 4 \mathcal{H} \partial_{i}\omega^{(1)} \partial^{i}\psi^{(1)} + 4 \partial_{i}\psi^{(1)} \partial^{i}\psi^{(1)} - 2 \mathcal{H} \partial_{i}\delta^{(1)} \partial^{i}v^{(1)} + 4 \mathcal{H} \partial_{i}\phi^{(1)} \partial^{i}v^{(1)} + 2 \mathcal{H} \partial_{i}\psi^{(1)} \partial^{i}v^{(1)} \nonumber \\ 
& - 2 \partial_{j}\partial^{i}v^{(1)} \partial^{j}\partial_{i}{\chi^{(1)}}^{\prime} - 2 \partial_{j}\partial^{i}\chi^{(1)} \partial^{j}\partial_{i}{v^{(1)}}^{\prime} - 2 \mathcal{H} \partial_{j}\partial^{i}\chi^{(1)} \partial^{j}\partial_{i}v^{(1)}
\end{align}

\section{Background quantities for the second-order DM kernel} \label{SEC:AppendixC}

In the following we give the explicit expression for the background functions $\g_i(a)$ found in the kernel (\ref{kernel}). They reads

\begin{flalign}
\g_{4}&(\tau)\equiv- \frac{5 \rho_m {f}^{2}}{2 {\Mpl}^{2}} + \frac{7 {c_3}^{6} {\rho_m}^{2} {{\pi}^{\prime}}^{12}}{64 {c_2}^{3} {M}^{18} {\Mpl}^{10} {\alpha}^{3} {\mathcal{H}}^{2} {a}^{10}} + \frac{3 {c_3}^{5} {\rho_m}^{2} {\alpha}^{\prime} {{\pi}^{\prime}}^{9}}{8 {c_2}^{3} {M}^{15} {\Mpl}^{8} {\alpha}^{4} {\mathcal{H}}^{2} {a}^{8}} \nonumber \\ 
& + \frac{7 {c_3}^{5} {\rho_m}^{2} {{\pi}^{\prime}}^{9}}{16 {c_2}^{3} {M}^{15} {\Mpl}^{8} {\alpha}^{3} \mathcal{H} {a}^{8}} -  \frac{3 {c_3}^{5} {\rho_m}^{2} {{\pi}^{\prime}}^{9} f}{8 {c_2}^{3} {M}^{15} {\Mpl}^{8} {\alpha}^{3} \mathcal{H} {a}^{8}} -  \frac{35 {c_3}^{4} {\rho_m}^{2} {{\pi}^{\prime}}^{6}}{2 {c_2}^{3} {M}^{12} {\Mpl}^{6} {\alpha}^{3} {a}^{6}} \nonumber \\ 
& + \frac{6 {c_3}^{4} {\rho_m}^{2} {{\pi}^{\prime}}^{6} f}{{c_2}^{3} {M}^{12} {\Mpl}^{6} {\alpha}^{3} {a}^{6}} -  \frac{{c_3}^{4} {\rho_m}^{2} {{\pi}^{\prime}}^{6} {f}^{2}}{2 {c_2}^{3} {M}^{12} {\Mpl}^{6} {\alpha}^{3} {a}^{6}} -  \frac{{c_3}^{4} {\rho_m}^{2} {{\alpha}^{\prime}}^{2} {{\pi}^{\prime}}^{6}}{2 {c_2}^{3} {M}^{12} {\Mpl}^{6} {\alpha}^{5} {\mathcal{H}}^{2} {a}^{6}} \nonumber \\ 
& + \frac{7 {c_3}^{4} {\rho_m}^{2} {{\pi}^{\prime}}^{8}}{32 {c_2}^{2} {M}^{12} {\Mpl}^{8} {\alpha}^{3} {\mathcal{H}}^{2} {a}^{6}} + \frac{9 {c_3}^{4} {\rho_m}^{2} {{\pi}^{\prime}}^{8}}{16 {c_2}^{2} {M}^{12} {\Mpl}^{8} {\alpha}^{2} {\mathcal{H}}^{2} {a}^{6}} + \frac{2 {f}^{2} {\mathcal{H}}^{2}}{{a}^{2}}\nonumber \\ 
& + \frac{{c_3}^{4} {\rho_m}^{2} {{\pi}^{\prime}}^{6} {\mathcal{H}}^{\prime}}{4 {c_2}^{3} {M}^{12} {\Mpl}^{6} {\alpha}^{3} {\mathcal{H}}^{2} {a}^{6}} -  \frac{6 {c_3}^{4} {\rho_m}^{2} {\alpha}^{\prime} {{\pi}^{\prime}}^{6}}{{c_2}^{3} {M}^{12} {\Mpl}^{6} {\alpha}^{4} \mathcal{H} {a}^{6}} + \frac{2 c_3 \rho_m {{\pi}^{\prime}}^{3} f}{{M}^{3} {\Mpl}^{4} \alpha \mathcal{H} {a}^{2}} \nonumber \\ 
& + \frac{{c_3}^{4} {\rho_m}^{2} {\alpha}^{\prime} {{\pi}^{\prime}}^{6} f}{{c_2}^{3} {M}^{12} {\Mpl}^{6} {\alpha}^{4} \mathcal{H} {a}^{6}} + \frac{{c_3}^{3} \rho_m {{\pi}^{\prime}}^{7} f}{c_2 {M}^{9} {\Mpl}^{6} \alpha \mathcal{H} {a}^{6}} -  \frac{8 {c_3}^{2} \rho_m {{\pi}^{\prime}}^{4} f}{c_2 {M}^{6} {\Mpl}^{4} \alpha {a}^{4}} \nonumber \\ 
& + \frac{5 {c_3}^{2} \rho_m {{\pi}^{\prime}}^{4} {f}^{2}}{4 c_2 {M}^{6} {\Mpl}^{4} \alpha {a}^{4}} + \frac{{c_3}^{3} {\rho_m}^{2} {\alpha}^{\prime} {{\pi}^{\prime}}^{5}}{{c_2}^{2} {M}^{9} {\Mpl}^{6} {\alpha}^{4} {\mathcal{H}}^{2} {a}^{4}} -  \frac{7 {c_3}^{3} {\rho_m}^{2} {\alpha}^{\prime} {{\pi}^{\prime}}^{5}}{4 {c_2}^{2} {M}^{9} {\Mpl}^{6} {\alpha}^{3} {\mathcal{H}}^{2} {a}^{4}} \nonumber \\ 
& + \frac{6 {c_3}^{3} {\rho_m}^{2} {{\pi}^{\prime}}^{5}}{{c_2}^{2} {M}^{9} {\Mpl}^{6} {\alpha}^{3} \mathcal{H} {a}^{4}} -  \frac{71 {c_3}^{3} {\rho_m}^{2} {{\pi}^{\prime}}^{5}}{8 {c_2}^{2} {M}^{9} {\Mpl}^{6} {\alpha}^{2} \mathcal{H} {a}^{4}} -  \frac{{c_3}^{2} \rho_m {\alpha}^{\prime} {{\pi}^{\prime}}^{4} f}{c_2 {M}^{6} {\Mpl}^{4} {\alpha}^{2} \mathcal{H} {a}^{4}} \nonumber \\ 
& -  \frac{{c_3}^{3} {\rho_m}^{2} {{\pi}^{\prime}}^{5} f}{{c_2}^{2} {M}^{9} {\Mpl}^{6} {\alpha}^{3} \mathcal{H} {a}^{4}} + \frac{3 {c_3}^{3} {\rho_m}^{2} {{\pi}^{\prime}}^{5} f}{4 {c_2}^{2} {M}^{9} {\Mpl}^{6} {\alpha}^{2} \mathcal{H} {a}^{4}} -  \frac{2 {\mathcal{H}}^{\prime} {f}^{2}}{{a}^{2}} -  \frac{2 c_3 \rho_m {{\pi}^{\prime}}^{3} f}{{M}^{3} {\Mpl}^{4} \mathcal{H} {a}^{2}} \nonumber \\ 
& -  \frac{{c_3}^{2} {\rho_m}^{2} {{\pi}^{\prime}}^{4}}{2 c_2 {M}^{6} {\Mpl}^{6} {\alpha}^{3} {\mathcal{H}}^{2} {a}^{2}} + \frac{25 {c_3}^{2} {\rho_m}^{2} {{\pi}^{\prime}}^{4}}{16 c_2 {M}^{6} {\Mpl}^{6} {\alpha}^{2} {\mathcal{H}}^{2} {a}^{2}} -  \frac{25 {c_3}^{2} {\rho_m}^{2} {{\pi}^{\prime}}^{4}}{16 c_2 {M}^{6} {\Mpl}^{6} \alpha {\mathcal{H}}^{2} {a}^{2}}
\end{flalign}

\begin{flalign}
\g_{5}&(\tau)\equiv\frac{3 \rho_m {f}^{2}}{2 {\Mpl}^{2}} + \frac{15 {c_3}^{6} {\rho_m}^{2} {{\pi}^{\prime}}^{12}}{64 {c_2}^{3} {M}^{18} {\Mpl}^{10} {\alpha}^{3} {\mathcal{H}}^{2} {a}^{10}} -  \frac{3 {c_3}^{5} {\rho_m}^{2} {\alpha}^{\prime} {{\pi}^{\prime}}^{9}}{8 {c_2}^{3} {M}^{15} {\Mpl}^{8} {\alpha}^{4} {\mathcal{H}}^{2} {a}^{8}} \nonumber \\ 
& -  \frac{33 {c_3}^{5} {\rho_m}^{2} {{\pi}^{\prime}}^{9}}{16 {c_2}^{3} {M}^{15} {\Mpl}^{8} {\alpha}^{3} \mathcal{H} {a}^{8}} + \frac{3 {c_3}^{5} {\rho_m}^{2} {{\pi}^{\prime}}^{9} f}{8 {c_2}^{3} {M}^{15} {\Mpl}^{8} {\alpha}^{3} \mathcal{H} {a}^{8}} + \frac{15 {c_3}^{4} {\rho_m}^{2} {{\pi}^{\prime}}^{8}}{32 {c_2}^{2} {M}^{12} {\Mpl}^{8} {\alpha}^{3} {\mathcal{H}}^{2} {a}^{6}} \nonumber \\ 
& -  \frac{3 {c_3}^{4} {\rho_m}^{2} {{\pi}^{\prime}}^{8}}{4 {c_2}^{2} {M}^{12} {\Mpl}^{8} {\alpha}^{2} {\mathcal{H}}^{2} {a}^{6}} -  \frac{3 {c_3}^{2} \rho_m {{\pi}^{\prime}}^{4} {f}^{2}}{4 c_2 {M}^{6} {\Mpl}^{4} \alpha {a}^{4}} + \frac{3 {c_3}^{3} {\rho_m}^{2} {\alpha}^{\prime} {{\pi}^{\prime}}^{5}}{4 {c_2}^{2} {M}^{9} {\Mpl}^{6} {\alpha}^{3} {\mathcal{H}}^{2} {a}^{4}} \nonumber \\ 
& + \frac{33 {c_3}^{3} {\rho_m}^{2} {{\pi}^{\prime}}^{5}}{8 {c_2}^{2} {M}^{9} {\Mpl}^{6} {\alpha}^{2} \mathcal{H} {a}^{4}} -  \frac{3 {c_3}^{3} {\rho_m}^{2} {{\pi}^{\prime}}^{5} f}{4 {c_2}^{2} {M}^{9} {\Mpl}^{6} {\alpha}^{2} \mathcal{H} {a}^{4}} -  \frac{15 {c_3}^{2} {\rho_m}^{2} {{\pi}^{\prime}}^{4}}{16 c_2 {M}^{6} {\Mpl}^{6} {\alpha}^{2} {\mathcal{H}}^{2} {a}^{2}} \nonumber \\ 
& + \frac{3 {c_3}^{2} {\rho_m}^{2} {{\pi}^{\prime}}^{4}}{16 c_2 {M}^{6} {\Mpl}^{6} \alpha {\mathcal{H}}^{2} {a}^{2}} + \frac{3 {\rho_m}^{2} {a}^{2}}{4 {\Mpl}^{4} {\mathcal{H}}^{2}}
\end{flalign}

\begin{flalign}
 \g_{6}&(\tau)\equiv\frac{15 \rho_m f}{2 {\Mpl}^{2}} -  \frac{15 \rho_m {f}^{2}}{4 {\Mpl}^{2}} -  \frac{9 c_2 \rho_m {{\pi}^{\prime}}^{2}}{2 {\Mpl}^{4} {\mathcal{H}}^{2}} + \frac{9 c_2 \rho_m {{\pi}^{\prime}}^{2}}{4 {\Mpl}^{4} \alpha {\mathcal{H}}^{2}} + \frac{9 c_2 \alpha \rho_m {{\pi}^{\prime}}^{2}}{4 {\Mpl}^{4} {\mathcal{H}}^{2}} \nonumber \\ 
& + \frac{3 \rho_m {\mathcal{H}}^{\prime}}{2 {\Mpl}^{2} {\mathcal{H}}^{2}} + \frac{59 {c_3}^{6} {\rho_m}^{2} {{\pi}^{\prime}}^{12}}{128 {c_2}^{3} {M}^{18} {\Mpl}^{10} {\alpha}^{3} {\mathcal{H}}^{2} {a}^{10}} -  \frac{9 {c_3}^{5} {\rho_m}^{2} {\alpha}^{\prime} {{\pi}^{\prime}}^{9}}{16 {c_2}^{3} {M}^{15} {\Mpl}^{8} {\alpha}^{4} {\mathcal{H}}^{2} {a}^{8}} \nonumber \\ 
& + \frac{21 {c_3}^{4} \rho_m {{\pi}^{\prime}}^{10}}{16 c_2 {M}^{12} {\Mpl}^{8} \alpha {\mathcal{H}}^{2} {a}^{8}} -  \frac{{c_3}^{6} {\rho_m}^{3} {{\pi}^{\prime}}^{10}}{16 {c_2}^{4} {M}^{18} {\Mpl}^{10} {\alpha}^{4} {\mathcal{H}}^{2} {a}^{8}} -  \frac{157 {c_3}^{5} {\rho_m}^{2} {{\pi}^{\prime}}^{9}}{32 {c_2}^{3} {M}^{15} {\Mpl}^{8} {\alpha}^{3} \mathcal{H} {a}^{8}} \nonumber \\ 
& + \frac{9 {c_3}^{5} {\rho_m}^{2} {{\pi}^{\prime}}^{9} f}{16 {c_2}^{3} {M}^{15} {\Mpl}^{8} {\alpha}^{3} \mathcal{H} {a}^{8}} + \frac{5 {c_3}^{4} {\rho_m}^{2} {{\pi}^{\prime}}^{6}}{{c_2}^{3} {M}^{12} {\Mpl}^{6} {\alpha}^{3} {a}^{6}} -  \frac{5 {c_3}^{4} {\rho_m}^{2} {{\pi}^{\prime}}^{6} f}{2 {c_2}^{3} {M}^{12} {\Mpl}^{6} {\alpha}^{3} {a}^{6}} \nonumber \\ 
& -  \frac{{c_3}^{4} {\alpha}^{\prime\prime} {\rho_m}^{2} {{\pi}^{\prime}}^{6}}{4 {c_2}^{3} {M}^{12} {\Mpl}^{6} {\alpha}^{4} {\mathcal{H}}^{2} {a}^{6}} -  \frac{3 {c_3}^{3} \rho_m {\alpha}^{\prime} {{\pi}^{\prime}}^{7}}{2 c_2 {M}^{9} {\Mpl}^{6} {\alpha}^{2} {\mathcal{H}}^{2} {a}^{6}} + \frac{79 {c_3}^{4} {\rho_m}^{2} {{\pi}^{\prime}}^{8}}{64 {c_2}^{2} {M}^{12} {\Mpl}^{8} {\alpha}^{3} {\mathcal{H}}^{2} {a}^{6}} \nonumber \\ 
& -  \frac{17 {c_3}^{4} {\rho_m}^{2} {{\pi}^{\prime}}^{8}}{16 {c_2}^{2} {M}^{12} {\Mpl}^{8} {\alpha}^{2} {\mathcal{H}}^{2} {a}^{6}} -  \frac{5 {c_3}^{4} {\rho_m}^{2} {{\pi}^{\prime}}^{6} {\mathcal{H}}^{\prime}}{4 {c_2}^{3} {M}^{12} {\Mpl}^{6} {\alpha}^{3} {\mathcal{H}}^{2} {a}^{6}} -  \frac{7 {\rho_m}^{2} {a}^{2}}{8 {\Mpl}^{4} {\mathcal{H}}^{2}}\nonumber \\ 
& + \frac{9 {c_3}^{4} {\rho_m}^{2} {\alpha}^{\prime} {{\pi}^{\prime}}^{6}}{4 {c_2}^{3} {M}^{12} {\Mpl}^{6} {\alpha}^{4} \mathcal{H} {a}^{6}} -  \frac{33 {c_3}^{3} \rho_m {{\pi}^{\prime}}^{7}}{2 c_2 {M}^{9} {\Mpl}^{6} \alpha \mathcal{H} {a}^{6}} -  \frac{{c_3}^{4} {\rho_m}^{2} {\alpha}^{\prime} {{\pi}^{\prime}}^{6} f}{2 {c_2}^{3} {M}^{12} {\Mpl}^{6} {\alpha}^{4} \mathcal{H} {a}^{6}} \nonumber \\ 
& + \frac{15 {c_3}^{3} \rho_m {{\pi}^{\prime}}^{7} f}{4 c_2 {M}^{9} {\Mpl}^{6} \alpha \mathcal{H} {a}^{6}} + \frac{54 {c_3}^{2} \rho_m {{\pi}^{\prime}}^{4}}{c_2 {M}^{6} {\Mpl}^{4} \alpha {a}^{4}} -  \frac{135 {c_3}^{2} \rho_m {{\pi}^{\prime}}^{4} f}{4 c_2 {M}^{6} {\Mpl}^{4} \alpha {a}^{4}} \nonumber \\ 
& + \frac{15 {c_3}^{2} \rho_m {{\pi}^{\prime}}^{4} {f}^{2}}{8 c_2 {M}^{6} {\Mpl}^{4} \alpha {a}^{4}} -  \frac{3 {c_3}^{2} {\alpha}^{\prime\prime} \rho_m {{\pi}^{\prime}}^{4}}{4 c_2 {M}^{6} {\Mpl}^{4} {\alpha}^{2} {\mathcal{H}}^{2} {a}^{4}} + \frac{3 {c_3}^{2} \rho_m {{\alpha}^{\prime}}^{2} {{\pi}^{\prime}}^{4}}{2 c_2 {M}^{6} {\Mpl}^{4} {\alpha}^{3} {\mathcal{H}}^{2} {a}^{4}} \nonumber \\ 
& -  \frac{{c_3}^{3} {\rho_m}^{2} {\alpha}^{\prime} {{\pi}^{\prime}}^{5}}{2 {c_2}^{2} {M}^{9} {\Mpl}^{6} {\alpha}^{4} {\mathcal{H}}^{2} {a}^{4}} + \frac{{c_3}^{3} {\rho_m}^{2} {\alpha}^{\prime} {{\pi}^{\prime}}^{5}}{8 {c_2}^{2} {M}^{9} {\Mpl}^{6} {\alpha}^{3} {\mathcal{H}}^{2} {a}^{4}} + \frac{15 c_3 \rho_m {{\pi}^{\prime}}^{3} f}{2 {M}^{3} {\Mpl}^{4} \alpha \mathcal{H} {a}^{2}}\nonumber \\ 
& -  \frac{15 {c_3}^{2} \rho_m {{\pi}^{\prime}}^{6}}{4 {M}^{6} {\Mpl}^{6} {\mathcal{H}}^{2} {a}^{4}} + \frac{15 {c_3}^{2} \rho_m {{\pi}^{\prime}}^{6}}{4 {M}^{6} {\Mpl}^{6} \alpha {\mathcal{H}}^{2} {a}^{4}} + \frac{{c_3}^{4} {\rho_m}^{3} {{\pi}^{\prime}}^{6}}{8 {c_2}^{3} {M}^{12} {\Mpl}^{8} {\alpha}^{3} {\mathcal{H}}^{2} {a}^{4}} \nonumber \\ 
& -  \frac{27 {c_3}^{2} \rho_m {{\pi}^{\prime}}^{4} {\mathcal{H}}^{\prime}}{4 c_2 {M}^{6} {\Mpl}^{4} \alpha {\mathcal{H}}^{2} {a}^{4}} + \frac{51 {c_3}^{2} \rho_m {\alpha}^{\prime} {{\pi}^{\prime}}^{4}}{4 c_2 {M}^{6} {\Mpl}^{4} {\alpha}^{2} \mathcal{H} {a}^{4}} -  \frac{15 c_3 \rho_m {{\pi}^{\prime}}^{3} f}{2 {M}^{3} {\Mpl}^{4} \mathcal{H} {a}^{2}}\nonumber \\ 
& -  \frac{3 {c_3}^{3} {\rho_m}^{2} {{\pi}^{\prime}}^{5}}{2 {c_2}^{2} {M}^{9} {\Mpl}^{6} {\alpha}^{3} \mathcal{H} {a}^{4}} + \frac{29 {c_3}^{3} {\rho_m}^{2} {{\pi}^{\prime}}^{5}}{16 {c_2}^{2} {M}^{9} {\Mpl}^{6} {\alpha}^{2} \mathcal{H} {a}^{4}} -  \frac{15 {c_3}^{2} \rho_m {\alpha}^{\prime} {{\pi}^{\prime}}^{4} f}{4 c_2 {M}^{6} {\Mpl}^{4} {\alpha}^{2} \mathcal{H} {a}^{4}} \nonumber \\ 
& + \frac{{c_3}^{3} {\rho_m}^{2} {{\pi}^{\prime}}^{5} f}{2 {c_2}^{2} {M}^{9} {\Mpl}^{6} {\alpha}^{3} \mathcal{H} {a}^{4}} -  \frac{9 {c_3}^{3} {\rho_m}^{2} {{\pi}^{\prime}}^{5} f}{8 {c_2}^{2} {M}^{9} {\Mpl}^{6} {\alpha}^{2} \mathcal{H} {a}^{4}} -  \frac{3 c_3 \rho_m {\alpha}^{\prime} {{\pi}^{\prime}}^{3}}{{M}^{3} {\Mpl}^{4} {\alpha}^{2} {\mathcal{H}}^{2} {a}^{2}} \nonumber \\ 
& + \frac{3 c_3 \rho_m {\alpha}^{\prime} {{\pi}^{\prime}}^{3}}{2 {M}^{3} {\Mpl}^{4} \alpha {\mathcal{H}}^{2} {a}^{2}} + \frac{{c_3}^{2} {\rho_m}^{2} {{\pi}^{\prime}}^{4}}{8 c_2 {M}^{6} {\Mpl}^{6} {\alpha}^{3} {\mathcal{H}}^{2} {a}^{2}} -  \frac{15 {c_3}^{2} {\rho_m}^{2} {{\pi}^{\prime}}^{4}}{32 c_2 {M}^{6} {\Mpl}^{6} {\alpha}^{2} {\mathcal{H}}^{2} {a}^{2}} + \frac{21 c_3 \rho_m {{\pi}^{\prime}}^{3}}{{M}^{3} {\Mpl}^{4} \mathcal{H} {a}^{2}}\nonumber \\ 
& + \frac{{c_3}^{4} {\rho_m}^{2} {{\alpha}^{\prime}}^{2} {{\pi}^{\prime}}^{6}}{2 {c_2}^{3} {M}^{12} {\Mpl}^{6} {\alpha}^{5} {\mathcal{H}}^{2} {a}^{6}} + \frac{23 {c_3}^{2} {\rho_m}^{2} {{\pi}^{\prime}}^{4}}{32 c_2 {M}^{6} {\Mpl}^{6} \alpha {\mathcal{H}}^{2} {a}^{2}} -  \frac{21 c_3 \rho_m {{\pi}^{\prime}}^{3}}{{M}^{3} {\Mpl}^{4} \alpha \mathcal{H} {a}^{2}}
\end{flalign}

\begin{flalign}
 \g_7&(\tau)\equiv- \frac{3 \rho_m {f}^{2}}{4 {\Mpl}^{2}} -  \frac{15 {c_3}^{6} {\pi^\prime}^{12} {\rho_m}^{2}}{128 {c_2}^{3} {M}^{18} {\Mpl}^{10} {\alpha}^{3} {\mathcal{H}}^{2} {a}^{10}} + \frac{3 {c_3}^{5} {\pi^\prime}^{9} {\rho_m}^{2} {\alpha}^{\prime}}{16 {c_2}^{3} {M}^{15} {\Mpl}^{8} {\alpha}^{4} {\mathcal{H}}^{2} {a}^{8}} \nonumber \\ 
& + \frac{33 {c_3}^{5} {\pi^\prime}^{9} {\rho_m}^{2}}{32 {c_2}^{3} {M}^{15} {\Mpl}^{8} {\alpha}^{3} \mathcal{H} {a}^{8}} -  \frac{3 {c_3}^{5} {\pi^\prime}^{9} {\rho_m}^{2} f}{16 {c_2}^{3} {M}^{15} {\Mpl}^{8} {\alpha}^{3} \mathcal{H} {a}^{8}} -  \frac{15 {c_3}^{4} {\pi^\prime}^{8} {\rho_m}^{2}}{64 {c_2}^{2} {M}^{12} {\Mpl}^{8} {\alpha}^{3} {\mathcal{H}}^{2} {a}^{6}} \nonumber \\ 
& + \frac{3 {c_3}^{4} {\pi^\prime}^{8} {\rho_m}^{2}}{8 {c_2}^{2} {M}^{12} {\Mpl}^{8} {\alpha}^{2} {\mathcal{H}}^{2} {a}^{6}} + \frac{3 {c_3}^{2} {\pi^\prime}^{4} \rho_m {f}^{2}}{8 c_2 {M}^{6} {\Mpl}^{4} \alpha {a}^{4}} -  \frac{3 {c_3}^{3} {\pi^\prime}^{5} {\rho_m}^{2} {\alpha}^{\prime}}{8 {c_2}^{2} {M}^{9} {\Mpl}^{6} {\alpha}^{3} {\mathcal{H}}^{2} {a}^{4}} \nonumber \\ 
& -  \frac{33 {c_3}^{3} {\pi^\prime}^{5} {\rho_m}^{2}}{16 {c_2}^{2} {M}^{9} {\Mpl}^{6} {\alpha}^{2} \mathcal{H} {a}^{4}} + \frac{3 {c_3}^{3} {\pi^\prime}^{5} {\rho_m}^{2} f}{8 {c_2}^{2} {M}^{9} {\Mpl}^{6} {\alpha}^{2} \mathcal{H} {a}^{4}} + \frac{15 {c_3}^{2} {\pi^\prime}^{4} {\rho_m}^{2}}{32 c_2 {M}^{6} {\Mpl}^{6} {\alpha}^{2} {\mathcal{H}}^{2} {a}^{2}} \nonumber \\ 
& -  \frac{3 {c_3}^{2} {\pi^\prime}^{4} {\rho_m}^{2}}{32 c_2 {M}^{6} {\Mpl}^{6} \alpha {\mathcal{H}}^{2} {a}^{2}} -  \frac{3 {\rho_m}^{2} {a}^{2}}{8 {\Mpl}^{4} {\mathcal{H}}^{2}}
\end{flalign}

\begin{flalign}
 \g_8&(\tau)\equiv\frac{69 c_2 {\pi^\prime}^{2} {\rho_m}^{2}}{8 {\Mpl}^{6} {\mathcal{H}}^{4}} + \frac{c_2 {\pi^\prime}^{2} {\rho_m}^{2}}{8 {\Mpl}^{6} {\alpha}^{2} {\mathcal{H}}^{4}} -  \frac{19 c_2 {\pi^\prime}^{2} {\rho_m}^{2}}{4 {\Mpl}^{6} \alpha {\mathcal{H}}^{4}} -  \frac{9 c_2 \alpha {\pi^\prime}^{2} {\rho_m}^{2}}{2 {\Mpl}^{6} {\mathcal{H}}^{4}} \nonumber \\ 
& -  \frac{3 {\rho_m}^{2} {\mathcal{H}}^{\prime}}{{\Mpl}^{4} {\mathcal{H}}^{4}} -  \frac{3 {\rho_m}^{2}}{4 {\Mpl}^{4} {\mathcal{H}}^{2}} -  \frac{21 {\rho_m}^{2} f}{2 {\Mpl}^{4} {\mathcal{H}}^{2}} + \frac{9 {\rho_m}^{2} {f}^{2}}{4 {\Mpl}^{4} {\mathcal{H}}^{2}}  + \frac{39 {c_3}^{8} {\pi^\prime}^{18} {\rho_m}^{2}}{16 {c_2}^{3} {M}^{24} {\Mpl}^{14} {\alpha}^{3} {\mathcal{H}}^{4} {a}^{16}}\nonumber \\ 
& -  \frac{3 {c_3}^{8} {\pi^\prime}^{16} {\rho_m}^{3}}{8 {c_2}^{4} {M}^{24} {\Mpl}^{14} {\alpha}^{4} {\mathcal{H}}^{4} {a}^{14}} -  \frac{165 {c_3}^{7} {\pi^\prime}^{15} {\rho_m}^{2} {\alpha}^{\prime}}{32 {c_2}^{3} {M}^{21} {\Mpl}^{12} {\alpha}^{4} {\mathcal{H}}^{4} {a}^{14}} -  \frac{141 {c_3}^{6} {\pi^\prime}^{14} {\rho_m}^{2}}{16 {c_2}^{2} {M}^{18} {\Mpl}^{12} {\alpha}^{2} {\mathcal{H}}^{4} {a}^{12}}\nonumber \\ 
& -  \frac{1227 {c_3}^{7} {\pi^\prime}^{15} {\rho_m}^{2}}{32 {c_2}^{3} {M}^{21} {\Mpl}^{12} {\alpha}^{3} {\mathcal{H}}^{3} {a}^{14}} + \frac{165 {c_3}^{7} {\pi^\prime}^{15} {\rho_m}^{2} f}{32 {c_2}^{3} {M}^{21} {\Mpl}^{12} {\alpha}^{3} {\mathcal{H}}^{3} {a}^{14}} + \frac{623 {c_3}^{6} {\pi^\prime}^{14} {\rho_m}^{2}}{64 {c_2}^{2} {M}^{18} {\Mpl}^{12} {\alpha}^{3} {\mathcal{H}}^{4} {a}^{12}} \nonumber \\ 
& -  \frac{3 {c_3}^{6} {\pi^\prime}^{12} {\alpha}^{\prime\prime} {\rho_m}^{2}}{2 {c_2}^{3} {M}^{18} {\Mpl}^{10} {\alpha}^{4} {\mathcal{H}}^{4} {a}^{12}} + \frac{3 {c_3}^{7} {\pi^\prime}^{13} {\rho_m}^{3} {\alpha}^{\prime}}{8 {c_2}^{4} {M}^{21} {\Mpl}^{12} {\alpha}^{5} {\mathcal{H}}^{4} {a}^{12}} + \frac{183 {c_3}^{6} {\pi^\prime}^{12} {\rho_m}^{2} {{\alpha}^{\prime}}^{2}}{32 {c_2}^{3} {M}^{18} {\Mpl}^{10} {\alpha}^{5} {\mathcal{H}}^{4} {a}^{12}} \nonumber \\ 
& -  \frac{105 {c_3}^{6} {\pi^\prime}^{12} {\rho_m}^{2} {\mathcal{H}}^{\prime}}{16 {c_2}^{3} {M}^{18} {\Mpl}^{10} {\alpha}^{3} {\mathcal{H}}^{4} {a}^{12}} + \frac{27 {c_3}^{7} {\pi^\prime}^{13} {\rho_m}^{3}}{16 {c_2}^{4} {M}^{21} {\Mpl}^{12} {\alpha}^{4} {\mathcal{H}}^{3} {a}^{12}} + \frac{453 {c_3}^{6} {\pi^\prime}^{12} {\rho_m}^{2} {\alpha}^{\prime}}{8 {c_2}^{3} {M}^{18} {\Mpl}^{10} {\alpha}^{4} {\mathcal{H}}^{3} {a}^{12}} \nonumber \\ 
& -  \frac{3 {c_3}^{7} {\pi^\prime}^{13} {\rho_m}^{3} f}{8 {c_2}^{4} {M}^{21} {\Mpl}^{12} {\alpha}^{4} {\mathcal{H}}^{3} {a}^{12}} -  \frac{135 {c_3}^{6} {\pi^\prime}^{12} {\rho_m}^{2} {\alpha}^{\prime} f}{16 {c_2}^{3} {M}^{18} {\Mpl}^{10} {\alpha}^{4} {\mathcal{H}}^{3} {a}^{12}} + \frac{5541 {c_3}^{6} {\pi^\prime}^{12} {\rho_m}^{2}}{32 {c_2}^{3} {M}^{18} {\Mpl}^{10} {\alpha}^{3} {\mathcal{H}}^{2} {a}^{12}} \nonumber \\ 
& -  \frac{465 {c_3}^{6} {\pi^\prime}^{12} {\rho_m}^{2} f}{8 {c_2}^{3} {M}^{18} {\Mpl}^{10} {\alpha}^{3} {\mathcal{H}}^{2} {a}^{12}} + \frac{87 {c_3}^{6} {\pi^\prime}^{12} {\rho_m}^{2} {f}^{2}}{32 {c_2}^{3} {M}^{18} {\Mpl}^{10} {\alpha}^{3} {\mathcal{H}}^{2} {a}^{12}} -  \frac{{c_3}^{6} {\pi^\prime}^{12} {\rho_m}^{3}}{2 {c_2}^{3} {M}^{18} {\Mpl}^{12} {\alpha}^{4} {\mathcal{H}}^{4} {a}^{10}} \nonumber \\ 
& + \frac{27 {c_3}^{6} {\pi^\prime}^{12} {\rho_m}^{3}}{16 {c_2}^{3} {M}^{18} {\Mpl}^{12} {\alpha}^{3} {\mathcal{H}}^{4} {a}^{10}} -  \frac{29 {c_3}^{5} {\pi^\prime}^{11} {\rho_m}^{2} {\alpha}^{\prime}}{2 {c_2}^{2} {M}^{15} {\Mpl}^{10} {\alpha}^{4} {\mathcal{H}}^{4} {a}^{10}} + \frac{81 {c_3}^{5} {\pi^\prime}^{11} {\rho_m}^{2} {\alpha}^{\prime}}{8 {c_2}^{2} {M}^{15} {\Mpl}^{10} {\alpha}^{3} {\mathcal{H}}^{4} {a}^{10}} \nonumber \\ 
& + \frac{3 {c_3}^{5} {\pi^\prime}^{9} {\alpha}^{\prime\prime} {\rho_m}^{2} {\alpha}^{\prime}}{2 {c_2}^{3} {M}^{15} {\Mpl}^{8} {\alpha}^{5} {\mathcal{H}}^{4} {a}^{10}} -  \frac{3 {c_3}^{5} {\pi^\prime}^{9} {\rho_m}^{2} {{\alpha}^{\prime}}^{3}}{{c_2}^{3} {M}^{15} {\Mpl}^{8} {\alpha}^{6} {\mathcal{H}}^{4} {a}^{10}} + \frac{27 {c_3}^{5} {\pi^\prime}^{9} {\rho_m}^{2} {\alpha}^{\prime} {\mathcal{H}}^{\prime}}{4 {c_2}^{3} {M}^{15} {\Mpl}^{8} {\alpha}^{4} {\mathcal{H}}^{4} {a}^{10}} \nonumber \\ 
& -  \frac{1375 {c_3}^{5} {\pi^\prime}^{11} {\rho_m}^{2}}{16 {c_2}^{2} {M}^{15} {\Mpl}^{10} {\alpha}^{3} {\mathcal{H}}^{3} {a}^{10}} + \frac{1107 {c_3}^{5} {\pi^\prime}^{11} {\rho_m}^{2}}{16 {c_2}^{2} {M}^{15} {\Mpl}^{10} {\alpha}^{2} {\mathcal{H}}^{3} {a}^{10}} + \frac{27 {c_3}^{5} {\pi^\prime}^{9} {\alpha}^{\prime\prime} {\rho_m}^{2}}{4 {c_2}^{3} {M}^{15} {\Mpl}^{8} {\alpha}^{4} {\mathcal{H}}^{3} {a}^{10}}\nonumber \\ 
& + \frac{3 {c_3}^{6} {\pi^\prime}^{10} {\rho_m}^{3} {\alpha}^{\prime}}{8 {c_2}^{4} {M}^{18} {\Mpl}^{10} {\alpha}^{5} {\mathcal{H}}^{3} {a}^{10}} -  \frac{243 {c_3}^{5} {\pi^\prime}^{9} {\rho_m}^{2} {{\alpha}^{\prime}}^{2}}{8 {c_2}^{3} {M}^{15} {\Mpl}^{8} {\alpha}^{5} {\mathcal{H}}^{3} {a}^{10}} + \frac{30 {c_3}^{5} {\pi^\prime}^{9} {\rho_m}^{2} {\mathcal{H}}^{\prime}}{{c_2}^{3} {M}^{15} {\Mpl}^{8} {\alpha}^{3} {\mathcal{H}}^{3} {a}^{10}} \nonumber \\ 
& + \frac{29 {c_3}^{5} {\pi^\prime}^{11} {\rho_m}^{2} f}{2 {c_2}^{2} {M}^{15} {\Mpl}^{10} {\alpha}^{3} {\mathcal{H}}^{3} {a}^{10}} -  \frac{261 {c_3}^{5} {\pi^\prime}^{11} {\rho_m}^{2} f}{16 {c_2}^{2} {M}^{15} {\Mpl}^{10} {\alpha}^{2} {\mathcal{H}}^{3} {a}^{10}} -  \frac{3 {c_3}^{5} {\pi^\prime}^{9} {\alpha}^{\prime\prime} {\rho_m}^{2} f}{2 {c_2}^{3} {M}^{15} {\Mpl}^{8} {\alpha}^{4} {\mathcal{H}}^{3} {a}^{10}} \nonumber \\ 
& + \frac{6 {c_3}^{5} {\pi^\prime}^{9} {\rho_m}^{2} {{\alpha}^{\prime}}^{2} f}{{c_2}^{3} {M}^{15} {\Mpl}^{8} {\alpha}^{5} {\mathcal{H}}^{3} {a}^{10}}-  \frac{27 {c_3}^{5} {\pi^\prime}^{9} {\rho_m}^{2} {\mathcal{H}}^{\prime} f}{4 {c_2}^{3} {M}^{15} {\Mpl}^{8} {\alpha}^{3} {\mathcal{H}}^{3} {a}^{10}} + \frac{15 {c_3}^{6} {\pi^\prime}^{10} {\rho_m}^{3}}{8 {c_2}^{4} {M}^{18} {\Mpl}^{10} {\alpha}^{4} {\mathcal{H}}^{2} {a}^{10}} \nonumber \\ 
& -  \frac{114 {c_3}^{5} {\pi^\prime}^{9} {\rho_m}^{2} {\alpha}^{\prime}}{{c_2}^{3} {M}^{15} {\Mpl}^{8} {\alpha}^{4} {\mathcal{H}}^{2} {a}^{10}} -  \frac{3 {c_3}^{6} {\pi^\prime}^{10} {\rho_m}^{3} f}{8 {c_2}^{4} {M}^{18} {\Mpl}^{10} {\alpha}^{4} {\mathcal{H}}^{2} {a}^{10}} + \frac{195 {c_3}^{5} {\pi^\prime}^{9} {\rho_m}^{2} {\alpha}^{\prime} f}{4 {c_2}^{3} {M}^{15} {\Mpl}^{8} {\alpha}^{4} {\mathcal{H}}^{2} {a}^{10}} \nonumber \\ 
& -  \frac{1335 {c_3}^{5} {\pi^\prime}^{9} {\rho_m}^{2}}{8 {c_2}^{3} {M}^{15} {\Mpl}^{8} {\alpha}^{3} \mathcal{H} {a}^{10}} + \frac{483 {c_3}^{5} {\pi^\prime}^{9} {\rho_m}^{2} f}{4 {c_2}^{3} {M}^{15} {\Mpl}^{8} {\alpha}^{3} \mathcal{H} {a}^{10}} -  \frac{147 {c_3}^{5} {\pi^\prime}^{9} {\rho_m}^{2} {f}^{2}}{8 {c_2}^{3} {M}^{15} {\Mpl}^{8} {\alpha}^{3} \mathcal{H} {a}^{10}} \nonumber \\ 
& -  \frac{375 {c_3}^{4} {\pi^\prime}^{6} {\rho_m}^{2}}{2 {c_2}^{3} {M}^{12} {\Mpl}^{6} {\alpha}^{3} {a}^{8}} + \frac{120 {c_3}^{4} {\pi^\prime}^{6} {\rho_m}^{2} f}{{c_2}^{3} {M}^{12} {\Mpl}^{6} {\alpha}^{3} {a}^{8}} -  \frac{33 {c_3}^{4} {\pi^\prime}^{6} {\rho_m}^{2} {f}^{2}}{2 {c_2}^{3} {M}^{12} {\Mpl}^{6} {\alpha}^{3} {a}^{8}} \nonumber \\ 
& + \frac{175 {c_3}^{4} {\pi^\prime}^{10} {\rho_m}^{2}}{16 c_2 {M}^{12} {\Mpl}^{10} {\alpha}^{3} {\mathcal{H}}^{4} {a}^{8}} -  \frac{601 {c_3}^{4} {\pi^\prime}^{10} {\rho_m}^{2}}{32 c_2 {M}^{12} {\Mpl}^{10} {\alpha}^{2} {\mathcal{H}}^{4} {a}^{8}} + \frac{39 {c_3}^{4} {\pi^\prime}^{10} {\rho_m}^{2}}{8 c_2 {M}^{12} {\Mpl}^{10} \alpha {\mathcal{H}}^{4} {a}^{8}} \nonumber \\ 
& -  \frac{2 {c_3}^{4} {\pi^\prime}^{8} {\alpha}^{\prime\prime} {\rho_m}^{2}}{{c_2}^{2} {M}^{12} {\Mpl}^{8} {\alpha}^{4} {\mathcal{H}}^{4} {a}^{8}} + \frac{9 {c_3}^{4} {\pi^\prime}^{8} {\alpha}^{\prime\prime} {\rho_m}^{2}}{4 {c_2}^{2} {M}^{12} {\Mpl}^{8} {\alpha}^{3} {\mathcal{H}}^{4} {a}^{8}} -  \frac{3 {c_3}^{5} {\pi^\prime}^{9} {\rho_m}^{2} {\alpha}^{\prime} {f}^{2}}{{c_2}^{3} {M}^{15} {\Mpl}^{8} {\alpha}^{4} {\mathcal{H}}^{2} {a}^{10}} \nonumber \\ 
& -  \frac{9 {c_3}^{5} {\pi^\prime}^{9} {\rho_m}^{3} {\alpha}^{\prime}}{8 {c_2}^{3} {M}^{15} {\Mpl}^{10} {\alpha}^{4} {\mathcal{H}}^{4} {a}^{8}} + \frac{135 {c_3}^{4} {\pi^\prime}^{8} {\rho_m}^{2} {{\alpha}^{\prime}}^{2}}{16 {c_2}^{2} {M}^{12} {\Mpl}^{8} {\alpha}^{5} {\mathcal{H}}^{4} {a}^{8}} -  \frac{87 {c_3}^{4} {\pi^\prime}^{8} {\rho_m}^{2} {{\alpha}^{\prime}}^{2}}{16 {c_2}^{2} {M}^{12} {\Mpl}^{8} {\alpha}^{4} {\mathcal{H}}^{4} {a}^{8}} \nonumber \\ 
& -  \frac{37 {c_3}^{4} {\pi^\prime}^{8} {\rho_m}^{2} {\mathcal{H}}^{\prime}}{4 {c_2}^{2} {M}^{12} {\Mpl}^{8} {\alpha}^{3} {\mathcal{H}}^{4} {a}^{8}} + \frac{3 {c_3}^{4} {\pi^\prime}^{6} {\rho_m}^{2} {{\alpha}^{\prime}}^{2} {\mathcal{H}}^{\prime}}{4 {c_2}^{3} {M}^{12} {\Mpl}^{6} {\alpha}^{5} {\mathcal{H}}^{4} {a}^{8}} -  \frac{3 {c_3}^{5} {\pi^\prime}^{9} {\rho_m}^{3}}{8 {c_2}^{3} {M}^{15} {\Mpl}^{10} {\alpha}^{4} {\mathcal{H}}^{3} {a}^{8}} \nonumber \\ 
& -  \frac{15 {c_3}^{5} {\pi^\prime}^{9} {\rho_m}^{3}}{4 {c_2}^{3} {M}^{15} {\Mpl}^{10} {\alpha}^{3} {\mathcal{H}}^{3} {a}^{8}} + \frac{479 {c_3}^{4} {\pi^\prime}^{8} {\rho_m}^{2} {\alpha}^{\prime}}{8 {c_2}^{2} {M}^{12} {\Mpl}^{8} {\alpha}^{4} {\mathcal{H}}^{3} {a}^{8}} -  \frac{33 {c_3}^{4} {\pi^\prime}^{8} {\rho_m}^{2} {\alpha}^{\prime}}{{c_2}^{2} {M}^{12} {\Mpl}^{8} {\alpha}^{3} {\mathcal{H}}^{3} {a}^{8}} \nonumber \\ 
& + \frac{3 {c_3}^{4} {\pi^\prime}^{6} {\alpha}^{\prime\prime} {\rho_m}^{2} {\alpha}^{\prime}}{2 {c_2}^{3} {M}^{12} {\Mpl}^{6} {\alpha}^{5} {\mathcal{H}}^{3} {a}^{8}}-  \frac{3 {c_3}^{4} {\pi^\prime}^{6} {\rho_m}^{2} {{\alpha}^{\prime}}^{3}}{{c_2}^{3} {M}^{12} {\Mpl}^{6} {\alpha}^{6} {\mathcal{H}}^{3} {a}^{8}} + \frac{15 {c_3}^{4} {\pi^\prime}^{6} {\rho_m}^{2} {\alpha}^{\prime} {\mathcal{H}}^{\prime}}{{c_2}^{3} {M}^{12} {\Mpl}^{6} {\alpha}^{4} {\mathcal{H}}^{3} {a}^{8}} \nonumber \\ 
& + \frac{3 {c_3}^{5} {\pi^\prime}^{9} {\rho_m}^{3} f}{2 {c_2}^{3} {M}^{15} {\Mpl}^{10} {\alpha}^{3} {\mathcal{H}}^{3} {a}^{8}} + \frac{51 {c_3}^{4} {\pi^\prime}^{8} {\rho_m}^{2} {\alpha}^{\prime} f}{4 {c_2}^{2} {M}^{12} {\Mpl}^{8} {\alpha}^{3} {\mathcal{H}}^{3} {a}^{8}} -  \frac{3 {c_3}^{4} {\pi^\prime}^{6} {\rho_m}^{2} {\alpha}^{\prime} {\mathcal{H}}^{\prime} f}{2 {c_2}^{3} {M}^{12} {\Mpl}^{6} {\alpha}^{4} {\mathcal{H}}^{3} {a}^{8}} \nonumber \\ 
& + \frac{2095 {c_3}^{4} {\pi^\prime}^{8} {\rho_m}^{2}}{16 {c_2}^{2} {M}^{12} {\Mpl}^{8} {\alpha}^{3} {\mathcal{H}}^{2} {a}^{8}} -  \frac{189 {c_3}^{4} {\pi^\prime}^{8} {\rho_m}^{2}}{16 {c_2}^{2} {M}^{12} {\Mpl}^{8} {\alpha}^{2} {\mathcal{H}}^{2} {a}^{8}} + \frac{15 {c_3}^{4} {\pi^\prime}^{6} {\alpha}^{\prime\prime} {\rho_m}^{2}}{2 {c_2}^{3} {M}^{12} {\Mpl}^{6} {\alpha}^{4} {\mathcal{H}}^{2} {a}^{8}} \nonumber \\ 
& -  \frac{30 {c_3}^{4} {\pi^\prime}^{6} {\rho_m}^{2} {{\alpha}^{\prime}}^{2}}{{c_2}^{3} {M}^{12} {\Mpl}^{6} {\alpha}^{5} {\mathcal{H}}^{2} {a}^{8}} + \frac{225 {c_3}^{4} {\pi^\prime}^{6} {\rho_m}^{2} {\mathcal{H}}^{\prime}}{4 {c_2}^{3} {M}^{12} {\Mpl}^{6} {\alpha}^{3} {\mathcal{H}}^{2} {a}^{8}} -  \frac{103 {c_3}^{4} {\pi^\prime}^{8} {\rho_m}^{2} {\alpha}^{\prime} f}{8 {c_2}^{2} {M}^{12} {\Mpl}^{8} {\alpha}^{4} {\mathcal{H}}^{3} {a}^{8}}\nonumber \\ 
& -  \frac{495 {c_3}^{4} {\pi^\prime}^{8} {\rho_m}^{2} f}{8 {c_2}^{2} {M}^{12} {\Mpl}^{8} {\alpha}^{3} {\mathcal{H}}^{2} {a}^{8}} + \frac{615 {c_3}^{4} {\pi^\prime}^{8} {\rho_m}^{2} f}{8 {c_2}^{2} {M}^{12} {\Mpl}^{8} {\alpha}^{2} {\mathcal{H}}^{2} {a}^{8}} -  \frac{3 {c_3}^{4} {\pi^\prime}^{6} {\alpha}^{\prime\prime} {\rho_m}^{2} f}{2 {c_2}^{3} {M}^{12} {\Mpl}^{6} {\alpha}^{4} {\mathcal{H}}^{2} {a}^{8}} \nonumber \\ 
& + \frac{6 {c_3}^{4} {\pi^\prime}^{6} {\rho_m}^{2} {{\alpha}^{\prime}}^{2} f}{{c_2}^{3} {M}^{12} {\Mpl}^{6} {\alpha}^{5} {\mathcal{H}}^{2} {a}^{8}} -  \frac{15 {c_3}^{4} {\pi^\prime}^{6} {\rho_m}^{2} {\mathcal{H}}^{\prime} f}{{c_2}^{3} {M}^{12} {\Mpl}^{6} {\alpha}^{3} {\mathcal{H}}^{2} {a}^{8}} + \frac{71 {c_3}^{4} {\pi^\prime}^{8} {\rho_m}^{2} {f}^{2}}{16 {c_2}^{2} {M}^{12} {\Mpl}^{8} {\alpha}^{3} {\mathcal{H}}^{2} {a}^{8}}\nonumber \\ 
& -  \frac{117 {c_3}^{4} {\pi^\prime}^{8} {\rho_m}^{2} {f}^{2}}{16 {c_2}^{2} {M}^{12} {\Mpl}^{8} {\alpha}^{2} {\mathcal{H}}^{2} {a}^{8}} + \frac{3 {c_3}^{4} {\pi^\prime}^{6} {\rho_m}^{2} {\mathcal{H}}^{\prime} {f}^{2}}{4 {c_2}^{3} {M}^{12} {\Mpl}^{6} {\alpha}^{3} {\mathcal{H}}^{2} {a}^{8}} -  \frac{225 {c_3}^{4} {\pi^\prime}^{6} {\rho_m}^{2} {\alpha}^{\prime}}{2 {c_2}^{3} {M}^{12} {\Mpl}^{6} {\alpha}^{4} \mathcal{H} {a}^{8}} \nonumber \\ 
& + \frac{5 {c_3}^{4} {\pi^\prime}^{8} {\rho_m}^{3}}{4 {c_2}^{2} {M}^{12} {\Mpl}^{10} {\alpha}^{3} {\mathcal{H}}^{4} {a}^{6}} -  \frac{15 {c_3}^{4} {\pi^\prime}^{8} {\rho_m}^{3}}{8 {c_2}^{2} {M}^{12} {\Mpl}^{10} {\alpha}^{2} {\mathcal{H}}^{4} {a}^{6}} -  \frac{63 {c_3}^{3} {\pi^\prime}^{7} {\rho_m}^{2} {\alpha}^{\prime}}{8 c_2 {M}^{9} {\Mpl}^{8} {\alpha}^{4} {\mathcal{H}}^{4} {a}^{6}} \nonumber \\ 
& + \frac{93 {c_3}^{4} {\pi^\prime}^{6} {\rho_m}^{2} {\alpha}^{\prime} f}{2 {c_2}^{3} {M}^{12} {\Mpl}^{6} {\alpha}^{4} \mathcal{H} {a}^{8}} + \frac{77 {c_3}^{3} {\pi^\prime}^{7} {\rho_m}^{2} {\alpha}^{\prime}}{8 c_2 {M}^{9} {\Mpl}^{8} {\alpha}^{3} {\mathcal{H}}^{4} {a}^{6}} + \frac{15 {c_3}^{3} {\pi^\prime}^{7} {\rho_m}^{2} {\alpha}^{\prime}}{8 c_2 {M}^{9} {\Mpl}^{8} {\alpha}^{2} {\mathcal{H}}^{4} {a}^{6}} \nonumber \\ 
& -  \frac{3 {c_3}^{4} {\pi^\prime}^{6} {\rho_m}^{2} {\alpha}^{\prime} {f}^{2}}{{c_2}^{3} {M}^{12} {\Mpl}^{6} {\alpha}^{4} \mathcal{H} {a}^{8}} -  \frac{3 {c_3}^{3} {\pi^\prime}^{5} {\rho_m}^{2} {\alpha}^{\prime} {\mathcal{H}}^{\prime}}{2 {c_2}^{2} {M}^{9} {\Mpl}^{6} {\alpha}^{4} {\mathcal{H}}^{4} {a}^{6}} + \frac{6 {c_3}^{3} {\pi^\prime}^{5} {\rho_m}^{2} {\alpha}^{\prime} {\mathcal{H}}^{\prime}}{{c_2}^{2} {M}^{9} {\Mpl}^{6} {\alpha}^{3} {\mathcal{H}}^{4} {a}^{6}} \nonumber \\ 
& -  \frac{261 {c_3}^{3} {\pi^\prime}^{7} {\rho_m}^{2}}{8 c_2 {M}^{9} {\Mpl}^{8} {\alpha}^{3} {\mathcal{H}}^{3} {a}^{6}} + \frac{22 {c_3}^{3} {\pi^\prime}^{7} {\rho_m}^{2}}{c_2 {M}^{9} {\Mpl}^{8} {\alpha}^{2} {\mathcal{H}}^{3} {a}^{6}} + \frac{291 {c_3}^{3} {\pi^\prime}^{7} {\rho_m}^{2}}{8 c_2 {M}^{9} {\Mpl}^{8} \alpha {\mathcal{H}}^{3} {a}^{6}} \nonumber \\ 
& -  \frac{3 {c_3}^{3} {\pi^\prime}^{5} {\alpha}^{\prime\prime} {\rho_m}^{2}}{2 {c_2}^{2} {M}^{9} {\Mpl}^{6} {\alpha}^{4} {\mathcal{H}}^{3} {a}^{6}} -  \frac{3 {c_3}^{4} {\pi^\prime}^{6} {\rho_m}^{3} {\alpha}^{\prime}}{4 {c_2}^{3} {M}^{12} {\Mpl}^{8} {\alpha}^{4} {\mathcal{H}}^{3} {a}^{6}} + \frac{6 {c_3}^{3} {\pi^\prime}^{5} {\rho_m}^{2} {{\alpha}^{\prime}}^{2}}{{c_2}^{2} {M}^{9} {\Mpl}^{6} {\alpha}^{5} {\mathcal{H}}^{3} {a}^{6}} \nonumber \\ 
& -  \frac{45 {c_3}^{3} {\pi^\prime}^{5} {\rho_m}^{2} {{\alpha}^{\prime}}^{2}}{4 {c_2}^{2} {M}^{9} {\Mpl}^{6} {\alpha}^{4} {\mathcal{H}}^{3} {a}^{6}} -  \frac{15 {c_3}^{3} {\pi^\prime}^{5} {\rho_m}^{2} {\mathcal{H}}^{\prime}}{{c_2}^{2} {M}^{9} {\Mpl}^{6} {\alpha}^{3} {\mathcal{H}}^{3} {a}^{6}} + \frac{6 {c_3}^{3} {\pi^\prime}^{5} {\alpha}^{\prime\prime} {\rho_m}^{2}}{{c_2}^{2} {M}^{9} {\Mpl}^{6} {\alpha}^{3} {\mathcal{H}}^{3} {a}^{6}} \nonumber \\ 
& + \frac{60 {c_3}^{3} {\pi^\prime}^{5} {\rho_m}^{2} {\mathcal{H}}^{\prime}}{{c_2}^{2} {M}^{9} {\Mpl}^{6} {\alpha}^{2} {\mathcal{H}}^{3} {a}^{6}} + \frac{63 {c_3}^{3} {\pi^\prime}^{7} {\rho_m}^{2} f}{8 c_2 {M}^{9} {\Mpl}^{8} {\alpha}^{3} {\mathcal{H}}^{3} {a}^{6}} -  \frac{83 {c_3}^{3} {\pi^\prime}^{7} {\rho_m}^{2} f}{4 c_2 {M}^{9} {\Mpl}^{8} {\alpha}^{2} {\mathcal{H}}^{3} {a}^{6}} \nonumber \\ 
& + \frac{51 {c_3}^{3} {\pi^\prime}^{7} {\rho_m}^{2} f}{8 c_2 {M}^{9} {\Mpl}^{8} \alpha {\mathcal{H}}^{3} {a}^{6}} + \frac{3 {c_3}^{3} {\pi^\prime}^{5} {\alpha}^{\prime\prime} {\rho_m}^{2} f}{2 {c_2}^{2} {M}^{9} {\Mpl}^{6} {\alpha}^{3} {\mathcal{H}}^{3} {a}^{6}} -  \frac{3 {c_3}^{3} {\pi^\prime}^{5} {\rho_m}^{2} {{\alpha}^{\prime}}^{2} f}{{c_2}^{2} {M}^{9} {\Mpl}^{6} {\alpha}^{4} {\mathcal{H}}^{3} {a}^{6}}\nonumber \\ 
& + \frac{3 {c_3}^{3} {\pi^\prime}^{5} {\rho_m}^{2} {\mathcal{H}}^{\prime} f}{2 {c_2}^{2} {M}^{9} {\Mpl}^{6} {\alpha}^{3} {\mathcal{H}}^{3} {a}^{6}}+ \frac{3 {c_3}^{3} {\pi^\prime}^{5} {\rho_m}^{2} {\mathcal{H}}^{\prime} f}{2 {c_2}^{2} {M}^{9} {\Mpl}^{6} {\alpha}^{2} {\mathcal{H}}^{3} {a}^{6}} -  \frac{15 {c_3}^{4} {\pi^\prime}^{6} {\rho_m}^{3}}{4 {c_2}^{3} {M}^{12} {\Mpl}^{8} {\alpha}^{3} {\mathcal{H}}^{2} {a}^{6}} \nonumber \\ 
& + \frac{81 {c_3}^{3} {\pi^\prime}^{5} {\rho_m}^{2} {\alpha}^{\prime}}{2 {c_2}^{2} {M}^{9} {\Mpl}^{6} {\alpha}^{4} {\mathcal{H}}^{2} {a}^{6}} -  \frac{135 {c_3}^{3} {\pi^\prime}^{5} {\rho_m}^{2} {\alpha}^{\prime}}{2 {c_2}^{2} {M}^{9} {\Mpl}^{6} {\alpha}^{3} {\mathcal{H}}^{2} {a}^{6}} + \frac{3 {c_3}^{4} {\pi^\prime}^{6} {\rho_m}^{3} f}{4 {c_2}^{3} {M}^{12} {\Mpl}^{8} {\alpha}^{3} {\mathcal{H}}^{2} {a}^{6}} \nonumber \\ 
& + \frac{3 {c_3}^{3} {\pi^\prime}^{5} {\rho_m}^{2} {\alpha}^{\prime} {f}^{2}}{{c_2}^{2} {M}^{9} {\Mpl}^{6} {\alpha}^{3} {\mathcal{H}}^{2} {a}^{6}} + \frac{90 {c_3}^{3} {\pi^\prime}^{5} {\rho_m}^{2}}{{c_2}^{2} {M}^{9} {\Mpl}^{6} {\alpha}^{3} \mathcal{H} {a}^{6}} -  \frac{825 {c_3}^{3} {\pi^\prime}^{5} {\rho_m}^{2}}{4 {c_2}^{2} {M}^{9} {\Mpl}^{6} {\alpha}^{2} \mathcal{H} {a}^{6}} \nonumber \\ 
& -  \frac{42 {c_3}^{3} {\pi^\prime}^{5} {\rho_m}^{2} f}{{c_2}^{2} {M}^{9} {\Mpl}^{6} {\alpha}^{3} \mathcal{H} {a}^{6}} + \frac{117 {c_3}^{3} {\pi^\prime}^{5} {\rho_m}^{2} f}{2 {c_2}^{2} {M}^{9} {\Mpl}^{6} {\alpha}^{2} \mathcal{H} {a}^{6}} + \frac{3 {c_3}^{3} {\pi^\prime}^{5} {\rho_m}^{2} {f}^{2}}{{c_2}^{2} {M}^{9} {\Mpl}^{6} {\alpha}^{3} \mathcal{H} {a}^{6}} \nonumber \\ 
& + \frac{51 {c_3}^{3} {\pi^\prime}^{5} {\rho_m}^{2} {f}^{2}}{4 {c_2}^{2} {M}^{9} {\Mpl}^{6} {\alpha}^{2} \mathcal{H} {a}^{6}} + \frac{33 {c_3}^{2} {\pi^\prime}^{6} {\rho_m}^{2}}{4 {M}^{6} {\Mpl}^{8} {\mathcal{H}}^{4} {a}^{4}} + \frac{43 {c_3}^{2} {\pi^\prime}^{6} {\rho_m}^{2}}{16 {M}^{6} {\Mpl}^{8} {\alpha}^{3} {\mathcal{H}}^{4} {a}^{4}} \nonumber \\ 
& -  \frac{9 {c_3}^{2} {\pi^\prime}^{6} {\rho_m}^{2}}{2 {M}^{6} {\Mpl}^{8} {\alpha}^{2} {\mathcal{H}}^{4} {a}^{4}} -  \frac{91 {c_3}^{2} {\pi^\prime}^{6} {\rho_m}^{2}}{16 {M}^{6} {\Mpl}^{8} \alpha {\mathcal{H}}^{4} {a}^{4}} -  \frac{{c_3}^{2} {\pi^\prime}^{4} {\alpha}^{\prime\prime} {\rho_m}^{2}}{2 c_2 {M}^{6} {\Mpl}^{6} {\alpha}^{3} {\mathcal{H}}^{4} {a}^{4}} \nonumber \\ 
& + \frac{3 {c_3}^{2} {\pi^\prime}^{4} {\alpha}^{\prime\prime} {\rho_m}^{2}}{2 c_2 {M}^{6} {\Mpl}^{6} {\alpha}^{2} {\mathcal{H}}^{4} {a}^{4}} + \frac{3 {c_3}^{3} {\pi^\prime}^{5} {\rho_m}^{3} {\alpha}^{\prime}}{4 {c_2}^{2} {M}^{9} {\Mpl}^{8} {\alpha}^{3} {\mathcal{H}}^{4} {a}^{4}} -  \frac{9 {c_3}^{3} {\pi^\prime}^{5} {\rho_m}^{2} {\alpha}^{\prime} f}{{c_2}^{2} {M}^{9} {\Mpl}^{6} {\alpha}^{4} {\mathcal{H}}^{2} {a}^{6}} \nonumber \\ 
& + \frac{7 {c_3}^{2} {\pi^\prime}^{4} {\rho_m}^{2} {{\alpha}^{\prime}}^{2}}{8 c_2 {M}^{6} {\Mpl}^{6} {\alpha}^{4} {\mathcal{H}}^{4} {a}^{4}} -  \frac{3 {c_3}^{2} {\pi^\prime}^{4} {\rho_m}^{2} {{\alpha}^{\prime}}^{2}}{c_2 {M}^{6} {\Mpl}^{6} {\alpha}^{3} {\mathcal{H}}^{4} {a}^{4}} -  \frac{21 c_3 {\pi^\prime}^{3} {\rho_m}^{2} f}{2 {M}^{3} {\Mpl}^{6} \alpha {\mathcal{H}}^{3} {a}^{2}}\nonumber \\ 
& + \frac{3 {c_3}^{2} {\pi^\prime}^{4} {\rho_m}^{2} {\mathcal{H}}^{\prime}}{4 c_2 {M}^{6} {\Mpl}^{6} {\alpha}^{3} {\mathcal{H}}^{4} {a}^{4}} -  \frac{17 {c_3}^{2} {\pi^\prime}^{4} {\rho_m}^{2} {\mathcal{H}}^{\prime}}{2 c_2 {M}^{6} {\Mpl}^{6} {\alpha}^{2} {\mathcal{H}}^{4} {a}^{4}} + \frac{c_3 {\pi^\prime}^{3} {\rho_m}^{2} f}{4 {M}^{3} {\Mpl}^{6} {\alpha}^{2} {\mathcal{H}}^{3} {a}^{2}}\nonumber \\ 
& + \frac{111 {c_3}^{2} {\pi^\prime}^{4} {\rho_m}^{2} {\mathcal{H}}^{\prime}}{4 c_2 {M}^{6} {\Mpl}^{6} \alpha {\mathcal{H}}^{4} {a}^{4}} + \frac{3 {c_3}^{3} {\pi^\prime}^{5} {\rho_m}^{3}}{4 {c_2}^{2} {M}^{9} {\Mpl}^{8} {\alpha}^{3} {\mathcal{H}}^{3} {a}^{4}} + \frac{3 {c_3}^{3} {\pi^\prime}^{5} {\rho_m}^{3}}{4 {c_2}^{2} {M}^{9} {\Mpl}^{8} {\alpha}^{2} {\mathcal{H}}^{3} {a}^{4}} \nonumber \\ 
& -  \frac{15 {c_3}^{2} {\pi^\prime}^{4} {\rho_m}^{2} {\alpha}^{\prime}}{4 c_2 {M}^{6} {\Mpl}^{6} {\alpha}^{4} {\mathcal{H}}^{3} {a}^{4}} + \frac{33 {c_3}^{2} {\pi^\prime}^{4} {\rho_m}^{2} {\alpha}^{\prime}}{2 c_2 {M}^{6} {\Mpl}^{6} {\alpha}^{3} {\mathcal{H}}^{3} {a}^{4}}  + \frac{45 c_3 {\pi^\prime}^{3} {\rho_m}^{2} f}{4 {M}^{3} {\Mpl}^{6} {\mathcal{H}}^{3} {a}^{2}}\nonumber \\ 
& -  \frac{51 {c_3}^{2} {\pi^\prime}^{4} {\rho_m}^{2} {\alpha}^{\prime}}{2 c_2 {M}^{6} {\Mpl}^{6} {\alpha}^{2} {\mathcal{H}}^{3} {a}^{4}} -  \frac{3 {c_3}^{3} {\pi^\prime}^{5} {\rho_m}^{3} f}{2 {c_2}^{2} {M}^{9} {\Mpl}^{8} {\alpha}^{2} {\mathcal{H}}^{3} {a}^{4}} + \frac{2 {c_3}^{2} {\pi^\prime}^{4} {\rho_m}^{2} {\alpha}^{\prime} f}{c_2 {M}^{6} {\Mpl}^{6} {\alpha}^{3} {\mathcal{H}}^{3} {a}^{4}} \nonumber \\ 
& + \frac{21 {c_3}^{2} {\pi^\prime}^{4} {\rho_m}^{2} {\alpha}^{\prime} f}{4 c_2 {M}^{6} {\Mpl}^{6} {\alpha}^{2} {\mathcal{H}}^{3} {a}^{4}} -  \frac{57 {c_3}^{2} {\pi^\prime}^{4} {\rho_m}^{2}}{4 c_2 {M}^{6} {\Mpl}^{6} {\alpha}^{3} {\mathcal{H}}^{2} {a}^{4}} + \frac{561 {c_3}^{2} {\pi^\prime}^{4} {\rho_m}^{2}}{8 c_2 {M}^{6} {\Mpl}^{6} {\alpha}^{2} {\mathcal{H}}^{2} {a}^{4}} \nonumber \\ 
& -  \frac{1317 {c_3}^{2} {\pi^\prime}^{4} {\rho_m}^{2}}{8 c_2 {M}^{6} {\Mpl}^{6} \alpha {\mathcal{H}}^{2} {a}^{4}} + \frac{15 {c_3}^{2} {\pi^\prime}^{4} {\rho_m}^{2} f}{4 c_2 {M}^{6} {\Mpl}^{6} {\alpha}^{3} {\mathcal{H}}^{2} {a}^{4}} -  \frac{49 {c_3}^{2} {\pi^\prime}^{4} {\rho_m}^{2} f}{4 c_2 {M}^{6} {\Mpl}^{6} {\alpha}^{2} {\mathcal{H}}^{2} {a}^{4}} \nonumber \\ 
& + \frac{60 {c_3}^{2} {\pi^\prime}^{4} {\rho_m}^{2} f}{c_2 {M}^{6} {\Mpl}^{6} \alpha {\mathcal{H}}^{2} {a}^{4}} -  \frac{23 {c_3}^{2} {\pi^\prime}^{4} {\rho_m}^{2} {f}^{2}}{8 c_2 {M}^{6} {\Mpl}^{6} {\alpha}^{2} {\mathcal{H}}^{2} {a}^{4}} + \frac{21 {c_3}^{2} {\pi^\prime}^{4} {\rho_m}^{2} {f}^{2}}{8 c_2 {M}^{6} {\Mpl}^{6} \alpha {\mathcal{H}}^{2} {a}^{4}} \nonumber \\ 
& -  \frac{{c_3}^{2} {\pi^\prime}^{4} {\rho_m}^{3}}{2 c_2 {M}^{6} {\Mpl}^{8} {\alpha}^{2} {\mathcal{H}}^{4} {a}^{2}} -  \frac{3 {c_3}^{2} {\pi^\prime}^{4} {\rho_m}^{3}}{4 c_2 {M}^{6} {\Mpl}^{8} \alpha {\mathcal{H}}^{4} {a}^{2}} -  \frac{3 c_3 {\pi^\prime}^{3} {\rho_m}^{2} {\alpha}^{\prime}}{4 {M}^{3} {\Mpl}^{6} {\alpha}^{3} {\mathcal{H}}^{4} {a}^{2}} \nonumber \\ 
& + \frac{23 c_3 {\pi^\prime}^{3} {\rho_m}^{2} {\alpha}^{\prime}}{4 {M}^{3} {\Mpl}^{6} {\alpha}^{2} {\mathcal{H}}^{4} {a}^{2}} -  \frac{3 c_3 {\pi^\prime}^{3} {\rho_m}^{2} {\alpha}^{\prime}}{{M}^{3} {\Mpl}^{6} \alpha {\mathcal{H}}^{4} {a}^{2}} -  \frac{171 c_3 {\pi^\prime}^{3} {\rho_m}^{2}}{4 {M}^{3} {\Mpl}^{6} {\mathcal{H}}^{3} {a}^{2}} + \frac{3 {\rho_m}^{3} {a}^{2}}{2 {\Mpl}^{6} {\mathcal{H}}^{4}}\nonumber \\ 
&+ \frac{3 c_3 {\pi^\prime}^{3} {\rho_m}^{2}}{4 {M}^{3} {\Mpl}^{6} {\alpha}^{3} {\mathcal{H}}^{3} {a}^{2}} -  \frac{27 c_3 {\pi^\prime}^{3} {\rho_m}^{2}}{4 {M}^{3} {\Mpl}^{6} {\alpha}^{2} {\mathcal{H}}^{3} {a}^{2}} + \frac{215 c_3 {\pi^\prime}^{3} {\rho_m}^{2}}{4 {M}^{3} {\Mpl}^{6} \alpha {\mathcal{H}}^{3} {a}^{2}}\,.
\end{flalign}

\end{document}